\documentclass[12pt]{article}
\pdfoutput=1
\usepackage{eurosym}
\usepackage{natbib}
\usepackage{amsmath}
\usepackage{amssymb}
\usepackage{mathtools}
\usepackage{dutchcal}
\usepackage[scr=rsfs]{mathalpha}
\usepackage{bm}
\usepackage{graphicx}
\usepackage[singlespacing]{setspace}
\usepackage[bottom,multiple]{footmisc}
\usepackage[justification=centering]{caption}
\usepackage{subcaption}
\usepackage{setspace}
\usepackage{color}
\usepackage{ragged2e}
\usepackage{bigstrut}
\usepackage{lscape}
\usepackage{arydshln}
\usepackage{booktabs}                                         
\usepackage{multirow}                                          
\usepackage[dvipsnames]{xcolor}
\usepackage{comment}
\usepackage[normalem]{ulem}                              
\usepackage{enumitem}
\definecolor{cranberry}{HTML}{DC143C}

\RequirePackage[breaklinks, colorlinks,citecolor=Mahogany,urlcolor=gray,linkcolor=cranberry]{hyperref}

\newtheorem{lemma}{Lemma}

\newtheorem{proposition}{Proposition}
\newtheorem{corollary}{Corollary}

\newenvironment{proof}[1][Proof]{\begin{trivlist}
\item[\hskip \labelsep {\bfseries #1}]}{\end{trivlist}}
\newtheorem{assumption}{Assumption}

\newcommand{\qed}{\nobreak \ifvmode \relax \else
      \ifdim\lastskip<1.5em \hskip-\lastskip
      \hskip1.5em plus0em minus0.5em \fi \nobreak
      \vrule height0.3em width0.5em depth0.25em\fi}
\newcommand{\vm}{\vspace{.1in}}

\def\toprule{\hline\hline}
\def\midrule{\hline}
\def\bottomrule{\hline\hline}
\def\tablenotes{\\ \noindent \justifying \small}

\def\toprule{\hline\hline\addlinespace[0.2cm]}
\def\midrule{\addlinespace[0.1cm]\hline\addlinespace[0.1cm]}
\def\bottomrule{\addlinespace[0.2cm]\hline\hline}
\def\tablenotes{\\ \vm \noindent \justifying \footnotesize Notes: }

\renewcommand{\today}{\ifcase \month \or January \or February \or March \or %
April \or May \or June \or July \or August \or September \or October \or November \or %
December \fi \number \year}

\usepackage{titling}

\begin{document}

\title{A Price-Regression Test of Collusive Conduct\thanks{This paper previously circulated under the title ``An Instrumental Variables Approach to Testing Firm Conduct.'' We thank Victor Aguiar, Jeff Gortmaker, Alex Gross, Jinsoo Han, Hiro Kasahara, Dongwoo Kim, Sungwon Lee, Aviv Nevo, Paul Schrimpf, Kevin Song, and Frank Verboven for their valuable comments. We are also grateful to seminar participants at the University of British Columbia, University of Montreal, Simon Fraser University, UC Riverside, Yonsei University, the 2025 Annual Meeting of Korea's Allied Economic Association, ASSA 2026 Annual Meeting, 2025 IIOC, and 2026 KASIO Antitrust Economics Conference. This work was supported by the Ministry of Education of the Republic of Korea and the National Research Foundation (NRF) of Korea (NRF-2025S1A5A8008930), the Research Grant Program funded by the Korea Fair Trade Mediation Agency, and the BK21 FOUR Program funded by the NRF.  All authors equally contributed to this paper.}}
\author{Youngjin Hong\thanks{%
Department of Economics, University of Michigan, Lorch Hall 611 Tappan Ave., Ann Arbor, MI 48109, E-mail: \texttt{yjinhong@umich.edu}.} \and
In Kyung Kim\thanks{%
Corresponding author, Department of Economics, Sogang University, 35 Baekbeom-ro Mapo-gu, Seoul 04107, South Korea, E-mail: \texttt{inkim@sogang.ac.kr}.} \and
Kyoo il Kim\thanks{%
Department of Economics, Michigan State University, 486 West Circle Drive, East Lansing, MI 48824, E-mail: \texttt{kyookim@msu.edu}.}}

\pagenumbering{roman}
\maketitle
\thispagestyle{empty}

\begin{abstract}

 We propose a test of competitive versus collusive conduct in differentiated-product markets where firms set prices. Because equilibrium prices respond differently to own and rival product characteristics across conduct, BLP-style instruments built under the correct conduct explain prices better. The test compares the fit of two price regressions, one with instruments built under competition and the other under collusion, via the Rivers-Vuong non-nested test, without estimating demand or supply. It is valid for general demand systems under collusion and, under logit demand, for either conduct when an observable market-share condition holds. Monte Carlo evidence and two applications demonstrate its performance.
 

\end{abstract}

\doublespacing
\clearpage

\section{Introduction}\label{sec: intro}
\pagenumbering{arabic}

 Understanding industry conduct, that is, the extent to which firms collude in a goods market, is one of the central issues in the study of industrial organization (IO). This issue also carries practical importance, as antitrust policies and regulations based on incorrect conclusions about industry conduct could harm consumers and reduce market efficiency. Yet conduct is not directly observed, and inferring it from data has required researchers to specify and estimate a model of demand and cost (e.g., \citealp{backus2021common}; \citealp{dearing2024learning}; \citealp{duarte2024testing}).
 
 In this article, we propose a new approach to testing competitive versus collusive conduct in differentiated product markets in which firms set prices. Our test builds on an observation of \citet[henceforth BLP]{Berry95} that Nash markups respond differently to the characteristics of own and rival products, so that instruments for price should distinguish between the two.\footnote{``[B]ecause Nash markups will respond differently to own and rival products, the optimal instruments will distinguish between the characteristics of products produced by the same multi-product firm versus the characteristics of products produced by rival firms'' \citep[p.~855]{Berry95}.} The logic of BLP extends to conduct: a firm's markup and price respond differently to changes in a rival's product characteristics depending on whether the two firms compete or collude. Therefore, the explanatory power of BLP-style instruments, which exploit the exchangeability of product characteristics within the firm and within the set of its rivals, hinges on whether the assumed model of firm conduct is correct. This dependence implies that we can infer whether firms compete or collude by comparing the explanatory power of instruments constructed under alternative conduct assumptions.

 We formalize this intuition by comparing the fit of two price regressions, each using one of two sets of BLP-style instruments: sums of the characteristics of the firm's other products (competition IVs), and sums that also include the characteristics of the products of \textit{suspected} colluding partners (collusion IVs). Assuming constant marginal costs and standard regularity conditions, we show that when firms collude, the pricing rule under any demand system treats the characteristics of partner products and those of the firm's own products symmetrically. Moreover, under random assignment of products to firms, this symmetry implies that own and partner characteristics receive identical coefficients in the price regression. Consequently, the collusion IVs provide a better fit whenever all firms belong to a suspected coalition, and, when only some do, as long as a firm does not alter its price response to its own products too sharply upon joining the coalition. When firms compete, the competition IVs fit better as long as the price response to a suspected partner product's characteristics is smaller in magnitude than an explicit fraction of the response to an own product's characteristics.

 Under logit demand, the conditions under which each set of IVs fits better under the conduct it presumes can be stated in terms of observable market shares. On the one hand, the collusion IVs fit better under collusion unless the colluding firms' combined market share exceeds one half while their products make up less than half of all products in the market. On the other hand, the competition IVs fit better under competition unless a suspected firm holds a dominant market share. Under more general demand systems, such as nested logit and random coefficients logit, these conditions can no longer be delineated in terms of market shares alone. Nevertheless, the collusion IVs continue to fit better under collusion given the conditions stated above, and the competition IVs continue to fit better under competition as long as each firm's share among the close substitutes for its products, such as its within-nest share under nested logit, is modest.
 
 We also argue that the test may fail when the assumption of random product assignment is violated: under collusion, when suspected partner products are systematically more distant substitutes for a product of the focal firm than the firm's own products are, and under competition, when they are systematically closer substitutes. The former failure is less consequential because, under weak substitutability between own and partner products, firms have little incentive to collude, and even if they do, collusion has little effect on prices. On the other hand, our testing framework can detect the latter failure, under which the coefficient on the collusion IVs takes the opposite sign from that on the competition IVs.

 Next, we construct the test statistic from the mean squared residuals of the two price regressions, following the non-nested model selection approach of \citet{vuong1989likelihood} and \citet{rivers2002model}, henceforth the RV test. A statistically significant positive (negative) value of the statistic provides evidence of collusion (competition) among the suspected firms. The primary advantage of our method over existing approaches built on equilibrium markup conditions is that it does not require estimating the demand and supply system, because it simply compares the fit of the two price regressions. We then evaluate the finite-sample performance of our proposed test statistic by conducting an extensive Monte Carlo study. 

 Our results show that the proposed test performs comparably to, or outperforms, the existing markup-based RV test in detecting collusion under various market conditions. Notably, the performance of our test is robust to endogenous product characteristics and missing cost shifters. Given its competitive testing power, minimal data requirements, and ease of implementation, our method serves as a practical tool for the preliminary diagnosis of firm conduct in differentiated product markets in which firms set prices. Once our test indicates collusion among suspected firms, researchers and regulatory authorities can proceed with existing approaches to determine which specific conduct model best fits the data. In this way, our test complements existing approaches.

 Our Monte Carlo study also reveals that, as expected, using collusion IVs instead of competition IVs improves the estimation performance of demand parameters under various collusive scenarios. This finding suggests that researchers can enhance the strength of BLP-style instruments by designing them to accurately reflect actual industry conduct. For instance, after finding suggestive evidence of collusion, researchers should incorporate the product characteristics of colluding partners into own-firm instruments while excluding them from other-firm instruments. This construction reflects the appropriate form of exchangeability in product characteristics implied by the correctly specified supply-side conduct.

 We apply our testing method to study firm conduct in two differentiated product markets in South Korea: the new passenger car market and the instant noodles market. Our test rejects the hypotheses of brand-level profit maximization and price collusion in favor of joint profit maximization at the parent company level in the car market. In addition, our results suggest that, despite the continued suspicions from the Korea Fair Trade Commission, instant noodle manufacturers did not engage in price collusion during the 2010s. Our findings align with anecdotal evidence of firm conduct in each market, court rulings, and previous empirical findings that instant noodle prices remain significantly below the collusive level \citep{kim2024collusion}.

 The convenience of our test comes at the potential cost of misleading conclusions if it is applied to settings in which firms do not set prices or marginal costs vary with output. Even when these assumptions hold, and beyond the conditions on market shares and product proximity discussed above, our test may fail in two further situations. First, if each firm owns a single product, then competition IVs cannot be constructed. Second, when firms share upstream input markets, a firm's marginal cost may depend on the characteristics of rival products, which generates a price response to rivals' characteristics even without collusion.\footnote{In such a case, alternative moment-based RV tests may also yield incorrect conclusions. If the cost function excludes the characteristics of other products while standard other-product characteristics are used as instruments, the moment condition may fail even under the true conduct model because the omitted characteristics are absorbed into the unobserved cost shock.} Therefore, researchers and regulatory authorities need to carefully examine whether the industry under consideration exhibits any of these conditions before applying our test.

\paragraph{Related literature}\label{sec: literature}

 This article is closely related to the empirical literature on implementing the RV approach for testing firm conduct.\footnote{As discussed in \cite{duarte2024testing}, alternative testing procedures include (i) the Estimation-Based (EB) test \citep{pakes2017empirical}, (ii) the Anderson-Rubin (AR) test \citep{anderson1949estimation}, and (iii) the Cox test \citep{cox1961tests, smith1992non}. Examples of empirical applications of these approaches include \cite{miller2017understanding} (EB test), \cite{bergquist2020competition} (AR test), and \cite{Villas-Boas07} (Cox test). An advantage of the RV test over these methods is that inference on conduct remains robust even under misspecification.} \cite{bonnet2010inference} investigated vertical relationships between manufacturers and retailers in the French bottled water market, concluding that manufacturers use two-part tariff contracts with resale price maintenance. In the premium ice cream market, \cite{sullivan2017ice} found evidence that Ben \& Jerry's and H\"{a}agen-Dazs exhibited behaviors consistent with full coordination on both pricing and product choice decisions.\footnote{To examine coordinated behavior in product choice decisions, \cite{sullivan2017ice} modeled endogenous product choices and estimated bounds for the fixed cost parameter using the partial identification approaches employed in \cite{eizenberg2014upstream} and \cite{wollmann2018trucks}. He then implemented the RV test extended by \cite{shi2015nondegenerate} within a moment inequalities framework.} \cite{hu2014collusion} examined price collusion within or across large corporate groups in the Chinese passenger-vehicle industry and found no evidence of collusive pricing behavior. These previous studies backed out marginal costs from demand estimates under each assumed firm conduct model, constructed moment-based objective functions, and determined which model provided a better fit.

 Recent studies on the RV test approach have built upon the falsifiable restrictions proposed by \cite{berry2014identification} to distinguish between alternative conduct models in differentiated goods markets. These studies have further examined the identification conditions required for testing, the selection of appropriate instruments, and the econometric properties of the RV approach. \cite{dearing2024learning} identified cost pass-through as a key determinant for selecting instruments to falsify models of firm behavior. \cite{duarte2024testing} analyzed the properties of the RV test statistic under weak instrument conditions and developed a robust inference framework for the RV approach in such scenarios. They also highlighted the advantages of the RV approach under model misspecification compared to other testing methods. Furthermore, \cite{backus2021common} refined the RV testing procedure by incorporating flexible nonparametric functional forms for marginal costs and proposing instrument functional forms that utilize scalar moment values to enhance the testing power of the RV framework. Our approach, while relying solely on first-stage price regressions, performs comparably to these methods, which are built upon equilibrium markup conditions, and thus offers a simple yet effective diagnostic for detecting collusive pricing behavior.

 This article also contributes to the broader empirical industrial organization literature on assessing firm conduct in differentiated product markets, a longstanding and central topic of inquiry in the field and among antitrust authorities. Examples include airlines \citep{ciliberto2014does}, automobiles \citep{bresnahan1987competition, verboven1996international, sudhir2001competitive, hu2014collusion}, beer \citep{hausman1994competitive, slade2004, rojas2008price, miller2017understanding, miller2021oligopolistic}, instant noodles \citep{kim2024collusion}, ready-to-eat (RTE) cereal \citep{Nevo01, backus2021common, michel2024estimating}, soft drinks \citep{gasmi1992econometric}, and ice cream \citep{sullivan2017ice}. These studies relied on structural models of demand and supply to identify the firm behavior that best explains the observed market outcomes. In contrast, our proposed method requires only a pair of price regressions, making it straightforward to implement.

 Finally, the literature on demand estimation with aggregate data in the discrete choice framework also bears on this article. Prior works (e.g., \citealp{berry1999voluntary}; \citealp{berry2014identification}; \citealp{reynaert2014improving}; \citealp{armstrong2016large}; \citealp{gandhi2019measuring}; \citealp{conlon2020best}) have addressed challenges arising from weak instruments in non-linear GMM estimation. Our work contributes to this field by enhancing the understanding of instruments and improving estimation performance, demonstrating that the strength of BLP-style instruments depends on the correctness of the imposed assumption on industry conduct.

 The remainder of this article is organized as follows. In Section~\ref{sec: conceptual framework}, we provide a theoretical framework for our testing procedure, which we formally propose in Section~\ref{sec: testing}. Using an extensive Monte Carlo study, we examine the performance of our testing method and compare it with that of existing procedures in Section~\ref{sec: mc}. In Section~\ref{sec: application}, we apply our method to study industry conduct in two differentiated product markets. Finally, we conclude in Section~\ref{sec: conclusion}.

\section{Testing framework}\label{sec: conceptual framework}

 In this section, we introduce an oligopoly supply-side model and clarify both the nature of price endogeneity and the relevance of BLP-style instruments that frame our testing problem. 

\subsection{Equilibrium prices and BLP-style instruments} \label{subsec: iv-construction}

 In each market $t=1,2,\ldots, T$, there are $J_t+1$ differentiated products, consisting of $J_t$ inside goods offered by $F_t$ firms and an outside option. Product $j$ has $R+1$ observed characteristics, consisting of price $p_{jt}$ and $R$ nonprice characteristics $\mathbf{x}_{jt}=(x_{jt}^{(1)},\ldots, x_{jt}^{(r)},\ldots, x_{jt}^{(R)})$, as well as an unobserved characteristic $\xi_{jt}$. Let $\mathscr{J}_t$ and $\mathscr{J}_{ft}$ denote the set of inside products and the set of products offered by firm $f = 1,2,\ldots, F_t$ in market $t$, respectively. In addition, let $f_c$ denote the coalition index, so that the mapping $f \mapsto f_c$ defines the coalition partition, which treats firms that collude, or are suspected of colluding, as a single entity. We index each coalition by its lowest-numbered member. For instance, if firms 1 and 2 collude and all other firms act independently, then $f_c = 1$ for both $f = 1$ and $f = 2$, and the coalitions are indexed by $f_c \in \{1, 3, \ldots, F_t\}$.
 
 Firms may engage in collusion, internalizing the profits of their colluding partners, while the true industry conduct is unknown to the researcher or regulatory authority. The variable profit of firm $f$ is given by:
    \begin{equation}\label{eqn: profit}
    \begin{aligned}
    	\Pi_{ft}(\mathbf{p}_t;\mathbf{x}_t,\bm{\xi}_t,\phi) & = \sum_{k\in\mathscr{J}_{ft}}(p_{kt}-mc_{kt})\cdot s_{kt}(\mathbf{p}_t;\mathbf{x}_t,\bm{\xi}_t)\cdot M_t \\
    	& +\phi\cdot\sum_{k\in\mathscr{J}_{f_ct}\setminus \mathscr{J}_{ft}}(p_{kt}-mc_{kt})\cdot s_{kt}(\mathbf{p}_t;\mathbf{x}_t,\bm{\xi}_t)\cdot M_t,
    \end{aligned}
    \end{equation}
 where $mc_{kt}$ is the marginal cost of product $k$, $s_{kt}(\mathbf{p}_t;\mathbf{x}_t,\bm{\xi}_t)$ is the product's market share, determined by consumer demand,\footnote{We keep the demand system unspecified until Section \ref{subsec: exchang_logit}.} $M_t$ is the market size (the number of consumers), and $\phi \in [0,1]$ measures the degree of profit internalization: firm $f$ weights the profits of its colluding partners uniformly by $\phi$ relative to its own, so that $\phi = 0$ corresponds to competition and $\phi = 1$ to joint profit maximization within the coalition. Intermediate values can arise under the common ownership hypothesis \citep{backus2021common} or under imperfect coordination \citep{miller2017understanding}.

 Firms set profit-maximizing prices simultaneously. Equilibrium prices, and consequently the vector of equilibrium markups, are determined by the following first-order conditions:
 \begin{equation}\label{eqn: foc bertrand}
 \mathbf{p}_{t} - \mathbf{mc}_{t} = - \left[\mathscr{H}_t \odot \frac{\partial \mathbf{s}_t(\mathbf{p}_t;\mathbf{x}_t,\bm{\xi}_t)'}{\partial \mathbf{p}_t} \right]^{-1} \mathbf{s}_t(\mathbf{p}_{t};\mathbf{x}_{t},\bm{\xi}_{t}),
 \end{equation}
 where $\mathbf{mc}_t$ is the vector of marginal costs, $\odot$ denotes component-wise multiplication, and $\mathscr{H}_t$ is an unknown (to the researcher) $J_t \times J_t$ ownership matrix. Specifically, $\mathscr{H}_{t,jk} = 1$ if $j$ and $k$ belong to the same firm, $\mathscr{H}_{t,jk} = \phi$ if they belong to different firms in the same coalition, and $\mathscr{H}_{t,jk} = 0$ otherwise. Competition ($\phi = 0$) and full collusion ($\phi = 1$ with $f_c=1$ for all $f$) are the cases in which $\mathscr{H}_t$ is block-diagonal by firm and a matrix of ones, respectively. As shown in the markup equation \eqref{eqn: foc bertrand}, the endogeneity of prices arises from the fact that firms set these prices based on demand, making them functions of the unobservable components~$\bm{\xi}$.
 
 The markup equation \eqref{eqn: foc bertrand} also shows that the equilibrium price of a product generally depends on the ownership structure. Therefore, interchanging a product owned by the focal firm with one owned by a rival may alter the pricing rule. This implies that the pricing rule is not invariant to arbitrary permutations across the characteristics of all other products.

 As BLP argue, when the Nash equilibrium is unique, the pricing rule retains a more limited form of permutation invariance. In particular, it is invariant to permutations of the focal firm's other products and to permutations of the product-level state vectors associated with rival firms. This observation motivates three distinct exchangeability restrictions for constructing instruments: (i) invariance to the ordering of rival firms; (ii) within each rival firm, invariance to the ordering of that firm's products; and (iii) within the focal firm, invariance to the ordering of its other products.

 Based on these exchangeability results, BLP used the sum of attributes across a firm's other products and the sum of attributes across other-firm products as excluded instruments for the product price, $p_{jt}$. Specifically, $\mathbf{z}_{jt}=(\mathbf{z}^{own}_{jt},\mathbf{z}^{other}_{jt})$, referred to as the BLP-style instruments throughout this article, is given by
\begin{equation}\label{eqn: blp sum iv}
\begin{aligned}
	\mathbf{z}_{jt}^{own} &= \left( \sum_{k \in \mathscr{J}_{ft} \setminus \{j\}} x_{kt}^{(1)},\ldots, \sum_{k \in \mathscr{J}_{ft} \setminus \{j\}} x_{kt}^{(R)} \right), \\
	\mathbf{z}_{jt}^{other} &= \left( \sum_{k \in \mathscr{J}_{t} \setminus \mathscr{J}_{ft}} x_{kt}^{(1)},\ldots, \sum_{k \in \mathscr{J}_{t} \setminus \mathscr{J}_{ft}}x_{kt}^{(R)} \right),
\end{aligned}
\end{equation}
 where $\mathscr{J}_{ft}\setminus\{j\}$ denotes the set of products, excluding $j$, offered by firm $f$ in market $t$.

 Importantly, BLP-style instruments must be constructed using the correct grouping of products whose prices are set to maximize joint profits. The three exchangeability restrictions above are valid only when the assumed coalition partition coincides with the true partition. This observation provides the basis for our testing framework, which compares alternative sets of instrumental variables constructed under different conduct hypotheses.

\subsection{Exchangeability under logit demand}\label{subsec: exchang_logit}

 To see how the coalition partition enters the pricing rule, consider a logit demand model in which the indirect utility of consumer $i$ from consuming product $j$ in market $t$ is given by
\begin{equation}\label{eqn: utility}
    u_{ijt} = \underbrace{\mathbf{x}_{jt}\beta - \alpha p_{jt} + \xi_{jt}}_{\delta_{jt}} + \varepsilon_{ijt},
\end{equation}
 where the idiosyncratic taste shock $\varepsilon_{ijt}$ is i.i.d. Type~I extreme value distributed. We also assume that marginal cost is constant in output:
\begin{equation*}
    mc_{jt} = \mathbf{x}_{jt}\gamma + \omega_{jt},
\end{equation*}
where $\omega_{jt}$ represents an unobserved cost component. Thus, the characteristics of other products affect $p_{jt}$ only through the markup, which is the channel our test exploits. We suppress the market subscript $t$ for the remainder of this section.

 We define a \emph{cell}, denoted by $\mathscr{C}$, as a set of products whose prices are set to maximize their joint profits. Each firm $f$ constitutes a cell under competition ($\phi = 0$), whereas the coalition partition $f_c$ defines cells under full profit internalization ($\phi = 1$).\footnote{For $\phi \in (0, 1)$, own and partner products are not exchangeable, and as a result, no cell structure exists (Appendix~\ref{app: lemma 1}). } Let $a_{j}$ denote the exponentiated mean utility of product $j$ evaluated at zero markup, and let $A_{\mathscr{C}}$ denote the corresponding cell aggregate:
\begin{equation}\label{eqn:a_index}
    a_{j} \equiv \exp \big( \mathbf{x}_{j}(\beta - \alpha\gamma) + \xi_{j} - \alpha\omega_{j} \big), \qquad A_{\mathscr{C}} \equiv \sum_{k \in \mathscr{C}} a_{k}.
\end{equation}
 We also define the aggregate market share of cell $\mathscr{C}$ as
\begin{equation*}
    s_{\mathscr{C}} \equiv \sum_{k \in \mathscr{C}} s_{k} = \dfrac{\sum_{k \in \mathscr{C}} \exp(\delta_{k})}{H}, \qquad H \equiv 1 + \sum_{k \in \mathscr{J}} \exp(\delta_{k}).
\end{equation*}

 Under the logit demand specification with $\phi \in \{0, 1\}$, the markup is uniform across all products within a given cell and admits the closed-form expression (see Appendix~\ref{app: lemma 1})
\begin{equation}\label{eqn: cell markup}
    \mu_k =p_k-mc_k= \mu_{\mathscr{C}} = \frac{1}{\alpha (1 - s_{\mathscr{C}})}, \quad \forall k \in \mathscr{C}.
\end{equation}
 Because the cell market share $s_{\mathscr{C}}$ depends on the full price vector, equation~\eqref{eqn: cell markup} defines an implicit system of equations in prices. The following lemma solves this system in terms of the model primitives.

\begin{lemma}[Equilibrium pricing rule under logit demand]\label{lem: logit}
Consider the logit demand model with conduct parameter $\phi \in \{0, 1\}$. The equilibrium price of product $j$ satisfies
\begin{equation}\label{eqn: exch rep}
    p_{j} = \mathbf{x}_{j}\gamma + m\big(A_{\mathscr{C}(j)}, H\big) + \omega_{j},
\end{equation}
 where the markup function $m(\cdot, \cdot)$ is strictly increasing in its first argument and strictly decreasing in its second argument, and $\mathscr{C}(j)$ denotes the cell containing product $j$. The market-wide inclusive value $H$ is the unique solution to
\begin{equation*}
    H = 1 + \sum_{\mathscr{C}} A_{\mathscr{C}} \exp \big(-\alpha\, m(A_{\mathscr{C}}, H)\big).
\end{equation*}
\end{lemma}

 \begin{proof}
    See Appendix \ref{app: lemma 1}.
\end{proof}

 Lemma~\ref{lem: logit} synthesizes established properties of the logit pricing game: within-firm uniform markups \citep[e.g.,][]{anderson1992discrete, konovalov2010price}, the uniqueness of the price equilibrium \citep{konovalov2010price}, and its aggregative game structure \citep{nocke2018multiproduct}. It yields several direct implications for the construction of price instruments. First, the pricing rule for product $j$ incorporates the characteristics of other products in its cell, $\mathscr{C}(j)$, solely through the symmetric sum $\sum_{k \in \mathscr{C}(j) \setminus \{j\}} a_{k}$, which enters both $A_{\mathscr{C}(j)}$ and $H$. Taking the conditional expectation of this sum given the characteristics of the other products in the cell and expanding the exponential function yields
 \begin{align}\label{eqn: sum expansion}
    & \mathbb{E} \left[ \sum_{k \in \mathscr{C}(j) \setminus \{j\}} a_{k} \;\middle|\; \{\mathbf{x}_{k}\}_{k \in \mathscr{C}(j) \setminus \{j\}} \right]
    = \mathbb{E}\big[e^{\xi_k - \alpha\omega_k}\big] \sum_{k \in \mathscr{C}(j) \setminus \{j\}} \exp(\mathbf{x}_{k}c) \nonumber \\
    & \qquad \qquad \qquad \qquad \qquad \qquad \propto \big( |\mathscr{C}(j)| - 1 \big) + \sum_{k \in \mathscr{C}(j) \setminus \{j\}} \mathbf{x}_{k}c + \frac{1}{2} \sum_{k \in \mathscr{C}(j) \setminus \{j\}} (\mathbf{x}_{k}c)^{2} + \cdots ,
 \end{align}
 where $c \equiv \beta - \alpha \gamma$ and the leading factor $\mathbb{E}\big[e^{\xi_k - \alpha\omega_k}\big]$ is a constant common to all products, which holds under the standard assumption that $\mathbf{x}_k$ is independent of $(\xi_k,\omega_k)$. The first-order term $\big(\sum_{k \in \mathscr{C}(j) \setminus \{j\}} \mathbf{x}_{k}\big) c$ is a linear combination of the standard BLP-style instruments constructed over the cell with weights $c$, while the second-order term is a weighted sum of the sums of squares and cross-products of the attributes. Entering these sums as separate regressors therefore captures the symmetric transmission channel through which cell-level characteristics shift the equilibrium price of the focal product, with the weights recovered by the projection coefficients rather than imposed.

 Second, Lemma~\ref{lem: logit} characterizes how individual products shift $p_{j}$. By equation~\eqref{eqn: cell markup}, the markup satisfies $m(A_{\mathscr{C}(j)}, H) = [\alpha(1 - s_{\mathscr{C}(j)})]^{-1}$, with cell market share $s_{\mathscr{C}(j)} = A_{\mathscr{C}(j)}e^{-\alpha\mu_{\mathscr{C}(j)}} / H$. Thus, the characteristics of any other product $k \neq j$ affect $p_{j}$ solely through $s_{\mathscr{C}(j)}$, and the relative price effects of any two products $k$ and $k'$ equal the ratio of the elasticities of $s_{\mathscr{C}(j)}$ with respect to $a_{k}$ and $a_{k'}$.\footnote{Note that
 \begin{align*}
 \frac{\partial p_{j}}{\partial \mathbf{x}_{k}} = \frac{\partial \mu_{\mathscr{C}(j)}}{\partial \log s_{\mathscr{C}(j)}} \frac{\partial \log s_{\mathscr{C}(j)}}{\partial \log a_{k}} \frac{\partial \log a_{k}}{\partial \mathbf{x}_{k}} = \left[s_{\mathscr{C}(j)}\,\frac{\mathrm{d}\mu_{\mathscr{C}(j)}}{\mathrm{d}s_{\mathscr{C}(j)}}\right] \frac{\partial\log s_{\mathscr{C}(j)}}{\partial\log a_{k}} c,
 \end{align*}
 where the bracketed term is a common multiplier identical across all products $k$ and $c \equiv \beta - \alpha\gamma$.} Holding markups fixed, this elasticity decomposes as follows:\footnote{To see this, recall that 
 \begin{equation*}
     s_{\mathscr{C}(j)} = \frac{A_{\mathscr{C}(j)}e^{-\alpha\mu_{\mathscr{C}(j)}}}{H}
     \implies
     \log  s_{\mathscr{C}(j)} = \log A_{\mathscr{C}(j)} - \alpha\mu_{\mathscr{C}(j)} - \log H.
 \end{equation*}}
\begin{equation*}
    \frac{\partial \log s_{\mathscr{C}(j)}}{\partial \log a_{k}} = \frac{\partial \log A_{\mathscr{C}(j)}}{\partial \log a_{k}} - \frac{\partial \log H}{\partial \log a_{k}},
\end{equation*}
where the first term equals $a_{k}/A_{\mathscr{C}(j)} = s_{k}/s_{\mathscr{C}(j)}$ for an inside product $k \in \mathscr{C}(j)$ and zero for an outside product, while the second term equals $a_{k}\exp\big(-\alpha\mu_{\mathscr{C}(k)}\big)/H = s_{k}$ for any product $k$. The following corollary formalizes this result, and shows that the characteristics of products outside the cell affect the focal product's price in the opposite direction from those of products within the cell, with a relative magnitude governed by the cell's market share.

\begin{corollary}[Relative price effects of inside- and outside-cell products]\label{cor: inside outside}
Under the conditions of Lemma~\ref{lem: logit}, holding markups fixed,
\begin{equation}\label{eqn: inside outside}
    \frac{\partial \log s_{\mathscr{C}(j)}}{\partial \log a_{k}}
    = \begin{cases}
        \;\; s_{k} \dfrac{1 - s_{\mathscr{C}(j)}}{s_{\mathscr{C}(j)}}, & \text{if } k \in \mathscr{C}(j), \\[10pt]
        - s_{k}, & \text{if } k \notin \mathscr{C}(j).
    \end{cases}
\end{equation}
Hence, for two products with equal market shares, the price effect of an outside-cell product on $p_{j}$ is opposite in sign to that of an inside-cell product and has a relative magnitude of $s_{\mathscr{C}(j)} / (1 - s_{\mathscr{C}(j)})$. 
\end{corollary}
 
When markups are allowed to adjust, the relative price effect remains invariant to first order in market shares (equations~\eqref{eqn: dpda inside approx} and~\eqref{eqn: dpda outside} in Appendix~\ref{app: elas H}). In sum, these results imply that under logit demand, the explanatory power of BLP-style instruments fundamentally depends on summing over the correct cell structure. Otherwise, the instruments misrepresent the substitution patterns that govern the pricing rule.

\subsection{Fit comparison under a general demand system} \label{subsec: exchang_general}

 We now extend our analysis to a general demand system and formalize the testing framework, which compares the relative fit of two price regressions estimated under the competing conduct hypotheses. Let $b\colon \mathbb{R}^{R} \to \mathbb{R}^{L}$ denote a row vector of basis functions of product characteristics used to construct the summation instruments. For instance, $b(\mathbf{x}) = \big(x^{(1)},\dots,x^{(R)}, (x^{(1)})^2,\dots,(x^{(R)})^2\big)$ with $L = 2R$. We impose the following assumptions throughout the analysis.

 \begin{assumption}[Specification]\label{ass: specification}
 \leavevmode
    \begin{enumerate}[noitemsep, label=(\arabic*)]
        \item The demand system is permutation-equivariant in product labels. That is, for any permutation of products in $\mathscr{J}$, permuting product labels together with their associated $(p_k,\mathbf{x}_k,\xi_k)$ induces the same permutation of the corresponding market shares.
        
        \item Marginal cost is product specific and constant in output. In particular, for each product $j$,
        \(
            mc_j = mc(\mathbf{x}_j,\omega_j)
        \) for some common function $mc(\cdot)$.
    \end{enumerate}
\end{assumption}

 \begin{assumption}[Sampling]\label{ass: sampling}
    In any given market, the product-level primitives $\{(\mathbf{x}_{k}, \xi_{k}, \omega_{k})\}_{k \in \mathscr{J}}$ are independent and identically distributed (i.i.d.) across products, and the assignment of products to firms and to coalitions is independent of the primitives.
 \end{assumption}

 \begin{assumption}[Regularity]\label{ass: regularity}
    \leavevmode
    \begin{enumerate}[noitemsep, label=(\arabic*)]
        \item The equilibrium of the pricing game is unique under each conduct hypothesis considered.
        \item The covariance matrix $\Sigma_{b} \equiv \mathrm{Var}[b(\mathbf{x}_{k})]$ is positive definite.
        \item $\bm{\lambda}_{o} \neq \mathbf{0}$, where $\bm{\lambda}_{o}$ is defined in equation~\eqref{eqn: linear rep} below.
    \end{enumerate}
\end{assumption}
  
 Assumption~\ref{ass: specification}(1) holds in standard random utility models, in which product-level heterogeneity enters only through $(p_k, \mathbf{x}_k, \xi_k)$. Assumption~\ref{ass: specification}(2) rules out economies of scale and scope in production. Assumption~\ref{ass: specification}, together with the uniqueness of equilibrium in Assumption~\ref{ass: regularity}(1), makes the price of a focal product a symmetric function of the primitives of the other products in its cell for $\phi \in \{0, 1\}$. 
 
 Assumption~\ref{ass: sampling} has two important features. First, firm and coalition membership carries no information about the distribution of product characteristics and unobservables, so that a firm's own products and its partner products are exchangeable within a coalition.\footnote{Exchangeability requires only that the primitives be identically distributed and that the assignment of products to firms and coalitions be independent of the primitives. Independence of primitives across different products is used to obtain the closed-form expressions in Lemma~\ref{lem: projection} and Proposition~\ref{prop: main}.} Second, although the validity of BLP-style instruments requires the observed product characteristics $\mathbf{x}_k$ to be uncorrelated with the unobserved demand component $\xi_k$, our proposed testing procedure does not rely on this exogeneity assumption. 
 
 For algebraic tractability, we also impose a symmetry assumption within this subsection and discuss at the end of the subsection and in Appendix~\ref{subsec: asymmetry} how the results change when this assumption is relaxed.
  \begin{assumption}[Symmetry]\label{ass: symmetry}
  Each firm offers the same number of products (at least two) in every market, and every suspected coalition consists of the same number of firms.
 \end{assumption}
 The second condition holds trivially when a single coalition is suspected, which is the most common case in practice.
 
 For each product $j \in \mathscr{J}_{f}$, we define the following three instrument vectors:
\begin{equation}\label{eqn: z decomposition}
    \mathbf{z}^{comp}_{j} = \sum_{k \in \mathscr{J}_{f} \setminus \{j\}} b(\mathbf{x}_{k}), \qquad
    \mathbf{z}^{part}_{j} = \sum_{k \in \mathscr{J}_{f_{c}} \setminus \mathscr{J}_{f}} b(\mathbf{x}_{k}), \qquad
    \mathbf{z}^{coll}_{j} = \sum_{k \in \mathscr{J}_{f_{c}} \setminus \{j\}} b(\mathbf{x}_{k}).
\end{equation}
 The first two vectors, $\mathbf{z}^{comp}_{j}$ and $\mathbf{z}^{part}_{j}$, represent the sums of basis functions evaluated at the characteristics of the firm's other products and of the suspected colluding partners' products, respectively. $\mathbf{z}^{coll}_{j}$ is the sum of these two vectors. We refer to $\mathbf{z}^{comp}_{j}$ and $\mathbf{z}^{coll}_{j}$ as the competition IVs and collusion IVs, respectively. We retain the term ``instrument'' to maintain consistency with the existing literature, although our framework, as stated in Assumption~\ref{ass: sampling}, does not require the observed product characteristics $\mathbf{x}_k$ to be uncorrelated with the unobserved demand component $\xi_k$.

 For a firm not suspected of collusion, $\mathscr{J}_{f_{c}} = \mathscr{J}_{f}$, which implies $\mathbf{z}^{part}_{j} = \mathbf{0}$ and $\mathbf{z}^{coll}_{j} = \mathbf{z}^{comp}_{j}$. Thus, the two instrument sets differ solely for the products of suspected firms. We denote by $D_{j} \in \{0, 1\}$ the indicator that product $j$ belongs to a firm in a suspected coalition, and let $\pi \equiv \mathbb{E}[D_{j}] = \Pr(D_{j} = 1)$. Since $D_{j}$ is a product-level indicator, $\pi$ is simply the fraction of products, by count, that belong to suspected firms; $\pi = 1$ when every firm is suspected. By Assumption~\ref{ass: sampling}, $D_{j}$ is independent of the primitives.
 
 Let $(\gamma_{0}, \gamma_{D}, \bm{\gamma}, \bm{\lambda}_{o}, \bm{\lambda}_{p})$ denote the coefficients from the population linear projection of $p_{j}$ onto $(1, D_{j}, \mathbf{x}_{j}, \mathbf{z}^{comp}_{j}, \mathbf{z}^{part}_{j})$ under the true conduct, with $\gamma_{D} \equiv 0$ when $\pi = 1$:
\begin{equation}\label{eqn: linear rep}
    p_{j} = \gamma_{0} + \gamma_{D} D_{j} + \mathbf{x}_{j}\bm{\gamma} + \mathbf{z}^{comp}_{j}\bm{\lambda}_{o} + \mathbf{z}^{part}_{j}\bm{\lambda}_{p} + e_{j},
\end{equation}
where $e_j$ denotes the population linear projection residual, which, by construction, satisfies
\begin{equation*}
    \mathbb{E}[e_{j}] = 0,\,\mathrm{Cov}(e_{j}, D_{j}) =0, \,\mathrm{Cov}(e_{j}, \mathbf{x}_{j}) =\textbf{0}, \;\text{and} \;\mathrm{Cov}(e_{j}, \mathbf{z}^{comp}_{j}) = \mathrm{Cov}(e_{j}, \mathbf{z}^{part}_{j}) = \mathbf{0}.
\end{equation*}
 This projection exists whenever second moments are finite and imposes no functional-form restrictions on the pricing rule. The residual $e_{j}$ subsumes any nonlinear variation not captured by the summation instruments. Combined with Assumptions~\ref{ass: specification} and~\ref{ass: sampling}, Assumption~\ref{ass: symmetry} ensures that products face the same population pricing environment within each conduct regime, and that the coefficients $\bm{\lambda}_o$ and $\bm{\lambda}_p$ in \eqref{eqn: linear rep} represent the price effects of own and partner products common to all firms, rather than averages of firm-specific effects. Under Assumption~\ref{ass: regularity}(3) that $\bm{\lambda}_{o} \neq \mathbf{0}$, the competition IVs have explanatory power for prices.\footnote{Under logit demand, this condition holds whenever $c \equiv \beta - \alpha \gamma \neq \mathbf{0}$ by equation~\eqref{eqn: sum expansion}.}

 The fit comparison turns on the magnitude of $\bm{\lambda}_{p}$ relative to that of $\bm{\lambda}_{o}$. Under full profit internalization ($\phi = 1$), the characteristics of partner products and those of the firm's own other products are exchangeable in the pricing rule, irrespective of the underlying demand system, as the following lemma formalizes.

 \begin{lemma}[Exchangeability under full profit internalization]
\label{lem: projection}
Suppose that Assumptions~\ref{ass: specification}--\ref{ass: symmetry}
hold and $\phi=1$. Then, for any choice of $b(\cdot)$,
\begin{equation}\label{eqn: lambda pooled}
    \bm{\lambda}_o = \Sigma_b^{-1}\big[\pi\,\mathbf{g}+(1-\pi)\,\mathbf{g}^N\big],
    \qquad
    \bm{\lambda}_p = \Sigma_b^{-1}\mathbf{g},
\end{equation}
where $\mathbf{g} \equiv \mathrm{Cov}\big(b(\mathbf{x}_k),p_j\big)$ for any product $j$ of a colluding firm and any other product $k \in \mathscr{J}_{f_c}\setminus\{j\}$ in its coalition, and $\mathbf{g}^N \equiv \mathrm{Cov}\big(b(\mathbf{x}_k),p_j\big)$ for any product $j$ of a non-colluding firm and any other product $k\in\mathscr{J}_f\setminus\{j\}$ of that firm. In particular, $\bm{\lambda}_o=\bm{\lambda}_p$
whenever $\pi=1$ or $\mathbf{g}^N=\mathbf{g}$.
\end{lemma}

 \begin{proof}
    See Appendix \ref{app: lemma 2}.
 \end{proof}
 
 Lemma~\ref{lem: projection} rests on the symmetry of the coalition's first-order conditions (Assumption~\ref{ass: specification}), which, given a unique equilibrium (Assumption~\ref{ass: regularity}(1)), makes $p_j$ a symmetric function of the primitives of the other products in its cell. It also relies on the identical distribution of own and partner primitives (Assumption~\ref{ass: sampling}). Combined, these two properties establish the equality of covariances for own and partner products within a coalition. The coefficient vector $\bm{\lambda}_o$ is a weighted average of this common covariance $\mathbf{g}$ and another common covariance $\mathbf{g}^N$ for non-colluding firms with weights $\pi$ and $1-\pi$. Consequently, $\bm{\lambda}_o = \bm{\lambda}_p$ when every firm colludes ($\pi = 1$). Otherwise, $\bm{\lambda}_o - \bm{\lambda}_p = (1-\pi)\,\Sigma_b^{-1}(\mathbf{g}^N - \mathbf{g})$, the magnitude of which depends on the underlying demand system. 

 There is no direct counterpart to the lemma under competition ($\phi = 0$). The extent to which a suspected partner's product characteristics affect $p_{j}$ depends on the demand system. Under logit demand, rival products enter only through $H$, so that $\bm{\lambda}_p = -[s_f/(1-s_f)]\,\bm{\lambda}_o$ to first order in market shares, where $s_f$ denotes a firm's expected market share (Corollary~\ref{cor: logit} below). In nested logit or random coefficients logit models, rival products located in the same nest or nearby in characteristic space shift price directly, so that $\bm{\lambda}_p$ is not necessarily small (Appendix~\ref{app: demand system}).

 Now, consider two price regressions:
\begin{align}
	& p_{j} = \gamma^{1}_{0} + \gamma^{1}_{D} D_{j} + \mathbf{x}_{j} \bm{\gamma}^1 + \mathbf{z}_{j}^{comp} \bm{\theta}^1 + e^1_{j} \label{eqn: comp reg} \\
	& p_{j} = \gamma^{2}_{0} + \gamma^{2}_{D} D_{j} + \mathbf{x}_{j} \bm{\gamma}^2 + \mathbf{z}_{j}^{coll} \bm{\theta}^2 + e^2_{j} \label{eqn: coll reg},
\end{align}
 where $\gamma^{1}_{D}$ and $\gamma^{2}_{D}$ are set to zero when $\pi = 1$.\footnote{Including $D_j$ in both regressions removes a level difference between the products of suspected and other firms, which is unrelated to the exchangeability that the fit comparison exploits, and it yields the closed-form threshold in Proposition~\ref{prop: main}(c). In the empirical applications of Section~\ref{sec: application}, the product fixed effects subsume $D_j$. If $D_j$ is omitted, parts (a) and (b) of Proposition~\ref{prop: main} continue to hold and only the value of the threshold changes; see the proof in Appendix~\ref{app: prop 1}.} The population mean squared residuals of the two regressions are
\begin{align*}
	Q_{comp} &= \min_{\gamma^{1}_{0}, \gamma^{1}_{D}, \bm{\gamma}^1, \bm{\theta}^1} \mathbb{E}\big[(p_{j} - \gamma^{1}_{0} - \gamma^{1}_{D} D_{j} - \mathbf{x}_{j}\bm{\gamma}^1 - \mathbf{z}^{comp}_{j}\bm{\theta}^1)^2\big], \\
	Q_{coll} &= \min_{\gamma^{2}_{0}, \gamma^{2}_{D}, \bm{\gamma}^2, \bm{\theta}^2} \mathbb{E}\big[(p_{j} - \gamma^{2}_{0} - \gamma^{2}_{D} D_{j} - \mathbf{x}_{j}\bm{\gamma}^2 - \mathbf{z}^{coll}_{j}\bm{\theta}^2)^2\big],
\end{align*}
 where the expectations are taken over all products in the market.

 Let $n_{o} \equiv |\mathscr{J}_{f}| - 1$ and $n_{p} \equiv |\mathscr{J}_{f_{c}} \setminus \mathscr{J}_{f}|$ denote the number of a firm's other products and the number of a collusive (or suspected) firm's partner products, respectively. We define $r$ as the expected number of partner products per own product, \(r \equiv \pi n_p / n_o \), and define the two thresholds
\begin{equation}\label{eqn: kappa star}
    \kappa_{+}
    \equiv
    \frac{1}{1+\sqrt{1+r}},
    \qquad
    \kappa_{-}
    \equiv
    -\frac{1}{\sqrt{1+r}-1},
    \qquad
    \kappa_{-}<0<\kappa_{+}<\frac{1}{2}.
\end{equation}
Letting $\Vert\mathbf{v}\Vert_{\Sigma_b} \equiv (\mathbf{v}'\Sigma_b\mathbf{v})^{1/2}$ denote the $\Sigma_b$-weighted norm of a vector $\mathbf{v}$, the following proposition provides an explicit bound on $\Vert\bm{\lambda}_p\Vert_{\Sigma_b}$ relative to $\Vert\bm{\lambda}_o\Vert_{\Sigma_b}$ under which the competition IVs fit price better than the collusion IVs.

\begin{proposition}[Sign of the population fit differential]\label{prop: main}
 Under Assumptions~\ref{ass: specification}--\ref{ass: symmetry},
\begin{equation}\label{eqn: Q diff general}
	Q_{comp} - Q_{coll} = n_o \left( \frac{\Vert \bm{\lambda}_o + r\bm{\lambda}_p\Vert_{\Sigma_b}^2}{1 + r} \;-\; \Vert\bm{\lambda}_o\Vert_{\Sigma_b}^2 \right) .
\end{equation}
 Consequently:
\begin{enumerate}
	\item[(a)] If $\bm{\lambda}_p = \bm{\lambda}_o$, then $Q_{comp} > Q_{coll}$.
	\item[(b)] If $\Vert\bm{\lambda}_p\Vert_{\Sigma_b} < \kappa_{+} \Vert\bm{\lambda}_o \Vert_{\Sigma_b}$, then $Q_{comp} < Q_{coll}$.
	\item[(c)] If $\bm{\lambda}_p = \kappa\bm{\lambda}_o$ for a scalar $\kappa$, then, writing $\bm{\lambda} = \bm{\lambda}_o$,
\begin{equation}\label{eqn: Q diff}
	Q_{comp} - Q_{coll} = 
    \bm{\lambda}'\Sigma_b\bm{\lambda} \cdot \frac{n_o r (r \kappa^2 + 2 \kappa - 1)}{1+r},
\end{equation}
 which, as a function of $\kappa$, is a convex quadratic with roots $\kappa_{-}$ and $\kappa_{+}$ as defined in \eqref{eqn: kappa star}; it is negative for $\kappa \in (\kappa_{-}, \kappa_{+})$, zero at the roots, and positive otherwise.
\end{enumerate}
\end{proposition}

 \begin{proof}
    See Appendix \ref{app: prop 1}.
 \end{proof}

 Parts (a) and (b) correspond to the two conduct regimes. Part (a), combined with Lemma~\ref{lem: projection}, ensures that the collusion IVs fit the price better when the suspected firms actually collude with full profit internalization ($\phi = 1$). More specifically, under collusive conduct, Lemma~\ref{lem: projection} yields $\bm{\lambda}_p = \bm{\lambda}_o$ whenever $\pi = 1$ or $\mathbf{g}^N = \mathbf{g}$, which implies $Q_{comp} > Q_{coll}$ by part (a). If instead $\pi < 1$ and $\mathbf{g}^{N} \neq \mathbf{g}$, then $\bm{\lambda}_o - \bm{\lambda}_p = (1-\pi)\,\Sigma_b^{-1}(\mathbf{g}^N - \mathbf{g})$. Because the fit differential $Q_{comp} - Q_{coll}$ in \eqref{eqn: Q diff general} is continuous and homogeneous of degree two in $(\bm{\lambda}_o, \bm{\lambda}_p)$, the inequality $Q_{comp} > Q_{coll}$ continues to hold whenever $\mathbf{g}^N - \mathbf{g}$ is sufficiently small relative to $\mathbf{g}$. Under logit demand, this holds if the market shares of both the coalition and non-colluding firms are sufficiently small (Corollary~\ref{cor: logit}); Appendix~\ref{app: demand system} discusses other demand systems.

 On the other hand, part (b) states that the competition IVs fit better than the collusion IVs as long as the standard deviation of a partner product's contribution to price, $b(\mathbf{x}_{k})\bm{\lambda}_{p}$, is less than a fraction $\kappa_{+}$ of that of an own product's contribution, $b(\mathbf{x}_k)\bm{\lambda}_o$. Intuitively, the collusion IVs constrain the coefficients on the own and partner sums to be identical. When partner products carry little price signal (as they do under competitive conduct, $\phi = 0$), this restriction pulls the coefficient on the own sum away from $\bm{\lambda}_o$. The resulting loss of fit on the own-product component of price then dominates any predictive gain contributed by the partner sum, ensuring that the competition IVs achieve a better fit. Whether the condition $\Vert\bm{\lambda}_p\Vert_{\Sigma_b} < \kappa_{+} \Vert\bm{\lambda}_o \Vert_{\Sigma_b}$ holds under competition depends on the demand system; Corollary~\ref{cor: logit} shows that under logit demand it reduces to a condition on the firm's market share, while Appendix~\ref{app: demand system} discusses other demand systems.
  
 Part (c) shows that the sign of $Q_{comp} - Q_{coll}$ depends on $\kappa$, the relative weight that the pricing rule places on partner products, and that this sign changes not at $\kappa = 0$, but at the two distinct roots $\kappa_{-}$ and $\kappa_{+}$. Specifically, when the pricing rule places modest positive weight on partner products ($0 < \kappa < \kappa_{+}$), the competition IVs provide a better fit.\footnote{Note that $\kappa_{+}$ decreases as $r$ increases: a smaller value of $\kappa$ suffices for the collusion IVs to fit better when suspected partners contribute more products relative to the firm's own product portfolio, and when a larger share of products belongs to the suspected coalition, as products outside the coalition carry no partner signal and thus dilute the comparison.} The negative root is consistent with competitive conduct because partner products shift price in the direction opposite to that of own products, leading to $\kappa < 0$. The competition IVs continue to achieve a better fit as long as $|\kappa| < |\kappa_{-}|$. Conversely, if suspected partner products exert such a strong negative price effect that $\kappa < \kappa_{-}$, the collusion IVs fit better even when firms actually compete.\footnote{This reversal requires an individual partner product to shift the price by more than $|\kappa_{-}|$ times the effect of an own product. Such an effect can arise under random coefficients logit if suspected partner products are substantially closer substitutes for product $j$ than are the firm's own products, a configuration that violates Assumption~\ref{ass: sampling} (see Section~\ref{subsec: demand system}).}
  
 For the logit model, the conditions of Proposition~\ref{prop: main} can be stated in terms of market shares.

\begin{corollary}[Logit]\label{cor: logit}
 Suppose demand is logit and Assumptions~\ref{ass: specification}--\ref{ass: symmetry} hold.  Let $s_f$ denote a firm's expected market share, where the expectation is taken over the product-level primitives. Under collusive conduct ($\phi = 1$), $Q_{comp} > Q_{coll}$ when $\pi = 1$, and also when $\pi < 1$ provided that the market shares of the coalition and of non-colluding firms are sufficiently small. Under competitive conduct ($\phi = 0$), $\bm{\lambda}_p = \kappa\bm{\lambda}_o$ with $\kappa = -s_f/(1-s_f)$ to first order in market shares, so that $Q_{comp} < Q_{coll}$ for sufficiently small market shares; to first order, this ranking holds if and only if $\kappa > \kappa_{-}$, that is, whenever the firm's market share satisfies
\begin{equation}\label{eqn: logit competition condition}
	s_f \;<\; 1/\sqrt{1+r}.
\end{equation}
\end{corollary}

 \begin{proof}
    See Appendix \ref{app: cor 2}.
 \end{proof}
Corollary~\ref{cor: logit} is the population counterpart of Corollary~\ref{cor: inside outside}: the relative price effects in \eqref{eqn: inside outside} involve the realized cell share in a given market, whereas $\bm{\lambda}_o$ and $\bm{\lambda}_p$ average these effects over the distribution of product primitives, so the shares in Corollary~\ref{cor: logit} are expected shares. In applications, they can be measured by average observed shares across markets. 

The condition in the first part of Corollary~\ref{cor: logit} is mild. Denoting the coalition's expected market share by $s_{f_c}$, the proof shows that, under collusive conduct, $\bm{\lambda}_p = \kappa\bm{\lambda}_o$, with $\kappa = 1 - (1-\pi)(s_{f_c} - s_f) > 0$ to first order, so that $Q_{comp} > Q_{coll}$ as long as $(1-\pi)(s_{f_c} - s_f) < 1 - \kappa_{+}$, where $1 - \kappa_{+} > 1/2$. The condition thus fails only if the coalition accounts for more than half of the market ($s_{f_c} > 1/2$) with fewer than half of the products ($\pi < 1/2$). In the Monte Carlo design of Section~\ref{sec: mc setup} with $F = 6$ firms each holding a market share of approximately $0.06$, a coalition of two to five firms ($F_c = 2,\ldots,5$) yields $s_{f_c} - s_f$ between $0.06$ and $0.24$ and $1-\pi$ between $1/6$ and $2/3$, so that $\kappa$ lies between $0.94$ and $0.96$, which is more than twice $\kappa_{+}$.

The second part shows that, under logit demand and competitive conduct, the condition for $Q_{comp} < Q_{coll}$ is likewise mild. Because $\kappa = -s_f/(1-s_f) < 0$, the relevant root in Proposition~\ref{prop: main}(c) is $\kappa_{-}$, and, to first order in market shares, the competition IVs fit better as long as $s_f < 1/\sqrt{1+r}$, that is, unless a suspected firm accounts for a sizable fraction of the market. In the Monte Carlo design of Section~\ref{sec: mc setup} with $F = 6$, $1/\sqrt{1+r}$ ranges from $0.45$ to $0.85$ for $F_c = 2,\ldots,5$, compared to an actual share of about $0.06$. The sufficient condition $s_f/(1-s_f) < \kappa_{+}$ implied by Proposition~\ref{prop: main}(b) is stricter but still mild: it requires $s_f < \kappa_{+}/(1+\kappa_{+})$, which lies between $0.24$ and $0.31$ in the same design.

For more general demand systems, the conditions of Proposition \ref{prop: main} no longer reduce to bounds on market shares as in Corollary~\ref{cor: logit} for logit demand. In Appendix \ref{app: demand system}, we argue that the fit comparison remains valid under nested logit or random coefficients logit demand: when all firms collude ($\pi=1$), the collusion IVs continue to fit better as long as Assumptions \ref{ass: specification}(1) and \ref{ass: sampling} hold; when firms compete, the competition IVs fit better as long as the firm's share among the close substitutes of the focal product is modest.
 
Finally, note that Assumption \ref{ass: symmetry} is imposed for analytical convenience and is not essential to the underlying exchangeability argument (see Appendix \ref{subsec: asymmetry}). When firms offer different numbers of products, the symmetry of the equilibrium pricing rule under full profit internalization continues to hold conditional on the coalition partition. However, the coefficients of the pooled price regressions generally become portfolio-size weighted averages of firm-specific price effects. To accommodate this heterogeneity, we can include the numbers of own-firm and partner products as additional controls in both price regressions. With these controls, Proposition \ref{prop: main} continues to hold after replacing the common number of own products with its product-level average and redefining the ratio of partner to own products accordingly. Under logit demand, the covariance approximations used in Corollary \ref{cor: logit} remain valid, but the resulting conditions involve appropriately weighted averages of firm market shares rather than a common market share. When all firms belong to a single collusive coalition, the equality of the own- and partner-instrument coefficients continues to hold. Thus, unequal portfolio sizes do not fundamentally alter the testing intuition, although the relevant thresholds and pooled price responses depend on the distribution of portfolio or coalition sizes.

\subsection{Type I and II errors under non-random product assignment}\label{subsec: demand system}

Assumption~\ref{ass: sampling} imposes random assignment of products to firms and coalitions. In practice, firms choose their product portfolios endogenously, and their incentives to collude may depend on the substitutability between their own products and those of potential partners. In this subsection, we consider configurations in which product placement, firm ownership, and coalition formation are correlated, violating Assumption~\ref{ass: sampling}. Such violations have two consequences. First, they may alter the second moments of the instruments that underlie the weights $\Sigma_b$ and the parameter $r$ in Proposition~\ref{prop: main}. This effect, however, is limited. Specifically, $Q_{comp} - Q_{coll}$ remains a continuous quadratic form in $(\bm{\lambda}_o, \bm{\lambda}_p)$, part~(a) of Proposition~\ref{prop: main} continues to hold, and part~(b) holds with thresholds that depend on these second moments rather than on $r$ alone.

Second, and more importantly, under non-random assignment, own and partner products need not exhibit the same degree of substitutability with the focal product. Consequently, substitution patterns affect the price effects of own and partner products asymmetrically, potentially reversing the sign of the fit differential in Proposition~\ref{prop: main} and producing either a Type~I error (detecting collusion when firms compete) or a Type~II error (failing to detect collusion when firms collude) in our testing framework. This channel, absent under logit demand, arises under richer demand systems in which substitution is governed by proximity. 



To see this, we return to the first-order condition for product $j\in\mathscr{J}_f$ under a given demand system:
\begin{equation}\label{eqn: foc general}
    \underbrace{
    s_j
    + \sum_{k\in\mathscr{J}_f}
    (p_k-mc_k)\frac{\partial s_k}{\partial p_j}
    }
    _
    {\text{demand channel}}
    + 
    \underbrace{
    \phi\sum_{\ell\in\mathscr{J}_{f_c}
        \setminus\mathscr{J}_f}
    (p_\ell-mc_\ell)\frac{\partial s_\ell}{\partial p_j}
    }
    _
    {\text{internalization channel}}
    = 0.
\end{equation}
A change in the characteristics of a partner product $\ell$ affects the equilibrium price of product $j$ through two channels. The \emph{demand channel} operates through the first two terms, as the characteristics of $\ell$ affect demand for the firm's own products. The \emph{internalization channel} operates only under collusion through the third term. Both channels depend on the substitutability between products, which is determined by the underlying demand system. See Appendix \ref{app: demand system} for more details.

\paragraph{Competitive conduct ($\phi=0$) and Type I error} Under competitive conduct, the internalization channel is absent and partner-product characteristics affect $p_j$ only through the demand channel. Suppose that in violation of Assumption~\ref{ass: sampling}, partner products are particularly close substitutes for product $j$.\footnote{This occurs when the firm's products are spread across nests (under nested logit) or in the characteristic space (under random coefficients logit), so that each is located in a nest, or a region of characteristic space, occupied by partner products rather than by the firm's other products.} Changes in their characteristics can substantially alter demand for the product and its equilibrium price, even without collusion. If these effects become sufficiently large relative to those of own-product characteristics, the collusion IVs may fit better despite competitive conduct, resulting in a Type I error.

We can formalize these intuitions using Proposition~\ref{prop: main}. Suppose that
$\bm{\lambda}_p=\kappa\bm{\lambda}_o$, as in Proposition~\ref{prop: main}(c). When firms compete, partner products typically shift $p_j$ in the direction opposite to that of own products,\footnote{See Corollary~\ref{cor: inside outside} for logit and Corollary~\ref{cor: nl inside outside} in Appendix~\ref{app: demand system} for nested logit.} implying $\kappa<0$. Greater substitutability between own and partner products tends to increase the magnitude of $\kappa$. By Proposition~\ref{prop: main}(c), the competition IVs fit better when $\kappa \in (\kappa_-, \kappa_+)$, whereas the collusion IVs fit better when $\kappa < \kappa_-$. Thus, sufficiently strong substitutability between own and suspected partner products can favor the collusion IVs even under competition (Type I error). Conversely, stronger within-firm substitutability tends to reduce the risk of falsely concluding collusion.

\paragraph{Collusive conduct ($\phi=1$) and Type II error}
When partner products are systematically weak substitutes for the products of a collusive firm, a change in their characteristics has little effect on demand for the firm's products. Moreover, because the cross-price derivative $\partial s_\ell/\partial p_j$ is small, the internalization channel is also weak, so that the firm has little incentive to adjust $p_j$ in response to changes in the characteristics of $\ell$. Consequently, partner-product characteristics may have relatively little explanatory power for prices even though firms collude. This problem is compounded when products within the same firm are close substitutes,\footnote{This occurs when the firm's own products share a nest while partner products lie in other nests under nested logit, and when the firm's own products are substantially closer to one another in characteristic space than to partner products under random coefficients logit.} since own-product characteristics may then affect $p_j$ substantially more than partner-product characteristics. In this case, the competition IVs may fit better despite collusive conduct, resulting in a Type II error. 

Under collusive conduct, stronger substitutability among a firm's own products relative to its partners' products tends to reduce $\kappa$, which is positive under collusion, by weakening the relative price effects of partner product characteristics. Proposition~\ref{prop: main}(c) shows that the population fit differential $Q_{comp}-Q_{coll}$ is increasing in $\kappa$ for $\kappa\geq0$, holding $r$ and $\bm{\lambda}_o$ fixed. Consequently, as $\kappa$ decreases below the threshold $\kappa_+$, the competition IVs fit better even under collusion (Type II error). Conversely, greater substitutability between own and partner products tends to increase $\kappa$, reducing the risk of failing to detect collusion.

\begin{figure}[!t]
    \centering
    \caption{Product substitutability and conduct
    classification errors}
    \begin{subfigure}[b]{0.47\textwidth}
        \centering
        \includegraphics[width=\textwidth]{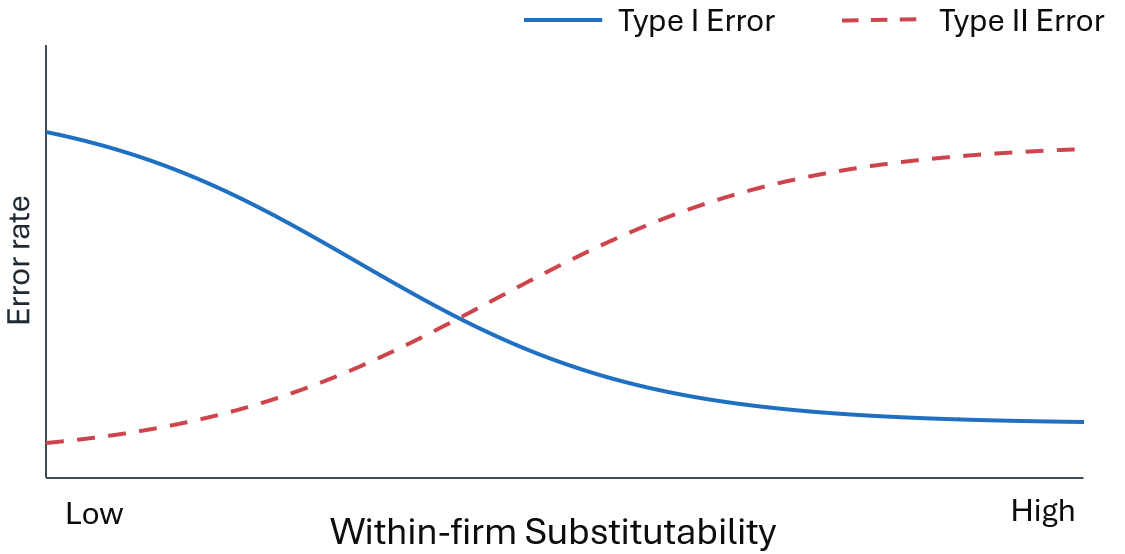}
        \caption{Within-firm substitutability}
    \end{subfigure}
    \hfill
    \begin{subfigure}[b]{0.47\textwidth}
        \centering
        \includegraphics[width=\textwidth]{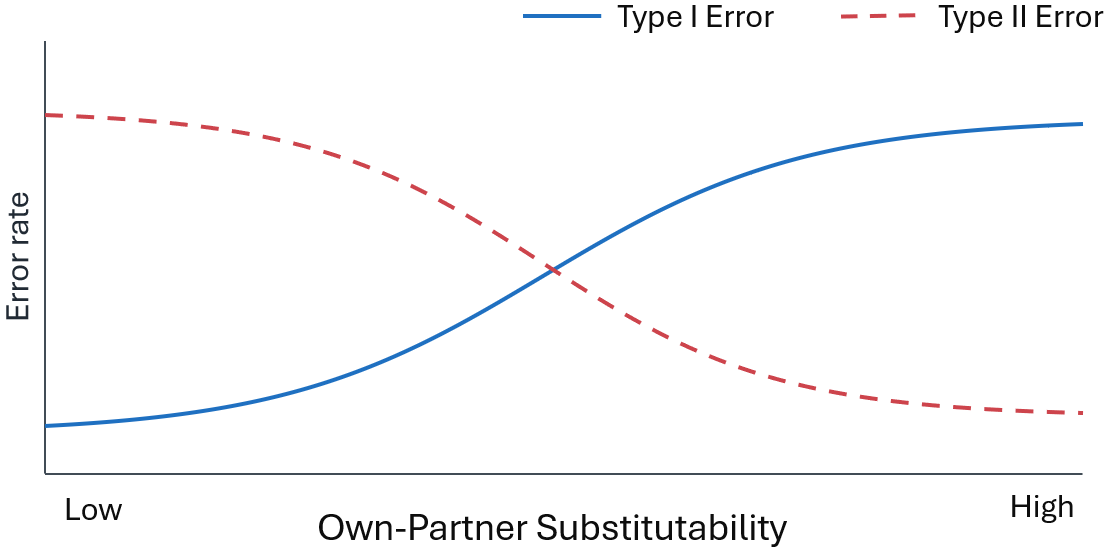}
        \caption{Own--partner substitutability}
    \end{subfigure}
    \label{fig: error rate}
\end{figure}

\paragraph{Implications for conduct testing}
Figure~\ref{fig: error rate} illustrates these opposing tendencies. Greater substitutability among a firm's own products tends to favor the competition IVs, potentially increasing the risk of Type II error while reducing the risk of Type I error. Greater substitutability between own and partner products tends to favor the collusion IVs, potentially reducing Type II error while increasing the risk of Type I error. 

 In our testing framework, a Type~I error is detectable in principle. Under competition, $\kappa < 0$, which implies that the coefficient on $\mathbf{z}^{coll}_j$ in the collusion regression (which equals $(1 + r\kappa)\bm{\lambda}_o/(1 + r)$, as shown in Appendix~\ref{app: prop 1}) has the opposite sign from the coefficient on $\mathbf{z}^{comp}_j$ in the competition regression whenever $\kappa < -1/r$, a region that contains $\kappa < \kappa_-$. A better fit of the collusion IVs accompanied by such a sign reversal therefore points to competition, whereas under collusion, $\kappa > 0$ and the two coefficients share the same sign. 

On the other hand, a Type~II error is less consequential. Weak substitutability between own and partner products limits the gains from collusion, so not only do firms have little incentive to collude in such a case, but collusion also has little effect on prices when they do; in the limiting case of vanishing cross-price effects, collusion leaves no trace in prices, and no price-based test can detect it (Appendix~\ref{app: demand system}). The existing markup-based RV test, described in Section~\ref{sec: markup testing}, also fails in this case because collusion leaves equilibrium markups unchanged, rendering the two conduct models observationally equivalent under the assumption of constant marginal costs.

\section{Testing procedure}\label{sec: testing}

In this section, we build on the results of Section~\ref{sec: conceptual framework} to propose a test of collusive behavior among firms that compares the fit of two price regressions. The competition IVs ($\mathbf{z}^{comp}$) sum the characteristics of the firm's other products, whereas the collusion IVs ($\mathbf{z}^{coll}$) also include those of the products of \textit{suspected} colluding partners. We estimate a price regression with each set of instruments and compare the relative explanatory power of the two regressions using the non-nested model selection test of \citet{rivers2002model}.\footnote{The implementation of the test of \citet{rivers2002model} in cross-sectional and panel regression settings can be inferred from \citet{wooldridge2010} (Section 13.11.2).}

\subsection{Proposed testing procedure based on exchangeability}\label{sec: iv testing}

 We first define the two BLP-style instruments for product $j\in\mathscr{J}_{ft}$ in market $t$, $\mathbf{z}_{jt}^{comp}$ and $\mathbf{z}_{jt}^{coll}$. These instruments are specified as follows, with the basis functions in Section~\ref{subsec: exchang_general} taken to be the characteristics and their squares:\footnote{In principle, the cross product of two distinct characteristics can also be included. We omit those cross-product terms for notational simplicity.}
\begin{equation}\label{eqn: iv testing}
\begin{aligned}
	& \mathbf{z}_{jt}^{comp} = \left(\sum_{k\in\mathscr{J}_{ft}\setminus\{j\}}x^{(1)}_{kt},\ldots, \sum_{k\in\mathscr{J}_{ft}\setminus\{j\}}x^{(R)}_{kt},~\sum_{k\in\mathscr{J}_{ft}\setminus\{j\}}\Big(x^{(1)}_{kt}\Big)^{2},\ldots,\sum_{k\in\mathscr{J}_{ft}\setminus\{j\}}\Big(x^{(R)}_{kt}\Big)^{2}\right) \\
    & \mathbf{z}_{jt}^{coll} = \left(\sum_{k\in\mathscr{J}_{f_ct}\setminus\{j\}}x^{(1)}_{kt},\ldots,\sum_{k\in\mathscr{J}_{f_ct}\setminus\{j\}}x^{(R)}_{kt},~\sum_{k\in\mathscr{J}_{f_ct}\setminus\{j\}}\Big(x^{(1)}_{kt}\Big)^{2},\ldots,\sum_{k\in\mathscr{J}_{f_ct}\setminus\{j\}}\Big(x^{(R)}_{kt}\Big)^{2} \right).
\end{aligned}
\end{equation}
 Note that $\mathscr{J}_{f_ct}$ denotes the set of products owned by firm $f$ and its suspected colluding partners, as $f_c$ indexes these firms as a single entity.

The choice of summation instruments, and hence the functional form in \eqref{eqn: iv testing}, follows the original BLP approach. Under logit demand, sums of powers of the characteristics approximate the pricing rule to any order (equation~\eqref{eqn: sum expansion}), and we include the first two powers. In practice, we may drop quadratic terms when constructing BLP-style instruments to avoid multicollinearity issues.

A functional form of the instruments can be flexibly specified, as illustrated by equation \eqref{eqn: z decomposition} and the results in Section \ref{subsec: exchang_general}. The core idea of our testing procedure is to exploit the exchangeability structure implied by the underlying conduct model, rather than a particular functional form of the demand system or of the basis functions $b(\cdot)$ used to construct the instruments.

 Next, we estimate the two price regressions \eqref{eqn: comp reg} and \eqref{eqn: coll reg}, which use $\mathbf{z}_{jt}^{comp}$ and $\mathbf{z}_{jt}^{coll}$, respectively, as the excluded instruments. Each set exploits the exchangeability restrictions on the observed attributes of other products implied by one of the two competing hypotheses about firm conduct: if the suspected firms collude, $\mathbf{z}_{jt}^{coll}$ should explain prices better than $\mathbf{z}_{jt}^{comp}$, whereas if they compete, the reverse holds. It is important to note that our test allows for potential misspecification of competing models. More specifically, it does not require either model \eqref{eqn: comp reg} or \eqref{eqn: coll reg} to be the true pricing equation, nor do they need to represent \textit{structural} pricing functions.\footnote{In general, a \textit{structural} pricing function cannot be estimated, as it emerges from the equilibrium behavior of firms and thus depends on numerous observed and unobserved factors.} Instead, the proposed test evaluates the relative fit of two alternative models, which makes the test easy to implement. 

 We now formalize our testing procedure by constructing the Rivers and Vuong (RV) model selection test statistic. Our RV test statistic, $T_{IV}^{RV}$, is given by:
\begin{equation}\label{eqn: iv stat}
    T_{IV}^{RV} = \frac{\sqrt{n}(\hat{Q}_{comp}-\hat{Q}_{coll})}{\hat{\sigma}_{IV}} \xrightarrow{d} N(0,1),
\end{equation}
 where $\hat{Q}_{comp}$ and $\hat{Q}_{coll}$ are the mean squared residuals of the linear regressions in \eqref{eqn: comp reg} and \eqref{eqn: coll reg}, respectively, and $\hat{\sigma}_{IV}$ is an estimator for the asymptotic standard deviation of $\sqrt{n}(\hat{Q}_{comp}-\hat{Q}_{coll})$.\footnote{In practice, we estimate the two linear regression models in \eqref{eqn: comp reg} and \eqref{eqn: coll reg} and obtain the two residuals, $\hat{e}^1_{jt}$ and $\hat{e}^2_{jt}$. We then regress the difference of the two squared residuals, $(\hat{e}_{jt}^1)^2-(\hat{e}_{jt}^2)^2$, on a constant term only. The $t$-statistic for this constant term is the RV test statistic in \eqref{eqn: iv stat}. Clustering is easily accommodated by specifying an appropriate cluster structure in the regression of the squared residual difference on the constant term.} Note that our test compares the fit of $\mathbf{z}_{jt}^{comp}$ versus $\mathbf{z}_{jt}^{coll}$, constructed under the two polar cases $\phi \in \{0, 1\}$, and does not estimate $\phi$. For intermediate values of $\phi$, own and partner products enter the pricing rule asymmetrically (Appendix~\ref{app: lemma 1}), so that neither set of instruments is exactly correct. The test then indicates which polar case better describes the observed pricing behavior; in Section~\ref{sec: mc}, we show that it favors collusion when $\phi$ is sufficiently large and competition when $\phi$ is small.

 The number of observations, $n$, is given by the number of product-market pairs, so the asymptotic sequence can take several forms: (i) $J_t$ grows while $T$ remains bounded; (ii) $T$ grows while $J_t$ remains bounded; (iii) both $J_t$ and $T$ grow. As \citet{armstrong2016large} demonstrates, BLP-style instruments can lose relevance in certain asymptotic regimes.\footnote{As an extreme example, consider a single market with an increasingly large number of products produced by single-product firms so that $n=F=J\to\infty$. In this case, markups converge to a common constant across products, and product characteristics therefore lose their explanatory power for cross-product variation in markups.} More specifically, while they remain relevant in case (ii) without additional restrictions, cases (i) and (iii) require further assumptions to preserve instrument relevance. In particular, cross-market or cross-firm variation in markups is required to maintain the relevance of BLP-style instruments in these cases (see Theorems 3 and 4 in \citet{armstrong2016large} for details). Because our test statistic becomes degenerate when BLP-style instruments lose their relevance, we focus on the asymptotic regimes under which these instruments remain relevant.

 Our RV test statistic is asymptotically normal under the null hypothesis that the two models, \eqref{eqn: comp reg} and \eqref{eqn: coll reg}, have the same fit: $Q_{comp} = Q_{coll}$, where $Q_{comp}$ and $Q_{coll}$ denote the population analogs of $\hat{Q}_{comp}$ and $\hat{Q}_{coll}$, defined in Section \ref{subsec: exchang_general}, respectively. The two alternative hypotheses are: (i) $H_1: Q_{comp}<Q_{coll}$ and (ii) $H_2: Q_{comp}>Q_{coll}$. By Lemma~\ref{lem: projection} and Proposition~\ref{prop: main}(a), $Q_{comp} > Q_{coll}$ under collusion and the conditions stated therein. Therefore, a statistically significant positive test statistic (e.g., 1.65 at the 0.05 significance level for a one-tailed test) can be interpreted as evidence of collusive behavior among suspected firms. In contrast, a statistically significant negative test statistic (e.g., -1.65 at the same significance level) suggests that firms engage in price competition. Under the conditions of Corollary~\ref{cor: logit}, $Q_{comp} - Q_{coll}$ is strictly negative under competition and strictly positive under full profit internalization ($\phi = 1$) in the logit model, so that $T_{IV}^{RV}$ diverges to $-\infty$ in the former case and to $+\infty$ in the latter: the test is consistent against either alternative.

\subsection{Existing testing procedure based on markups}\label{sec: markup testing}

 As elaborated in Section \ref{sec: intro}, there is a growing body of literature on firm conduct testing procedures that employ moment-based RV tests. These methods construct moments using the equilibrium markup conditions under the two alternative firm conduct models, and then compare the GMM objective functions of the two models to determine which one provides a better fit. This approach assumes that the researcher correctly specifies the demand system and the marginal cost function. Consequently, estimating demand parameters and specifying the functional form of marginal costs are necessary prerequisites. In this subsection, we briefly introduce the RV methodology in the existing literature. For expository purposes, we present a simplified form of this methodology, while extended discussions on this framework are provided in the footnotes accompanying the main text.

 The marginal cost function is given by:
\begin{equation}\label{eqn: marginal cost fcn}
	mc_{jt} = h(\mathbf{x}_{jt},\mathbf{w}_{jt}) + \omega_{jt},
\end{equation}
 where $\mathbf{x}_{jt}$ and $\mathbf{w}_{jt}$ represent observed product attributes and cost shifters excluded from the demand function, respectively, while $\omega_{jt}$ denotes an unobserved cost component. To simplify the discussion while conveying the main idea, we impose the following two assumptions: (i) the marginal cost function, $h(\mathbf{x}_{jt}, \mathbf{w}_{jt})$, which must be pre-specified by researchers to implement the testing procedure, is linear in $\mathbf{x}_{jt}$ and $\mathbf{w}_{jt}$; and (ii)  marginal cost is constant in output.\footnote{\cite{backus2021common} proposed a testing procedure that incorporates a flexible specification of the cost function. Additionally, the marginal cost can be nonconstant with respect to quantities produced. In such cases, the instruments are required to satisfy additional conditions to distinguish between alternative firm conduct models. See \cite{duarte2025conduct} for details and applications.}

 Let $\mathcal{D}(\mathbf{x}_t)$ denote a known demand system, where $\mathbf{x}_t$ represents a full set of exogenous product characteristics. Given demand estimates, the equilibrium markup for product $j$ under market-invariant firm conduct model $m$ (e.g., competition, collusion, partial collusion, common ownership), $\eta_{jt}^m$, is also known from the profit-maximizing conditions. The marginal costs of products can then be recovered as follows:
\begin{equation}\label{eqn: marginal cost}
	mc_{jt}^m = p_{jt} - \eta_{jt}^m(\mathbf{s}_t,\mathbf{p}_t,\mathcal{D}(\mathbf{x}_t)),
\end{equation}
 where $\mathbf{s}_t = (s_{0t}, s_{1t}, \ldots, s_{Jt})$ denotes the vector of observed market shares in the market. Under the \textit{true} conduct model $m=0$, assuming no specification error, the following moment condition holds:
\begin{equation*}
    \mathbb{E}[\omega_{jt}^0\vert\mathbf{z}_{jt}]=0,
\end{equation*}
 where $\omega_{jt}^0 = p_{jt} - \eta_{jt}^0 - h(\mathbf{x}_{jt},\mathbf{w}_{jt})$. A vector of excluded instruments, denoted by $\mathbf{z}_{jt}$, typically consists of BLP-style instruments (functions of exogenous product attributes) and cost shifters excluded from the demand specification, $\mathbf{w}_{jt}$.\footnote{\cite{backus2021common} proposed a functional form for BLP-style instruments to enhance the power of the RV testing framework, drawing on the literature on optimal instruments in nonlinear GMM settings \citep{chamberlain1987asymptotic}. This functional form results in a scalar moment that is independent of the choice of the weighting matrix used to form the objective function. Additionally, the choice of instruments (beyond BLP-style instruments) to satisfy falsifiable conditions to distinguish between two alternative conduct models affects the validity of the testing procedure. See \cite{dearing2024learning} for discussions on falsifiable conditions and empirical applications.}

 The existing methods choose between two non-nested conduct models, $m=1$ and $m=2$. More specifically, the linear parameters in the marginal cost function $h(\mathbf{x}_{jt}, \mathbf{w}_{jt})$ are estimated by running an OLS regression of the marginal cost under conduct model $m$:
\begin{equation*}
    p_{jt} - \eta_{jt}^m = h(\mathbf{x}_{jt}, \mathbf{w}_{jt}) + \omega_{jt}^m,
\end{equation*}
resulting in different estimates across conduct models $m=1$ and $m=2$.\footnote{In a more flexible framework, the marginal cost function can be estimated using non-parametric regression. For example, \cite{backus2021common} employed random forest regression \citep{breiman2001random} to better capture nonlinear relationships.} After obtaining the residual $\hat{\omega}_{jt}^m$, the GMM objective function, $\hat{Q}_m$, is constructed from the equilibrium markup condition under conduct model $m$ as follows:
\begin{equation*}
    \hat{Q}_m = \hat{g}_m' \hat{W} \hat{g}_m,
\end{equation*}
where, given the total number of observations $n$,
\begin{equation*}
    \hat{g}_m = n^{-1} \sum_{t} \sum_{j} \mathbf{z}_{jt}'\hat{\omega}_{jt}^m \quad \textrm{and} \quad \hat{W} = n \cdot \left(\sum_t \sum_j \mathbf{z}_{jt}' \mathbf{z}_{jt}\right)^{-1}.
\end{equation*}

 Finally, the RV test statistic based on equilibrium markup conditions, $T_{markup}^{RV}$, is given by:
\begin{equation}\label{eqn: markup stat}
    T_{markup}^{RV} = \frac{\sqrt{n}(\hat{Q}_{1}-\hat{Q}_{2})}{\hat{\sigma}_{markup}} \xrightarrow{d} N(0,1),
\end{equation}
 where $\hat{\sigma}_{markup}$ is an estimator for the asymptotic standard deviation of $\sqrt{n}(\hat{Q}_1-\hat{Q}_2)$. The exact form of $\hat{\sigma}_{markup}$, along with a detailed discussion on adjustments for clustering and two-step estimation errors arising from demand estimation, is provided in \cite{duarte2024testing}. This RV statistic can be computed using the Python package \texttt{PyRVtest} \citep{pyrvtest}.\footnote{Additionally, diagnostics for weak instruments in this testing framework, as illustrated in \cite{duarte2024testing}, can be executed using this package.}

\subsection{Comparison of the two test procedures}

 The basic idea behind our proposed conduct testing procedure aligns with that of existing methods: to select the firm conduct model that best fits the observed market outcomes. In our approach, model fit is evaluated using the price regressions, whereas the existing markup-based RV test relies on the GMM objective function.

 The advantage of the existing statistic, $T_{markup}^{RV}$, is that researchers can test any model of conduct, provided the first-order profit-maximizing conditions can be derived from it, and determine which model best fits the observed data based on the moment restrictions imposed on the supply-side model. Unlike these methods, our testing procedure is tailored to comparing competitive and collusive conduct among price-setting firms. However, it does not require demand estimation or marginal cost specification, making it straightforward to implement. Furthermore, our procedure is data-efficient, as it can be performed without market share data or additional instruments including exogenous cost shifters. Importantly, the exogeneity assumption on observed characteristics is not required. As demonstrated in the following section, the performance of our testing method is comparable to that of the existing markup-based RV test.

 Consequently, these two approaches can complement each other, creating a more practical, efficient, and powerful framework for testing firm conduct. More specifically, our method can serve as a preliminary tool for researchers and regulatory authorities to diagnose collusive behavior among suspected firms in the market. Once preliminary results indicate the presence of collusion, researchers or regulatory authorities can proceed with existing tests on industry conduct.

\section{Monte Carlo evidence}\label{sec: mc}

 In this section, we conduct an extensive Monte Carlo study to demonstrate the validity of our proposed test and show that it performs comparably to, or better than, the existing markup-based RV test in detecting collusion across a variety of scenarios. 

\subsection{Monte Carlo set-up}\label{sec: mc setup}

We begin by introducing the Monte Carlo design and the computation of the test statistics.

\paragraph{Data-generating process}

 Our data-generating process (DGP) builds upon the framework in \cite{armstrong2016large} and \cite{conlon2020best}, with a focus on modeling various collusive behaviors among price-setting firms. For each Monte Carlo configuration, the number of products ($J$) and the number of firms ($F$) are fixed across markets ($t=1,\ldots,T$). Each firm produces $J/F$ products.\footnote{When $J/F$ is not an integer, some firms are randomly assigned to produce $\lceil J/F \rceil$ products, while others produce $\lfloor J/F \rfloor$ products, ensuring that the total number of products remains $J$. Here, $\lceil \cdot \rceil$ denotes the ceiling function, which rounds a number up to the nearest integer, and $\lfloor \cdot \rfloor$ denotes the floor function, which rounds a number down to the nearest integer.} This allocation results in nearly symmetric market shares across firms and markets under the DGP described below.

 The indirect utility of consumer $i$ in market $t$ from consuming product $j$, and the marginal cost of product $j$ in market $t$, are specified as follows:
\begin{align}
	&u_{ijt} = \beta_1 + \beta_2x_{jt} - \alpha p_{jt} + \xi_{jt} + \sigma_xv_ix_{jt} + \varepsilon_{ijt}, \label{eqn: dgp utility} \\
	&mc_{jt} = \gamma_1 + \gamma_2 x_{jt} + \gamma_3 w_{jt} + \omega_{jt}. \label{eqn: dgp mc}
\end{align}
 There are two exogenous product characteristics, a constant term and \(x_{jt}\), and one exogenous cost shifter, \(w_{jt}\), where both \(x_{jt}\) and \(w_{jt}\) are randomly drawn from a standard uniform distribution. Price, $p_{jt}$, is an endogenous product characteristic determined by equilibrium conditions in a differentiated goods market in which firms set prices. The true linear demand parameters are given by $\alpha=1$, $\beta_1=-4.5$, and $\beta_2=6$. When included, the heterogeneous component of demand is given by $\sigma_xv_ix_{jt}$, where $\sigma_x=3$ and $v_i$ follows a standard normal distribution. Cost parameters are given by $\gamma_1=2$, $\gamma_2=1$, and $\gamma_3=0.2$. The unobservable components of demand and cost, $(\xi_{jt},\omega_{jt})$, are randomly drawn from a mean-zero bivariate normal distribution with variances $\sigma^2_{\xi}=0.2$, $\sigma^2_{\omega}=0.2$, and covariance $\sigma_{\xi\omega}=0.1$.

 Firms may engage in collusion, internalizing the profits of their colluding partners as part of their own. The degree of collusion is measured by $F/F_c$, where $F_c \in \{1, 2, \ldots, F-1\}$ denotes the effective number of competitors, i.e., the number of distinct competitive entities after treating all colluding firms as a single entity. We assume that the first $F-F_c+1$ firms either collude or are suspected of colluding by the regulatory authority. For example, suppose $F = 6$ and $F_c = 4$. In this case, the degree of collusion is 1.5, with the first three firms ($f = 1, 2, 3$) colluding or suspected of colluding to jointly set profit-maximizing prices, leaving the effective competitors indexed as $f_c = 1, 4, 5, 6$. The degree of profit internalization, represented by the conduct parameter $\phi \in [0,1]$, quantifies the weight assigned to the profits of colluding partners, as shown in equation \eqref{eqn: profit}. Continuing with the previous example, when $F_c = 4$ but $\phi = 0$, this configuration represents a scenario in which regulatory authorities falsely suspect three firms of collusion when, in fact, they are engaged in price competition.

 Equilibrium prices in each market are endogenously determined by the first-order conditions \eqref{eqn: foc bertrand}. These prices are solved using the fixed-point algorithm proposed by \cite{morrow2011fixed} and implemented in \texttt{PyBLP} \citep{conlon2020best}. We use nine Gauss-Hermite quadrature nodes to numerically integrate the choice probabilities when a random coefficient is included in the indirect utility function.

 We also consider the case in which the product attribute is correlated with the unobservable demand component, $\xi_{jt}$, making it endogenous. We denote the endogenous attribute as $x_{jt}^{endo}$ and derive it as follows. First, we compute \( x_{jt}^{\text{unscaled}} = x_{jt} + \rho \cdot \xi_{jt} \), where \( x_{jt} \) is drawn from a standard uniform distribution and \( \xi_{jt} \) from a bivariate normal distribution, as described earlier. The parameter \( \rho \in \{-10, -5, -1, 0, 1, 5, 10\} \) controls the degree and direction of endogeneity. Next, given that the variation of $x_{jt}^{unscaled}$ depends on $\rho$, we apply min-max normalization to define $x_{jt}^{endo} = \frac{x_{jt}^{unscaled} - x_t^{\min}}{x_t^{\max} - x_t^{\min}}$, where $x_t^{\min}$ and $x_t^{\max}$ are the minimum and maximum values of $x_{jt}^{unscaled}$ in market $t$. In this way, we fix the support of the endogenous attribute: $x_{jt}^{endo} \in [0, 1]$ for any $\rho$.

 Each Monte Carlo configuration is represented by a unique five-tuple $(J, F, T, \phi, F_c)$. Specifically, $J = 36$, $F \in \{1, 2, \ldots, 36\}$, $T = 100$, $\phi \in \{0, 0.1, \ldots, 1\}$, and $F_c \in \{1, 2, \ldots, F-1\}$.\footnote{When the DGP includes an endogenous product attribute, $\phi$ is fixed at 1, and $\rho$ is varied to generate unique five-tuples $(J, F, T, \rho, F_c)$.} For each configuration, we generate 500 simulated datasets ($S=500$) to evaluate the finite-sample performance of the proposed test statistic, $T_{IV}^{RV}$, and compare it with the performance of the test based on equilibrium markup conditions, $T_{markup}^{RV}$. We also investigate the performance of $\mathbf{z}^{comp}$ and $\mathbf{z}^{coll}$ as instruments for demand estimation. Furthermore, in Appendix \ref{sec: own vs other}, we focus on cases with varying $F$ and fixed $J$ but without collusive behavior ($\phi = 0$ and $F_c = F$) to evaluate the performance of BLP-style IVs, following the approach of \cite{armstrong2016large}.

\paragraph{Computation of test statistics}

 For each dataset with configuration $(J, F, T, \phi, F_c)$, we construct $\mathbf{z}^{comp}$ and $\mathbf{z}^{coll}$ as outlined in Section~\ref{sec: iv testing}.\footnote{Note that the summation of the constant term, often referred to as the product counts IV, cannot be utilized in our DGP setting because it fails to generate cross-firm or cross-market variations in the number of products. Moreover, the inclusion of second-order polynomial IVs in \eqref{eqn: iv testing} ensures that the rank condition for the GMM framework is satisfied in our DGP when a random coefficient is incorporated into the indirect utility function \eqref{eqn: dgp utility}.} Note that even when $\phi = 0$, we still construct $\mathbf{z}^{coll}$ based on a scenario in which the researcher falsely suspects collusion among the firms of interest. To assess the relevance of these two instrument sets under a given true conduct model, we compute two F-statistics associated with (i) $\bm{\theta}^1 = 0$ (Fstat$_1$) and (ii) $\bm{\theta}^2 = 0$ (Fstat$_2$) from equations \eqref{eqn: comp reg} and \eqref{eqn: coll reg}, respectively. Our proposed test statistic, $T_{IV}^{RV}$, is then calculated using equation \eqref{eqn: iv stat}. The indicator $D_j$ included in both price regressions \eqref{eqn: comp reg} and \eqref{eqn: coll reg} absorbs level differences in prices between suspected and non-suspected products. Consequently, the fit comparison exploits only the exchangeability of own and partner products within the suspected coalition.\footnote{Omitting $D_j$ substantially raises the power of our test because $\mathbf z^{part}_j$ then also picks up the price premium of suspected products (Appendix~\ref{app: prop 1}). That premium, however, may reflect differences in costs, quality, or brand rather than collusion, in which case omitting $D_j$ would lead to false detection under competition. We therefore include $D_j$ in the baseline. A specification without $D_j$ may serve as a supplementary diagnostic when the suspected coalition is small and suspected and non-suspected firms appear comparable, but its results should be interpreted with caution.}  Heteroskedasticity-robust standard errors are used to compute $\hat{\sigma}_{IV}$ in equation \eqref{eqn: iv stat}.

 To compute $T_{markup}^{RV}$, we first estimate the demand parameters using the standard own- and other-firm IVs, defined as:
\begin{equation}\label{eqn: own and other iv}
    \mathbf{z}_{jt}^{own} = \left( \sum_{k \in \mathscr{J}_{ft} \setminus \{j\}} x_{kt}, \sum_{k \in \mathscr{J}_{ft} \setminus \{j\}} x_{kt}^2 \right), \quad
    \mathbf{z}_{jt}^{other} = \left( \sum_{k \in \mathscr{J}_{t} \setminus \mathscr{J}_{ft}} x_{kt}, \sum_{k \in \mathscr{J}_{t} \setminus \mathscr{J}_{ft}} x_{kt}^2 \right),
\end{equation}
 as excluded instruments $\mathbf{z}_{jt} = \left(\mathbf{z}_{jt}^{own}, \mathbf{z}_{jt}^{other}\right)$. Note that these instruments are constructed based on \textit{observed} ownership rather than suspected ownership. Next, we specify the marginal cost function as a linear function of the constant term and $x_{jt}$ only, even though the true marginal cost \eqref{eqn: dgp mc} also depends on the cost shifter $w_{jt}$.\footnote{Accordingly, we do not include $w_{jt}$ as an instrument when estimating the demand.} Thus, we consider a scenario in which the cost function is potentially misspecified, either due to data limitations on cost shifters or the omission of a relevant cost shifter. This approach enables us to evaluate the performance of the two testing procedures without using additional data. We then follow the procedure outlined in Section \ref{sec: markup testing} to construct $T_{markup}^{RV}$ under two alternative firm conduct models: one with $\phi = 0$ (competition) and the other with $\phi = 1$ (full profit internalization within the suspected coalition, i.e., the effective-firm structure of the DGP).\footnote{Although the testing procedure built on equilibrium markup conditions allows us to compare conduct models with other values of $\phi$, such as $\phi = 0.2$ versus $\phi = 0.8$, the primary interest of researchers and regulatory authorities is often to test full profit internalization ($\phi = 1$) against competition ($\phi = 0$). Therefore, we compute $T_{markup}^{RV}$ under these two conduct models.} The standard error $\hat{\sigma}_{markup}$ in equation \eqref{eqn: markup stat} is adjusted for two-step demand estimation errors and is heteroskedasticity-robust. The computation of $T_{markup}^{RV}$ is carried out with \texttt{PyRVtest} \citep{pyrvtest}.

\subsection{Strength of instruments and estimation performance}\label{sec: estimation performance}

 Before comparing the testing performance of $T_{IV}^{RV}$ and $T_{markup}^{RV}$, we first evaluate the performance of the two instrument sets: $\mathbf{z}^{comp}$ and $\mathbf{z}^{coll}$. As a baseline scenario, we consider a case in which four of the six firms collude ($F = 6, F_c = 3$). We generate Monte Carlo datasets, following the design in Section \ref{sec: mc setup}, for various values of $\phi \in [0, 1]$. The top panel of Table \ref{tab: Fstat evidence} presents the results when a random coefficient is excluded from the indirect utility function \eqref{eqn: dgp utility} in the DGP, while the bottom panel shows the results when it is included.

 First, we examine and compare the two F-statistics: Fstat$_1$, associated with the null hypothesis $H_0: \bm{\theta}^1 = 0$ in equation \eqref{eqn: comp reg}, and Fstat$_2$, associated with $H_0: \bm{\theta}^2 = 0$ in equation \eqref{eqn: coll reg}. The results indicate that as the degree of profit internalization increases, Fstat$_1$ decreases moderately, whereas Fstat$_2$ increases sharply. As a result, the median of Fstat$_2$ exceeds that of Fstat$_1$ for $\phi \geq 0.5$ without a random coefficient and for $\phi \geq 0.7$ with a random coefficient, and Fstat$_2$ exceeds Fstat$_1$ in more than 95 percent of the simulated datasets at $\phi \geq 0.9$ in both cases. These findings align with the intuition behind our testing procedure: $\mathbf{z}^{coll}$ is expected to yield a stronger price regression when firms collude, whereas $\mathbf{z}^{comp}$ should perform better under competition. Taken together, the results provide preliminary evidence in favor of our testing procedure, which compares the model fit of the two price regressions.

 Results from more extensive simulations with varying numbers of firms ($F$) and effective competitors ($F_c$) are illustrated in Figure \ref{fig: Fstat evidence} in the Appendix. The figure graphically depicts the share of cases where Fstat$_2 >$ Fstat$_1$ in 500 simulation results for each Monte Carlo configuration. For a given degree of collusion (measured by the ratio $F/F_c$), the relative strength of $\mathbf{z}^{coll}$ always increases as the degree of profit internalization $\phi$ rises. Moreover, once the degree of profit internalization is sufficiently high, the more collusive the firms in the market are (higher $F/F_c$), the stronger the relevance of $\mathbf{z}^{coll}$.
 
\begin{table}[!t]
\small
  \centering
  \caption{Comparison of IV performance: $\mathbf{z}^{comp}$ vs $\mathbf{z}^{coll}$ ($F=6$, $F_c=3$, $T=100$)}
    \begin{tabular}{lrrrrrrrrr}
    \toprule
          & \multicolumn{3}{c}{$\mathbf{z}^{comp}$ } &       & \multicolumn{3}{c}{$\mathbf{z}^{coll}$} &       &  \\
\cmidrule{2-4}\cmidrule{6-8}     & \multicolumn{7}{c}{Median values ($S=500$) of}        & \multicolumn{1}{r}{\% of Fstat$_2$} & Median \\
    $\phi$ & \multicolumn{1}{c}{$\vert\alpha-\hat{\alpha}\vert$} & \multicolumn{1}{c}{$\vert\sigma_x-\hat{\sigma}_x\vert$} & \multicolumn{1}{c}{$\text{Fstat}_1$} &       & \multicolumn{1}{c}{$\vert\alpha-\hat{\alpha}\vert$} & \multicolumn{1}{c}{$\vert\sigma_x-\hat{\sigma}_x\vert$} & \multicolumn{1}{c}{$\text{Fstat}_2$} & \multicolumn{1}{r}{ $>$ Fstat$_1$} & \multicolumn{1}{c}{of $s_0$} \\
    \midrule
    \multicolumn{10}{c}{\underline{\textit{Panel A: without a random coefficient}}} \\
    \\
    0  & 0.174 &        & 7.124 &        & 0.362 &        & 1.886 & 0.040 & 0.653 \\
    0.1 & 0.171 &        & 7.032 &        & 0.261 &        & 2.650 & 0.070 & 0.655 \\
    0.2 & 0.175 &        & 6.961 &        & 0.194 &        & 3.601 & 0.136 & 0.658 \\
    0.3 & 0.176 &        & 6.844 &        & 0.142 &        & 4.778 & 0.264 & 0.661 \\
    0.4 & 0.178 &        & 6.749 &        & 0.112 &        & 6.100 & 0.396 & 0.664 \\
    0.5 & 0.178 &        & 6.605 &        & 0.092 &        & 7.513 & 0.598 & 0.666 \\
    0.6 & 0.177 &        & 6.550 &        & 0.080 &        & 9.152 & 0.726 & 0.669 \\
    0.7 & 0.177 &        & 6.427 &        & 0.069 &        & 10.865 & 0.848 & 0.672 \\
    0.8 & 0.177 &        & 6.367 &        & 0.062 &        & 12.841 & 0.934 & 0.674 \\
    0.9 & 0.177 &        & 6.269 &        & 0.055 &        & 14.935 & 0.976 & 0.677 \\
    1 & 0.178 &        & 6.201 &        & 0.050 &        & 17.219 & 0.996 & 0.679 \\
    \midrule
    \multicolumn{10}{c}{\underline{\textit{Panel B: with a random coefficient}}} \\
    \\
    0  & 0.504 & 1.989 & 17.770 &        & 0.918 & 0.401 & 1.737 & 0.000 & 0.595 \\
    0.1 & 0.504 & 1.901 & 17.236 &        & 0.353 & 0.304 & 2.582 & 0.002 & 0.599 \\
    0.2 & 0.505 & 1.736 & 16.801 &        & 0.167 & 0.245 & 3.858 & 0.002 & 0.602 \\
    0.3 & 0.519 & 1.708 & 16.284 &        & 0.108 & 0.227 & 5.442 & 0.004 & 0.606 \\
    0.4 & 0.516 & 1.689 & 15.841 &        & 0.079 & 0.228 & 7.543 & 0.030 & 0.610 \\
    0.5 & 0.504 & 1.673 & 15.326 &        & 0.062 & 0.234 & 10.069 & 0.152 & 0.614 \\
    0.6 & 0.501 & 1.610 & 14.828 &        & 0.050 & 0.238 & 13.197 & 0.372 & 0.618 \\
    0.7 & 0.490 & 1.527 & 14.286 &        & 0.043 & 0.241 & 16.979 & 0.660 & 0.622 \\
    0.8 & 0.483 & 1.464 & 13.691 &        & 0.037 & 0.247 & 21.319 & 0.866 & 0.626 \\
    0.9 & 0.483 & 1.440 & 13.119 &        & 0.033 & 0.252 & 26.182 & 0.970 & 0.630 \\
    1 & 0.467 & 1.368 & 12.399 &        & 0.030 & 0.259 & 31.066 & 0.992 & 0.634 \\
    \bottomrule
    \end{tabular}%
  \label{tab: Fstat evidence}%
\tablenotes The table compares the median absolute errors of the estimated price and nonlinear coefficients, as well as the median F-statistics, across 500 simulated datasets for each Monte Carlo configuration ($J=36, F=6, T=100, \phi, F_c=3$), obtained using $\mathbf{z}^{comp}$ and $\mathbf{z}^{coll}$ as instruments individually. The top panel presents the results when a random coefficient is excluded from the indirect utility function \eqref{eqn: dgp utility} in the DGP, while the bottom panel presents the results when it is included. We note that under the random coefficient specification, $\mathbf{z}^{comp}$ yields large median absolute errors for $\hat{\alpha}$ and $\hat{\sigma}_x$ even when firms engage in price competition. In Appendix \ref{sec: own vs other}, we show that including other-firm instruments greatly improves estimation performance.
\end{table}%

 Our simulation results presented in Table \ref{tab: Fstat evidence} also show that the median absolute error of the estimated price coefficient $\hat{\alpha}$ decreases significantly in response to a rise in $\phi$ when using $\mathbf{z}^{coll}$ as instruments. As a result, $\mathbf{z}^{coll}$ outperforms $\mathbf{z}^{comp}$ in estimating $\alpha$ when firms collude, even with a low degree of profit internalization ($\phi \geq 0.3$ without a random coefficient and $\phi \geq 0.1$ with a random coefficient). Additionally, under the random coefficient specification, $\mathbf{z}^{comp}$ yields large median absolute errors for $\hat{\alpha}$ even when firms engage in price competition. While our focus here is primarily on comparing the strength of $\mathbf{z}^{comp}$ and $\mathbf{z}^{coll}$ in the price regression, in Appendix~\ref{sec: own vs other}, we show that the inclusion of other-firm instruments greatly improves estimation performance.

 As $\phi$ rises, the median absolute error of the estimated non-linear coefficient $\hat{\sigma}_x$ obtained using $\mathbf{z}^{coll}$ declines initially but changes little thereafter. Moreover, for any $\phi$, the median absolute error obtained using $\mathbf{z}^{coll}$ as instruments is smaller than that obtained using $\mathbf{z}^{comp}$. This finding does not contradict our earlier observation that, when the degree of profit internalization is low ($\phi \leq 0.6$ with a random coefficient), using $\mathbf{z}^{coll}$ instead of $\mathbf{z}^{comp}$ leads to lower F-statistics in the price regression. This is because identifying $\sigma_x$ requires instruments that are correlated not only with price but also with market shares.\footnote{As a simple illustration, consider a nested logit specification, a type of random coefficient model. After inverting a demand system \citep{Berry94}, the resulting estimation equation using a standard linear instrumental variables approach is as follows:
\begin{equation*}
    \log \frac{s_{jt}}{s_{0t}} = \beta x_{jt} - \alpha p_{jt} + \sigma \log s_{j\vert g,t} + \xi_{jt},
\end{equation*}
 where $s_{j\vert g,t}$ denotes the market share of product $j$ within group $g$. Identifying this random coefficient model requires instruments for both the endogenous price and the endogenous market share $s_{j\vert g,t}$. See \cite{berry2014identification} for the instrument requirements under general nonparametric settings.} We revisit this point in Appendix \ref{sec: own vs other}; we show that own-firm instruments are more relevant in the price regression and more informative about the price coefficient, whereas other-firm instruments are more useful for identifying the nonlinear coefficient. Consequently, $\mathbf{z}^{coll}$ yields smaller errors for $\hat{\sigma}_x$ even at low values of $\phi$.

\subsection{Comparison of testing power}

 Now, we compute and compare the two test statistics, $T_{IV}^{RV}$ and $T_{markup}^{RV}$, to investigate how the degree of profit internalization ($\phi$) and the degree of collusion ($F/F_c$) influence their values. Recall that for both statistics, a statistically significant positive value (greater than 1.65 at the 0.05 significance level) indicates collusion, while a statistically significant negative value (less than -1.65 at the same level) indicates competition.

 Table \ref{tab: exo vuong} reports the median values of these two statistics under various Monte Carlo configurations elaborated in Section \ref{sec: mc setup}. We first examine the results without a random coefficient presented in the upper panel. Overall, our test statistic, $T_{IV}^{RV}$, performs comparably to or better than the existing statistic, $T_{markup}^{RV}$, in detecting collusion.\footnote{When $F_c=1$, there is no within-market variation in the collusion IVs, and their relevance relies solely on cross-market variations. This suggests that the performance of our test may improve as the number of available markets increases.} For example, when all six firms collude ($F_c = 1$) and fully internalize the profits of other firms ($\phi = 1$), both tests reject the competition hypothesis with median statistics of 3.309 for $T_{IV}^{RV}$ and 3.564 for $T_{markup}^{RV}$. As the degree of profit internalization decreases, the median of $T_{IV}^{RV}$ also declines but remains above 1.65 for $\phi \geq 0.7$, while the median of $T_{markup}^{RV}$ exceeds 1.65 only for $\phi \geq 0.9$. Our testing procedure continues to perform better in detecting collusion than the markup-based test when only a subset of firms is suspected ($F_c \geq 2$). In contrast, the markup-based test tends to outperform our test in rejecting collusion in favor of competition at lower values of $\phi$. For example, when $F_c = 1$ and $\phi = 0.1$, only $T_{markup}^{RV}$ has a median below -1.65 (-6.088), allowing the rejection of collusion.

\begin{table}[!t]
\scriptsize
  \centering
  \caption{(Exogenous) $F=6$ and $T=100$}
    \begin{tabular}{lrrrrrrrrrrr}
    \toprule
     & \multicolumn{5}{c}{$T_{IV}^{RV}$ ($\mathbf{z}^{comp}$ vs $\mathbf{z}^{coll}$)}&       & \multicolumn{5}{c}{$T_{markup}^{RV}$ ($\phi=0$ vs $\phi=1$)} \\
\cmidrule{2-6}\cmidrule{8-12}    $\phi$ & \multicolumn{1}{c}{$F_c=1$} & \multicolumn{1}{c}{$F_c=2$} & \multicolumn{1}{c}{$F_c=3$} & \multicolumn{1}{c}{$F_c=4$} & \multicolumn{1}{c}{$F_c=5$} &       & \multicolumn{1}{c}{$F_c=1$} & \multicolumn{1}{c}{$F_c=2$} & \multicolumn{1}{c}{$F_c=3$} & \multicolumn{1}{c}{$F_c=4$} & \multicolumn{1}{c}{$F_c=5$} \\
    \midrule
    \multicolumn{12}{c}{\underline{\textit{Panel A: without a random coefficient}}} \\
    \\
    0 (competition) & -1.657 & -1.594 & -1.537 & -1.319 & -0.993 &       & -7.133 & -3.872 & -2.143 & -1.033 & -0.385 \\
    0.1   & -1.458 & -1.371 & -1.328 & -1.131 & -0.818 &       & -6.088 & -3.360 & -1.839 & -0.880 & -0.326 \\
    0.2   & -1.112 & -1.068 & -1.023 & -0.882 & -0.636 &       & -5.106 & -2.819 & -1.524 & -0.719 & -0.271 \\
    0.3   & -0.604 & -0.627 & -0.669 & -0.613 & -0.456 &       & -4.083 & -2.228 & -1.175 & -0.556 & -0.227 \\
    0.4   & 0.011 & -0.149 & -0.257 & -0.321 & -0.269 &       & -2.930 & -1.614 & -0.820 & -0.393 & -0.165 \\
    0.5   & 0.650 & 0.399 & 0.175 & 0.006 & -0.081 &       & -1.758 & -0.958 & -0.453 & -0.221 & -0.131 \\
    0.6   & 1.272 & 0.929 & 0.590 & 0.326 & 0.107 &       & -0.644 & -0.300 & -0.081 & -0.055 & -0.075 \\
    0.7   & 1.849 & 1.455 & 1.012 & 0.652 & 0.298 &       & 0.477 & 0.380 & 0.277 & 0.119 & -0.011 \\
    0.8   & 2.372 & 1.908 & 1.420 & 0.964 & 0.492 &       & 1.553 & 1.010 & 0.593 & 0.305 & 0.043 \\
    0.9   & 2.851 & 2.333 & 1.798 & 1.258 & 0.692 &       & 2.597 & 1.622 & 0.944 & 0.469 & 0.097 \\
    1     & 3.309 & 2.735 & 2.156 & 1.543 & 0.891 &       & 3.564 & 2.208 & 1.283 & 0.643 & 0.162 \\
    \midrule
    \multicolumn{12}{c}{\underline{\textit{Panel B: with a random coefficient}}} \\
    \\
    0 (competition) & -2.737 & -2.884 & -2.950 & -2.746 & -2.138 &       & -8.601 & -2.536 & -1.480 & -0.754 & -0.383 \\
    0.1   & -2.636 & -2.829 & -2.786 & -2.502 & -1.872 &       & -8.092 & -2.353 & -1.322 & -0.640 & -0.318 \\
    0.2   & -2.549 & -2.680 & -2.549 & -2.190 & -1.566 &       & -7.535 & -2.126 & -1.158 & -0.512 & -0.255 \\
    0.3   & -2.460 & -2.468 & -2.208 & -1.820 & -1.248 &       & -6.894 & -1.844 & -0.944 & -0.354 & -0.181 \\
    0.4   & -2.320 & -2.076 & -1.722 & -1.366 & -0.899 &       & -6.091 & -1.504 & -0.676 & -0.200 & -0.083 \\
    0.5   & -2.109 & -1.471 & -1.089 & -0.839 & -0.545 &       & -5.143 & -1.090 & -0.412 & -0.037 & 0.004 \\
    0.6   & -1.659 & -0.652 & -0.336 & -0.263 & -0.188 &       & -4.004 & -0.559 & -0.036 & 0.110 & 0.071 \\
    0.7   & -0.732 & 0.361 & 0.467 & 0.333 & 0.199 &       & -2.480 & 0.009 & 0.276 & 0.280 & 0.149 \\
    0.8   & 0.721 & 1.382 & 1.272 & 0.934 & 0.566 &       & -0.971 & 0.686 & 0.556 & 0.428 & 0.199 \\
    0.9   & 2.290 & 2.348 & 2.020 & 1.530 & 0.931 &       & -0.833 & 1.299 & 0.818 & 0.548 & 0.271 \\
    1     & 3.620 & 3.206 & 2.716 & 2.046 & 1.287 &       & -0.459 & 1.738 & 1.075 & 0.708 & 0.338 \\
    \bottomrule
    \end{tabular}%
  \label{tab: exo vuong}%
\tablenotes The table reports the median values of the two test statistics, $T_{IV}^{RV}$ and $T_{markup}^{RV}$, across 500 simulated datasets for each Monte Carlo configuration ($J=36, F=6, T=100, \phi, F_c$). $T_{markup}^{RV}$ is constructed under the two alternative firm conduct models: one with $\phi = 0$ (competition) and the other with $\phi = 1$ (full profit internalization within the suspected coalition, i.e., the effective-firm structure of the DGP). The top panel presents the results when a random coefficient is excluded from the indirect utility function \eqref{eqn: dgp utility} in the DGP, while the bottom panel presents the results when it is included.
\end{table}%

 The relative testing power of our method is higher when a random coefficient is included in the consumer utility function.\footnote{This higher testing power suggests that overall, the price effect of product attributes is more sensitive to industry conduct under the random coefficient specification than under the logit specification employed in our Monte Carlo study.} For every $F_c$, it rejects collusion in favor of competition at low values of $\phi$, while detecting collusion at high values of $\phi$ unless only two of the six firms are suspected ($F_c = 5$). Conversely, the inclusion of a random coefficient tends to reduce the testing power of $T_{markup}^{RV}$. These results highlight the advantage of our approach: it performs comparably to, or outperforms, the existing method under various collusive scenarios, while its implementation does not require estimating the demand and supply system.

In our simulations, both tests lose power as the coalition becomes smaller, most notably when only two of the six firms collude ($F_c = 5$). At $\phi = 1$, the median of $T^{RV}_{IV}$ is 0.891 (1.287 with a random coefficient), compared with 0.162 (0.338) for $T^{RV}_{markup}$. This is because a small coalition has only a modest effect on prices, and few products carry a partner signal ($r = 0.4$, compared with $r = 6$ when $F_c = 1$ where $r$ denotes the expected number of partner products per own product), so the fit differential in Proposition \ref{prop: main} is small. 

Figure \ref{fig: p-value iv vuong} in the Appendix presents the median p-values of $T_{IV}^{RV}$ under the alternative hypothesis, $H_2: Q_{comp} > Q_{coll}$, derived from extensive additional simulations. Clearly, there is a positive monotonic relationship between testing power and the degree of profit internalization across all $(F, F_c)$ configurations considered in our Monte Carlo study. The degree of collusion ($F/F_c$) also influences testing power; once the degree of profit internalization is sufficiently high, a higher degree of collusion corresponds to greater power in detecting collusive behavior among firms.

 So far, we have treated the product characteristic, $x_{jt}$, as an exogenous variable. However, this assumption may not hold in situations where firms determine product attributes and prices simultaneously with the contemporaneous unobservable components of demand, $\xi_{jt}$. While BLP-style instruments fail to satisfy the exclusion restriction in such cases, our test does not require the instruments to be exogenous; it compares the fit of two price regressions and never uses them to estimate demand.
  
 To evaluate the testing power of the two statistics when $x_{jt}$ is endogenous, we generate endogenous product attributes as described in Section \ref{sec: mc setup}. The true profit internalization parameter, $\phi$, is fixed at 1, while the direction and degree of endogeneity are parameterized by $\rho \in \{-10, -5, -1, 0, 1, 5, 10\}$ as explained in Section \ref{sec: mc setup}.

 The top panel of Table \ref{tab: endo vuong} reports the results obtained without a random coefficient in the indirect utility specification \eqref{eqn: dgp utility} in the DGP. While the power of both test statistics remains robust across various endogeneity scenarios, the direction of correlation between $\xi_{jt}$ and $x_{jt}$ affects the testing power. When the correlation is negative, the testing power declines; conversely, when the correlation is positive, the testing power improves. Results obtained with a random coefficient in the utility specification, presented in the bottom panel of the table, indicate that the testing power of both statistics tends to be higher when $x_{jt}$ is endogenous.
 
 In sum, while both statistics remain informative under endogenous product characteristics, our test statistic, $T_{IV}^{RV}$, exhibits higher power than the existing statistic, $T_{markup}^{RV}$, across all specifications other than $F_c = 1$, where both statistics remain well above 1.65 (except for $T_{markup}^{RV}$ in the exogenous case with a random coefficient).
 
\begin{table}[!t]
\scriptsize
  \centering
  \caption{(Endogenous) $F=6$ and $T=100$}
    \begin{tabular}{lrrrrrrrrrrr}
    \toprule
      & \multicolumn{5}{c}{$T_{IV}^{RV}$ ($\mathbf{z}^{comp}$ vs $\mathbf{z}^{coll}$)} &       & \multicolumn{5}{c}{$T_{markup}^{RV}$ ($\phi=0$ vs $\phi=1$)} \\
\cmidrule{2-6}\cmidrule{8-12}    $\rho$ & \multicolumn{1}{c}{$F_c=1$} & \multicolumn{1}{c}{$F_c=2$} & \multicolumn{1}{c}{$F_c=3$} & \multicolumn{1}{c}{$F_c=4$} & \multicolumn{1}{c}{$F_c=5$} &       & \multicolumn{1}{c}{$F_c=1$} & \multicolumn{1}{c}{$F_c=2$} & \multicolumn{1}{c}{$F_c=3$} & \multicolumn{1}{c}{$F_c=4$} & \multicolumn{1}{c}{$F_c=5$} \\
    \midrule
    \multicolumn{12}{c}{\underline{\textit{Panel A: without a random coefficient}}} \\
    Exogenous &       &       &       &       &       &       &       &       &       &       &  \\
    \hspace{0.1in}$\rho=0$ & 3.546 & 2.985 & 2.336 & 1.652 & 0.959 &       & 3.905 & 2.469 & 1.489 & 0.739 & 0.243 \\
    Endogenous $(-)$ &       &       &       &       &       &       &       &       &       &       &  \\
    \hspace{0.1in}$\rho=-1$ & 2.523 & 2.075 & 1.620 & 1.145 & 0.664 &       & 2.853 & 1.825 & 1.053 & 0.528 & 0.193 \\
    \hspace{0.1in}$\rho=-5$ & 2.446 & 2.014 & 1.568 & 1.070 & 0.653 &       & 2.782 & 1.755 & 1.011 & 0.484 & 0.139 \\
    \hspace{0.1in}$\rho=-10$ & 2.440 & 2.020 & 1.581 & 1.069 & 0.638 &       & 2.818 & 1.763 & 1.038 & 0.489 & 0.111 \\
    Endogenous $(+)$ &       &       &       &       &       &       &       &       &       &       &  \\
    \hspace{0.1in}$\rho=1$ & 4.494 & 3.740 & 2.983 & 2.162 & 1.290 &       & 4.858 & 3.060 & 1.853 & 0.967 & 0.366 \\
    \hspace{0.1in}$\rho=5$ & 4.849 & 4.048 & 3.251 & 2.349 & 1.417 &       & 5.253 & 3.327 & 2.022 & 1.110 & 0.492 \\
    \hspace{0.1in}$\rho=10$ & 4.857 & 4.058 & 3.253 & 2.351 & 1.414 &       & 5.130 & 3.195 & 1.913 & 0.979 & 0.359 \\
    \midrule
    \multicolumn{12}{c}{\underline{\textit{Panel B: with a random coefficient}}} \\
    Exogenous &       &       &       &       &       &       &       &       &       &       &  \\
    \hspace{0.1in}$\rho=0$ & 3.650 & 3.217 & 2.772 & 2.128 & 1.285 &       & 0.265 & 1.712 & 1.033 & 0.720 & 0.326 \\
    Endogenous $(-)$ &       &       &       &       &       &       &       &       &       &       &  \\
    \hspace{0.1in}$\rho=-1$ & 4.005 & 3.583 & 3.062 & 2.366 & 1.545 &       & 3.953 & 2.047 & 1.212 & 0.697 & 0.321 \\
    \hspace{0.1in}$\rho=-5$ & 4.057 & 3.775 & 3.131 & 2.410 & 1.614 &       & 4.839 & 2.061 & 1.145 & 0.677 & 0.297 \\
    \hspace{0.1in}$\rho=-10$ & 4.065 & 3.748 & 3.161 & 2.434 & 1.592 &       & 4.907 & 2.090 & 1.173 & 0.709 & 0.308 \\
    Endogenous $(+)$ &       &       &       &       &       &       &       &       &       &       &  \\
    \hspace{0.1in}$\rho=1$ & 5.126 & 4.438 & 3.677 & 2.941 & 1.940 &       & 5.299 & 2.426 & 1.319 & 0.894 & 0.421 \\
    \hspace{0.1in}$\rho=5$ & 5.446 & 4.611 & 3.908 & 3.152 & 2.041 &       & 6.463 & 2.499 & 1.380 & 0.835 & 0.406 \\
    \hspace{0.1in}$\rho=10$ & 5.441 & 4.602 & 3.917 & 3.168 & 2.088 &       & 6.502 & 2.424 & 1.348 & 0.847 & 0.358 \\
    \bottomrule
    \end{tabular}%
  \label{tab: endo vuong}%
\tablenotes The table reports the median values of the two test statistics, $T_{IV}^{RV}$ and $T_{markup}^{RV}$, across 500 simulated datasets for each Monte Carlo configuration ($J=36, F=6, T=100, \rho, F_c$). The product attribute $x_{jt}$ is endogenous in the DGP; when computing $T_{markup}^{RV}$, demand is estimated as in Table \ref{tab: exo vuong}, treating $x_{jt}$ as exogenous. The direction and degree of endogeneity are parameterized by $\rho \in \{-10, -5, -1, 0, 1, 5, 10\}$, while the true profit internalization parameter $\phi$ is fixed at 1. $T_{markup}^{RV}$ is constructed under the two alternative firm conduct models: one with $\phi = 0$ (competition) and the other with $\phi = 1$ (full profit internalization within the suspected coalition, i.e., the effective-firm structure of the DGP). The top panel presents the results when a random coefficient is excluded from the indirect utility function \eqref{eqn: dgp utility} in the DGP, while the bottom panel presents the results when it is included. The exogenous case ($\rho = 0$) differs slightly from the $\phi = 1$ row of Table \ref{tab: exo vuong} only because $x_{jt}$ is min-max normalized here.

\end{table}%

\subsection{Discussion}\label{sec: monte carlo discussion}

 The most notable difference between our proposed testing procedure and the existing one is that ours does not require researchers to estimate the demand system or impose a marginal cost specification. Therefore, $T_{IV}^{RV}$ does not rely on the statistical properties of demand estimates, circumventing potential challenges associated with estimating the model by non-linear GMM and the inputs required for this task.\footnote{Moreover, our approach is expected to be less dependent on the functional form of choice probabilities such as nested logit, multivariate probit, and multiple-choice models compared to existing methods that require demand estimation as a prerequisite. This is because the exchangeability it exploits holds for any demand system under full profit internalization (Lemma~\ref{lem: projection}).} In contrast, reliable and powerful testing with $T_{markup}^{RV}$ depends on obtaining consistent and efficient demand estimates. Furthermore, since our testing method simply compares the model fit of two linear price regressions, it can be implemented in data-limited settings where researchers lack information on product market shares.

 In addition to its preliminary role in detecting collusive behavior, our proposed testing procedure can also be used to construct stronger BLP-style instruments. For instance, once researchers observe evidence of collusive behavior, they can construct $\mathbf{z}^{coll}$ based on the colluding firm index and use it in demand estimation instead of $\mathbf{z}^{comp}$. More precise initial demand estimates, achieved through the use of instruments that properly capture true firm conduct behavior, would ultimately enhance performance in the second-stage estimation, where (feasible) optimal instruments are employed (e.g., \citealp{reynaert2014improving}; \citealp{gandhi2019measuring}; \citealp{conlon2020best}).

\section{Empirical application}\label{sec: application}

 In this section, we apply our testing method to study industry conduct in two differentiated product markets in South Korea: the new passenger car market and the instant noodles market. We then examine whether the results align with anecdotal and analytical evidence of industry conduct in each market. Namely, automobile manufacturers owned by the same parent company jointly determine the prices of their car models, while instant noodle prices remain substantially below the collusive level despite suspicions from the Korea Fair Trade Commission (KFTC).

 \subsection{The new passenger car market}

 South Korea is one of the largest car markets in the world, with approximately 1.6 million vehicles sold in 2022.\footnote{\url{https://www.statista.com/statistics/265891/vehicles-sales-in-selected-countries/}} The two largest firms, Hyundai and Kia, together accounted for about 65 percent of total vehicle sales over the past decade. Since both firms are owned by their parent company, the Hyundai Motor Group, following Kia's merger into the group in 1998, pricing is likely coordinated between the two firms (brands).\footnote{We use the terms ``firm'' and ``brand'' interchangeably.} We test this joint profit maximization by the parent company against own-profit maximization, in which each brand sets prices independently.\footnote{Refer to Table \ref{tab: auto mkt structure} in the Appendix for the list of brands and their parent companies in our sample dataset.} We also consider three progressively broader collusion scenarios: joint profit maximization among German brands, collusion among domestic automakers and among foreign automakers, and full collusion.\footnote{In February 2023, the KFTC imposed a total fine of 42.3 billion Won (approximately 33 million US dollars) on the four German automakers (Mercedes-Benz, BMW, Audi, and Volkswagen) for colluding on emission reduction technology applied to their diesel passenger cars. Here, we investigate whether the German automakers also colluded on prices. The KFTC's ruling is available at \url{https://www.ftc.go.kr/www/selectBbsNttView.do?pageUnit=10&pageIndex=92&searchCnd=all&key=12&bordCd=3&searchCtgry=01,02&nttSn=42758}.}

 Drawing on province/year/product-level data on prices and attributes for 776 products from 13 brands between 2012 and 2023,\footnote{We define a product as a unique combination of nameplate and fuel type. Appendix \ref{app: data appendix} provides details on the raw data and sample construction.} we conduct pairwise hypothesis tests by running the following price regressions:
\begin{equation}
	p_{jt} = \mathbf{x}_{jt}\bm{\gamma}^{h} + \mathbf{z}_{jt}^{h}\bm{\theta}^h + \psi_j + \psi_{y}^{\textrm{fuel}} + \psi_r + e^h_{jt}, \label{eqn: application}
\end{equation}
 where $\mathbf{x}_{jt}$ represents four exogenous characteristics of product $j$ in market (year-province) $t$: fuel economy (km per 1,000 Won), acceleration (horsepower/weight), size (width$\times$length$\times$height), and a constant term. The regression model \eqref{eqn: application} also includes controls for product, year-fuel type (to account for differential trends in government policies regarding fuel types), and province, denoted by $\psi_j$, $\psi_{y}^{\textrm{fuel}}$, and $\psi_r$, respectively.

 As for the excluded instruments under conduct hypothesis $h \in \{H_0, H_1\}$, denoted by $\mathbf{z}_{jt}^{h}$, we use the four observed product attributes to construct four summation IVs (sums over own‐firm products based on the coalition index $f_c$). In addition, we incorporate a two‐level nesting structure based on the car segment (small/compact, midsize, mid luxury, large/luxury, small SUV, standard SUV, large/luxury SUV) and fuel type (gasoline, diesel, LPG, EV, HEV),\footnote{Note that while the nesting order is important for demand estimation, it is irrelevant to our testing procedure.} and create 12 additional summation IVs: sums over own‐firm products in the same segment, sums over own‐firm products with the same fuel type, and sums over own‐firm products with the same fuel type in the same segment.\footnote{Entering the sums over same-nest products separately from the sums over all own-firm products gives same-nest and different-nest products separate coefficients, which raises the explanatory power of the IVs to the extent that the pricing rule weights the two groups differently. Under nested logit demand, the exchangeability of Section~\ref{sec: conceptual framework} holds among products of the same cell within the same nest and, separately, across nests (Corollary~\ref{cor: nl inside outside} in Appendix~\ref{app: demand system}), so that summing within nests applies the same construction to this finer partition.} 

 We report the testing results (clustered by market) in the upper panel of Table \ref{tab: empirical application} where a positive value greater than 1.65 (a negative value less than -1.65) constitutes evidence for the row (column) conduct model. The results indicate that, as expected, joint profit maximization by the parent company is more consistent with the data than own-profit maximization. Moreover, our test favors parent-company pricing over each collusive hypothesis. These results, robust to alternative choices of fixed effects and clustering, align with the fact that no antitrust cases have been filed against the car manufacturers in South Korea in recent years, except for the non-price collusion case described above.

\begin{table}[!t]
  \centering
  \caption{Testing firm conduct in the two differentiated product markets}
    \begin{tabular}{lrrrrr}
    \toprule
    Firm conduct & \multicolumn{1}{c}{1.} & \multicolumn{1}{c}{2.} & \multicolumn{1}{c}{3.} & \multicolumn{1}{c}{4.} & \multicolumn{1}{c}{5.} \bigstrut\\
    \hline
          &       &       &       &       &  \\
\multicolumn{6}{c}{\underline{\textit{Panel A: new passenger car market}}} \bigstrut[b] \\
    1. Brand ($F=13$) &       & -3.718 & 1.013 & 0.335 & 4.940 \\
    2. Parent company ($F_c=9$) & 3.718 &       & 5.404 & 2.229 & 7.104 \\
    3. German automakers ($F_c=7$) & -1.013 & -5.404 &       & -0.343 & 4.954 \\
    4. Domestic/Foreign ($F_c=2$) & -0.335 & -2.229 & 0.343 &       & 4.168 \\
    5. Full collusion ($F_c=1$) & -4.940 & -7.104 & -4.954 & -4.168 &  \\
          &       &       &       &       &  \\
    \multicolumn{6}{c}{\underline{\textit{Panel B: instant noodles market}}} \bigstrut[b] \\
    1. Competition ($F=4$) &       & 18.727 & 17.546 & 18.411 & 2.619 \\
    2. Full collusion ($F_c=1$) & -18.727 &       & -6.324 & -5.444 & -22.873 \\
    3. N-O-P ($F_c=2$) & -17.546 & 6.324 &       & 3.747 & 18.316 \\
    4. N-O-S ($F_c=2$) & -18.411 & 5.444 & -3.747 &       & 21.808 \\
    5. N-P-S ($F_c=2$) & -2.619 & 22.873 & -18.316 & -21.808 &  \\
    \bottomrule
    \end{tabular}%
  \label{tab: empirical application}%
\tablenotes The table reports the results of pairwise hypothesis tests using our testing framework. Results for the new passenger car market are in the upper panel, while those for the instant noodles market are in the bottom panel. The names of the four \textit{ramen} producers (Nongshim, Ottogi, Paldo, and Samyang) are abbreviated using their initial letters: N, O, P, and S. Entry $(i,k)$ reports $\sqrt{n}(\hat{Q}_k-\hat{Q}_i)/\hat{\sigma}_{IV}$, so that a positive (negative) value supports the row conduct model $i$ (the column model $k$, respectively).
\end{table}%

 \subsection{Instant noodles market}

 The South Korean instant noodles (\textit{ramen}) market is highly concentrated, with four firms (Nongshim, Ottogi, Samyang, and Paldo) accounting for over 90 percent of total sales. In March 2012, the KFTC fined them 135 billion Won (approximately 120 million US dollars) for price collusion. In particular, Nongshim, the market leader with more than a 50\% sales share, was fined 100 billion Won for leading the collusion. However, this ruling was overturned by the Supreme Court of Korea in December 2015.\footnote{The rulings of the KFTC and the Supreme Court are available at \url{https://www.ftc.go.kr/www/selectBbsNttView.do?pageUnit=10&pageIndex=507&searchCnd=all&key=12&bordCd=3&searchCtgry=01,02&nttSn=37962} and \url{https://www.scourt.go.kr/supreme/news/NewsViewAction2.work?seqnum=5081&gubun=4&searchOption=&searchWord=}, respectively.} Given that the KFTC remains suspicious of these firms and has been closely monitoring \textit{ramen} prices,\footnote{For example, the KFTC considered investigating collusion among the \textit{ramen} manufacturers in 2023, citing that the \textit{ramen} prices, which increased in the aftermath of the COVID-19 outbreak, did not decrease despite falling flour prices since late 2022: \url{https://www.newsis.com/view/NISX20230623_0002349874}.} we test various collusion hypotheses, with Nongshim as a member of every coalition, against the hypothesis of own-profit maximization. Specifically, we consider full collusion and three cases in which Nongshim coordinates prices with two other firms.

 The data used for our test comprise region/year-month/product-level prices and product-level attributes of the 70 best-selling instant noodle products from 2010 to 2019.\footnote{A region consists of adjacent provinces, except that Seoul forms a region of its own. See Appendix \ref{app: data appendix} for data details.} Our sample period largely postdates the alleged collusion (2001 to 2010), so that our test examines conduct in the 2010s, during which the KFTC continued to monitor the industry. 
 
 For each of the seven product attributes --- cholesterol, calorie, sugar, fat, protein, sodium contents, and a constant term --- we calculate the sum over the other products in the firm's cell under hypothesis $h$, and use them as excluded instruments. In addition, given that the \textit{ramen} market is differentiated along two dimensions \citep{hong2023does, kim2024collusion}, that is, package (pouch vs. cup) and soup type (red soup, white soup, soupless), we construct 21 additional IVs as in the car application: sums over own‐firm products that share the same package, sums over own‐firm products that share the same soup type, and sums over own‐firm products that share both the same package and soup type. Then, we regress the price on product, time (year-month), and region dummy variables, along with the 28 IVs, for each conduct scenario.\footnote{Product attributes are absorbed by the product dummies.}

 According to the testing results (clustered by market) presented in the bottom panel of Table~\ref{tab: empirical application}, our test rejects all collusion hypotheses considered in favor of the own-profit maximization.\footnote{We also consider other collusion scenarios in which Nongshim, along with another firm, coordinates prices, and obtain results in favor of price competition.} These results are consistent with the Supreme Court's finding of insufficient evidence of collusion and with \citet{kim2024collusion}, who found that the observed markups are too low to support any collusive behavior.

 \section{Conclusion}\label{sec: conclusion}

 In this article, we propose a practical procedure for testing firm conduct that simply compares the performance of two BLP-style instrument sets (competition IVs and collusion IVs) in price regressions.

 First, we show that, for any demand system, the collusion IVs fit better under collusion whenever colluding and non-colluding firms exhibit sufficiently similar price responses to their own product characteristics. When firms compete, the competition IVs fit better as long as a firm's price response to the characteristics of suspected partner products is smaller in magnitude than an explicit fraction of the response to its own product characteristics. Under logit demand, the conditions under which each set of IVs fits better under its presumed conduct can be stated in terms of observable market shares. For more general demand systems, the fit comparison remains valid under analogous conditions as long as products are randomly assigned to firms. When they are not, however, the proximity of partner products to a firm's own products governs whether the test may fail.

 Next, we construct the RV test statistic from the mean squared residuals of the two price regressions; a statistically significant test statistic is interpreted as evidence of either competitive or collusive behavior among firms, depending on its sign. Through extensive Monte Carlo simulations, we evaluate the finite-sample performance of our test statistic under various market conditions, characterized by different numbers of firms, degrees of collusion, and degrees of profit internalization. The results show that our method is robust to endogenous product characteristics and missing cost shifters, performing comparably to, or better than, the existing markup-based RV test in detecting collusion across various collusive scenarios.

 The simplicity of our approach, combined with its high testing power, makes it a practical tool for the preliminary diagnosis of industry conduct. By complementing existing methods, our testing framework provides researchers and regulatory authorities with an efficient and effective way to assess firm behavior and inform antitrust investigations. Our test also has limitations; in particular, when marginal costs vary with output, the characteristics of other products enter a firm's price through the cost as well as through the markup. Consequently, rivals' characteristics carry an additional price signal under competition, rendering the fit comparison potentially misleading. Extending the framework to accommodate such cost structures is left for future research.

 Finally, our procedure can aid in designing more effective BLP-style instruments. Once collusion is detected, researchers are advised to incorporate the product characteristics of colluding partners when building own-firm instruments while excluding them from other-firm instruments. This approach strengthens the instruments in demand estimation.

\newpage

\bibliographystyle{apalike}
\bibliography{reference}

\renewcommand{\thefigure}{\Alph{section}\arabic{figure}}
\renewcommand{\thetable}{\Alph{section}\arabic{table}}
\renewcommand{\theequation}{\Alph{section}\arabic{equation}}
\renewcommand{\theHfigure}{\Alph{section}\arabic{figure}}
\renewcommand{\theHtable}{\Alph{section}\arabic{table}}
\renewcommand{\theHequation}{\Alph{section}\arabic{equation}}
\counterwithin*{figure}{section}
\counterwithin*{table}{section}
\counterwithin*{equation}{section}
\setcounter{table}{0}
\setcounter{figure}{0}%
\setcounter{equation}{0}

\renewcommand{\thesection}{\Alph{section}}
\renewcommand{\thesubsection}{\thesection.\arabic{subsection}}

\clearpage
\setcounter{page}{1}
\renewcommand{\thepage}{A-\arabic{page}}
\appendix

\begin{center}
\textbf{\huge{Online Appendix}}
\end{center}
\doublespacing

\section{Proofs}\label{app: proofs}

Throughout this appendix, the market subscript $t$ is suppressed. We write $O(\cdot)$ for a term whose absolute value is bounded by a constant multiple of that of its argument whenever the argument is sufficiently small. A statement that holds ``to first order in market shares'' neglects terms that are quadratic or higher order in market shares.

\subsection{Proof of Lemma \ref{lem: logit}}\label{app: lemma 1}

 We first establish that the markup is uniform within a cell. Differentiating the profit function \eqref{eqn: profit} with respect to $p_j$ for $j\in\mathscr{J}_f$ and dividing by $M$ yields the following first-order condition:
 \begin{equation*}
	s_j + \sum_{k\in\mathscr{J}_f}\mu_k\frac{\partial s_k}{\partial p_j} + \phi\sum_{k\in\mathscr{J}_{f_c}\setminus\mathscr{J}_f}\mu_k\frac{\partial s_k}{\partial p_j} = 0 .
 \end{equation*}
 Under logit demand, the choice probability derivatives satisfy
 \begin{equation}\label{eqn: logit deriv}
	\frac{\partial s_k}{\partial p_j} = \begin{cases} -\alpha\, s_j(1 - s_j) & k = j, \\ \alpha\, s_j s_k & k \neq j. \end{cases}
 \end{equation}
 Substituting \eqref{eqn: logit deriv} into the first-order condition and dividing by $\alpha s_j > 0$, we obtain
 \begin{equation}\label{eqn: foc general logit}
	\mu_j = \frac{1}{\alpha} + \sum_{k\in\mathscr{J}_f}\mu_k s_k + \phi\sum_{k\in\mathscr{J}_{f_c}\setminus\mathscr{J}_f}\mu_k s_k.
 \end{equation}
 Because the right-hand side of \eqref{eqn: foc general logit} does not depend on $j$, $\mu_j = \mu_f$ for all $j\in\mathscr{J}_f$ under any $\phi$. Therefore, under logit demand, markups are uniform within each firm for any $\phi \in [0, 1]$:
 \begin{equation}\label{eqn: foc uniform general}
 \begin{aligned}
	\mu_j = \mu_f &= \frac{1}{\alpha} + \sum_{k\in\mathscr{J}_f}\mu_f s_k + \phi\sum_{k\in\mathscr{J}_{f_c}\setminus\mathscr{J}_f}\mu_{f(k)} s_k \\
          &= \frac{1}{\alpha} + \mu_f s_f + \phi\sum_{f' \in \mathscr{F}_{f_c}\setminus\{f\}} \mu_{f'} s_{f'},
 \end{aligned}
 \end{equation}
 where $f(k)$ denotes the firm owning product $k$, and $\mathscr{F}_{f_c}$ denotes the set of firms in the coalition $f_c$. For $\phi \in (0, 1)$, markups remain uniform within each firm but generally differ across firms within a coalition, because partner firms' markups and market shares enter firm $f$'s first-order condition scaled by $\phi$.

 Now consider $\phi\in\{0,1\}$, and let $\mathscr{C}$ denote the cell containing product $j$, so that $\mathscr{C} = \mathscr{J}_f$ if $\phi = 0$ and $\mathscr{C} = \mathscr{J}_{f_c}$ if $\phi = 1$. In either case, the last two sums in \eqref{eqn: foc general logit} combine into a single sum over $\mathscr{C}$, $\sum_{k\in\mathscr{C}}\mu_k s_k$, which is independent of $j$. Consequently, within-firm uniformity of markups extends to within-cell uniformity:
 \begin{equation}\label{eqn: foc uniform}
	\mu_{\mathscr{C}} = \frac{1}{\alpha} + \mu_{\mathscr{C}}\sum_{k\in\mathscr{C}} s_k = \frac{1}{\alpha} + \mu_{\mathscr{C}}\, s_{\mathscr{C}} .
 \end{equation}
 Solving for $\mu_{\mathscr{C}}$ yields $\mu_{\mathscr{C}} = [\alpha(1 - s_{\mathscr{C}})]^{-1}$, establishing \eqref{eqn: cell markup}.

 Substituting $p_k = mc_k + \mu_{\mathscr{C}} = \mathbf{x}_k\gamma + \omega_k + \mu_{\mathscr{C}}$ for $k \in \mathscr{C}$, we obtain
 \begin{equation}\label{eqn: a decomposition}	\sum_{k\in\mathscr{C}} \exp(\delta_k) = \sum_{k\in\mathscr{C}} \exp \big(\mathbf{x}_k\beta - \alpha \mathbf{x}_k\gamma - \alpha\omega_k + \xi_k\big)\exp(-\alpha \mu_{\mathscr{C}}) = A_{\mathscr{C}} e^{-\alpha\mu_{\mathscr{C}}},
\end{equation}
so that
 \begin{equation}\label{eqn: s_C and H}
 s_{\mathscr{C}} = \frac{A_{\mathscr{C}} e^{ -\alpha\mu_{\mathscr{C}} }}{H}, \quad H = 1 + \sum_{\mathscr{C}'} A_{\mathscr{C}'} e^{-\alpha\mu_{\mathscr{C}'}}.
 \end{equation}
 Substituting $s_{\mathscr{C}}$ into \eqref{eqn: cell markup} shows that the cell markup $\mu_{\mathscr{C}}$ solves
 \begin{equation}\label{eqn: psi}
	\chi(\mu;A_{\mathscr{C}},H) \equiv \mu - \frac{1}{\alpha}\Big(1 - \frac{A_{\mathscr{C}}\,e^{-\alpha\mu}}{H}\Big)^{-1} = 0 .
 \end{equation}

 Fix $H$ for now. Since $0 < s_{\mathscr{C}} < 1$, the domain of $\mu$ is $(\alpha^{-1}\ln(A_{\mathscr{C}}/H), \infty)$ from \eqref{eqn: s_C and H}. Writing $s = A_{\mathscr{C}}e^{-\alpha\mu}/H$ on this domain and using $\partial s/\partial\mu = -\alpha s$, we obtain the following derivatives:
 \begin{equation}\label{eqn: psi derivs}
	\frac{\partial\chi}{\partial\mu} = 1 + \frac{s}{(1-s)^2} > 0, \qquad
	\frac{\partial\chi}{\partial A_{\mathscr{C}}} = -\frac{1}{\alpha}\,\frac{e^{-\alpha\mu}/H}{(1-s)^2} < 0, \qquad
	\frac{\partial\chi}{\partial H} = \frac{1}{\alpha}\,\frac{A_{\mathscr{C}}e^{-\alpha\mu}/H^2}{(1-s)^2} > 0 .
 \end{equation}
 Because $\chi$ is continuous and strictly increasing in $\mu$ on this domain, with $\chi \to -\infty$ as $\mu \to \alpha^{-1}\ln(A_{\mathscr{C}}/H)$ (or equivalently, as $s \to 1$) and $\chi \to +\infty$ as $\mu \to \infty$ (or equivalently, as $s \to 0$), equation~\eqref{eqn: psi} has a unique solution, denoted by $\mu_{\mathscr{C}} = m(A_{\mathscr{C}},H)$. This uniform cell markup depends on the rest of the market outside the cell exclusively through $H$.

 By the implicit function theorem and \eqref{eqn: psi derivs},
\begin{equation}\label{eqn: m derivs}
	\frac{\partial m}{\partial A_{\mathscr{C}}} = -\frac{\chi_A}{\chi_\mu} = \frac{e^{-\alpha\mu}/(\alpha H)}{(1-s)^2 + s} > 0, \qquad
	\frac{\partial m}{\partial H} = -\frac{\chi_H}{\chi_\mu} = -\frac{s}{\alpha H\big[(1-s)^2 + s\big]} < 0 ,
\end{equation}
 evaluated at $\mu = m(A_{\mathscr{C}},H)$. Thus, $m$ is strictly increasing in the cell's own aggregate and strictly decreasing in the market aggregate. 

 In equilibrium, the aggregate $H$ must be generated by the best responses of all cells. Therefore, as stated in Lemma~\ref{lem: logit}, $H$ solves the one-dimensional fixed-point condition
    \begin{equation}\label{eqn: H fixed point}
	\mathcal{F}(H) \equiv 1 + \sum_{\mathscr{C}} A_{\mathscr{C}} \exp \big(-\alpha\, m(A_{\mathscr{C}},H)\big) - H = 0.
    \end{equation}
 To establish the existence of a solution to \eqref{eqn: H fixed point}, observe that $\mathcal{F}(1) > 0$, whereas as $H\to\infty$, $s_{\mathscr{C}}(H)\to 0$, $m(A_{\mathscr{C}},H)\to 1/\alpha$, and $\mathcal{F}(H)\to-\infty$. By the intermediate value theorem, $\mathcal{F}$ has at least one root. To show that the solution is unique, observe that
 \begin{equation}\label{eqn: Gamma deriv}
	\frac{d}{d H} A_{\mathscr{C}}e^{-\alpha m(A_{\mathscr{C}},H)}
	= -\alpha  H s_{\mathscr{C}} \frac{\partial m}{\partial H}
	= \frac{s_{\mathscr{C}}^2}{(1-s_{\mathscr{C}})^2 + s_{\mathscr{C}}}
    <  s_{\mathscr{C}},
 \end{equation}
 so that, using $\sum_{\mathscr{C}'} s_{\mathscr{C}'}(H) = (\mathcal{F}(H) + H - 1)/H$, 
 \begin{equation}\label{eqn: F deriv}
 \mathcal{F}'(H) = \sum_{\mathscr{C}'} \frac{s_{\mathscr{C}'}^2}{(1-s_{\mathscr{C}'})^2 + s_{\mathscr{C}'}} - 1 < \sum_{\mathscr{C}'} s_{\mathscr{C}'} - 1 = \frac{\mathcal{F}(H) + H - 1}{H} - 1 = \frac{\mathcal{F}(H)-1}{H}.
 \end{equation}
 Because any solution $H$ to equation \eqref{eqn: H fixed point} satisfies $\mathcal{F}(H) = 0$, it follows from \eqref{eqn: F deriv} that $\mathcal{F}'(H) < -H^{-1} < 0$ at any such solution. A continuously differentiable function whose derivative is strictly negative at every root has at most one root.

 Consequently, \eqref{eqn: H fixed point} has a unique solution $H$, and $\mu_{\mathscr{C}} = m(A_{\mathscr{C}},H)$ is the unique equilibrium markup, yielding \(p_j = mc_j + \mu_{\mathscr{C}(j)} = \mathbf{x}_j\gamma + \omega_j + m\big(A_{\mathscr{C}(j)},H\big)\), which establishes \eqref{eqn: exch rep}. \qed

 \subsection{Derivatives of \texorpdfstring{$H$}{H} and \texorpdfstring{$p_j$}{pj} with respect to \texorpdfstring{$a_k$}{ak}}\label{app: elas H}
 
 Let product $k$ belong to cell $\mathscr{C}$. Differentiating both sides of equation \eqref{eqn: H fixed point} with respect to $a_k$ yields
 \begin{equation}\label{eqn: dHda 1}
	\frac{\partial H}{\partial a_k} =
    e^{-\alpha \mu_{\mathscr{C}}} - \alpha A_{\mathscr{C}}e^{-\alpha \mu_{\mathscr{C}}} \frac{\partial m}{\partial A_{\mathscr{C}}}
    - \alpha \sum_{\mathscr{C}'} A_{\mathscr{C}'}e^{-\alpha \mu_{\mathscr{C}'}} \frac{\partial m}{\partial H} \frac{\partial H}{\partial a_k} .
 \end{equation}
 Substituting $\partial m/\partial A_{\mathscr{C}}$ and $\partial m/\partial H$ for each cell from \eqref{eqn: m derivs}, and using $A_{\mathscr{C}'}e^{-\alpha\mu_{\mathscr{C}'}}/H = s_{\mathscr{C}'}$ from \eqref{eqn: s_C and H}, we obtain
 \begin{equation}\label{eqn: dHda}
	\frac{\partial H}{\partial a_k} = e^{-\alpha\mu_{\mathscr{C}}} \cdot \frac{ (1 - s_{\mathscr{C}})^2 }{ (1 - s_{\mathscr{C}})^2 + s_{\mathscr{C}} } \cdot
	\left( 1 - \sum_{\mathscr{C}'} \frac{ s_{\mathscr{C}'}^2 }{ (1 - s_{\mathscr{C}'})^2 + s_{\mathscr{C}'} } \right)^{-1}.    
 \end{equation}
 Multiplying by $a_k/H$ and using $s_k = a_k e^{-\alpha\mu_{\mathscr{C}}}/H$ yields the elasticity of $H$ with respect to $a_k$, denoted by $\partial\log H/\partial\log a_k$:
 \begin{equation}\label{eqn: dHda 2}
 \frac{\partial\log H}{\partial\log a_k} = s_k \cdot \frac{ (1 - s_{\mathscr{C}})^2 }{ (1 - s_{\mathscr{C}})^2 + s_{\mathscr{C}} }  \cdot
 \left( 1 - \sum_{\mathscr{C}'} \frac{ s_{\mathscr{C}'}^2 }{ (1 - s_{\mathscr{C}'})^2 + s_{\mathscr{C}'} } \right)^{-1} .
 \end{equation}
 The first factor, $s_k$, is the elasticity holding all markups fixed. The second factor captures the increase in cell $\mathscr{C}$'s own markup in response to the rise in $A_{\mathscr{C}}$ and equals $1 - s_{\mathscr{C}} + O(s_{\mathscr{C}}^2)$. The third factor captures the adjustment of all cells' markups to the change in $H$ and equals $1 + O(\sum_{\mathscr{C}'} s_{\mathscr{C}'}^2)$. Hence, $\partial\log H/\partial\log a_k \approx s_k$ whenever cell market shares are sufficiently small, for any product $k$ in the market.

 Combining \eqref{eqn: dHda} with \eqref{eqn: m derivs} gives the total derivative of $p_j$ with respect to $a_k$ for a product $k$ in cell $\mathscr{C}(j)$ with markups adjusting. Since $p_j = \mathbf{x}_j\gamma + \omega_j + m(A_{\mathscr{C}(j)}, H)$,
 \begin{equation}
 \begin{aligned}\label{eqn: dpda inside}
	\frac{\partial p_j}{\partial a_k}
	&= \frac{\partial m}{\partial A_{\mathscr{C}(j)}} + \frac{\partial m}{\partial H}\,\frac{\partial H}{\partial a_k} \\
&= \frac{s_k}{\alpha a_k}\,
	  \frac{1}{(1-s_{\mathscr{C}(j)})^2 + s_{\mathscr{C}(j)}}
	  \left[ 1 - \frac{ s_{\mathscr{C}(j)}\,(1-s_{\mathscr{C}(j)})^2 }
	  { \big[(1-s_{\mathscr{C}(j)})^2 + s_{\mathscr{C}(j)}\big]
	    \big( 1 - \sum_{\mathscr{C}'} \frac{ s_{\mathscr{C}'}^2 }{ (1 - s_{\mathscr{C}'})^2 + s_{\mathscr{C}'} } \big) } \right],
 \end{aligned}
 \end{equation}
 where we used $e^{-\alpha\mu_{\mathscr{C}(j)}}/H = s_k/a_k$ for $k \in \mathscr{C}(j)$. The product of the last two factors equals $1 + O(s^2)$, where $O(s^k)$ collects terms of order $k$ in cell market shares. Hence,
 \begin{equation}\label{eqn: dpda inside approx}
 \frac{\partial p_j}{\partial a_k} = \frac{s_k}{\alpha a_k}\big(1 + O(s^2)\big)
 = \frac{e^{-1}}{\alpha H} (1 - s_{\mathscr{C}(j)})\big(1 + O(s^2)\big),
 \end{equation}
 where the second equality uses $s_k/a_k = e^{-\alpha\mu_{\mathscr{C}(j)}}/H = e^{-1/(1-s_{\mathscr{C}(j)})}/H = e^{-1}(1 - s_{\mathscr{C}(j)})\big(1 + O(s^2)\big)/H$. Thus, to first order in market shares, this derivative depends on the cell only through the factor $1 - s_{\mathscr{C}(j)}$.

 For a product $k$ outside cell $\mathscr{C}(j)$, $a_k$ affects $p_j$ only through $H$, so that
 \begin{equation}\label{eqn: dpda outside}
 \begin{aligned}
 	\frac{\partial p_j}{\partial a_k} &= \frac{\partial m}{\partial H}\,\frac{\partial H}{\partial a_k} \\
 	&= -\frac{s_k}{\alpha  a_k}\,\frac{s_{\mathscr{C}(j)}}{(1-s_{\mathscr{C}(j)})^2 + s_{\mathscr{C}(j)}}
 	\cdot \frac{(1-s_{\mathscr{C}(k)})^2}{(1-s_{\mathscr{C}(k)})^2 + s_{\mathscr{C}(k)}}
 	\cdot \Big( 1 - \sum_{\mathscr{C}'} \frac{ s_{\mathscr{C}'}^2 }{ (1 - s_{\mathscr{C}'})^2 + s_{\mathscr{C}'} } \Big)^{-1}.
    \end{aligned}
 \end{equation}
 
 To isolate the role of markup adjustment, note that, holding markups fixed, $a_k$ affects $p_j$ only through $s_{\mathscr{C}(j)}$, so that $\partial p_j/\partial a_k = a_k^{-1}\big[s_{\mathscr{C}(j)}\,d\mu_{\mathscr{C}(j)}/ds_{\mathscr{C}(j)}\big]\,\partial\log s_{\mathscr{C}(j)}/\partial\log a_k$ with $d\mu_{\mathscr{C}}/ds_{\mathscr{C}} = 1/[\alpha(1-s_{\mathscr{C}})^2]$. By Corollary~\ref{cor: inside outside}, 
 \begin{equation*}
    \frac{\partial p_j}{\partial a_{k}}
    = \begin{cases}
        \;\; \dfrac{s_k}{\alpha  a_k} \dfrac{1}{1-s_{\mathscr{C}(j)}}, & \text{if } k \in \mathscr{C}(j), \\[10pt]
        - \dfrac{s_k}{\alpha  a_k} \dfrac{s_{\mathscr{C}(j)}}{(1-s_{\mathscr{C}(j)})^2}, & \text{if } k \notin \mathscr{C}(j).
    \end{cases}
\end{equation*}
 Comparing these values with \eqref{eqn: dpda inside approx} and \eqref{eqn: dpda outside}, allowing markups to adjust multiplies the inside-cell derivative by $1 - s_{\mathscr{C}(j)} + O(s^2)$ and the outside-cell derivative by $\big(1 - s_{\mathscr{C}(j)} + O(s^2)\big)\big(1 - s_{\mathscr{C}(k)} + O(s^2)\big)\big(1 + O(\sum_{\mathscr{C}'} s_{\mathscr{C}'}^2)\big)$. Both factors are of the form $1 + O(s)$, so that the ratio of the two derivatives is unchanged to first order in market shares.

\subsection{Proof of Lemma~\ref{lem: projection}}\label{app: lemma 2}

 Let product $j$ belong to cell $\mathscr{C}$. Under $\phi = 1$, and because demand and costs depend on products only through their characteristics (Assumption~\ref{ass: specification}), the system of first-order conditions \eqref{eqn: foc bertrand} is invariant to any permutation of the products in $\mathscr{C} \setminus \{j\}$. Because the equilibrium is unique (Assumption~\ref{ass: regularity}(1)), the equilibrium price $p_{j}$ is a symmetric function of $\{(\mathbf{x}_{k}, \xi_{k}, \omega_{k})\}_{k \in \mathscr{C} \setminus \{j\}}$, holding fixed $(\mathbf{x}_{j}, \xi_{j}, \omega_{j})$ as well as the primitives of products outside cell $\mathscr{C}$. Furthermore, because the primitives are i.i.d.\ across products (Assumption~\ref{ass: sampling}), the joint distribution of $\big(b(\mathbf{x}_{k}), p_{j}\big)$ is identical for all $k \in \mathscr{C} \setminus \{j\}$, which implies that $\mathrm{Cov}(b(\mathbf{x}_{k}), p_{j})$ equals a common vector $\mathbf{g}$. Because $(n_o, n_p)$ are constant (Assumption~\ref{ass: symmetry}) and the primitives are i.i.d., this distribution does not depend on $j$ either, so $\mathbf{g}$ is common to all products of colluding firms. An analogous argument applied to a non-colluding firm demonstrates that $\mathrm{Cov}(b(\mathbf{x}_{k}), p_{j})$ for $k \in \mathscr{J}_{f} \setminus \{j\}$ equals a common vector $\mathbf{g}^{N}$. Neither result depends on the functional form of $b(\cdot)$.

 To obtain the projection coefficients, we adopt the following convention throughout the proofs: for any variable $y$, $\tilde{y} \equiv y - L[y \mid 1, D_{j}]$ denotes its deviation from its population linear projection onto $(1, D_{j})$. Because $D_{j}$ is binary, $L[y \mid 1, D_{j}] = \mathbb{E}[y \mid D_{j}]$. In particular, $\tilde{\mathbf{z}}^{part}_{j} = \mathbf{z}^{part}_{j} - \mathbb{E}[\mathbf{z}^{part}_{j} \mid D_{j}] = D_{j}\big(\mathbf{z}^{part}_{j} - n_{p}\bm{\mu}_b\big)$, where $\bm{\mu}_b \equiv \mathbb{E}[b(\mathbf{x}_{k})]$, represents the partner instrument vector after partialling out $D_{j}$. 
 
 By construction, $\mathbf{x}_j$, $\mathbf{z}^{comp}_j$, and $\mathbf{z}^{part}_j$ are functions of disjoint sets of products, which are mutually independent and independent of $D_j$ under Assumption~\ref{ass: sampling}. Moreover, $\tilde{\mathbf{z}}^{part}_j$ has mean zero conditional on $D_j$, and is therefore uncorrelated with $D_j$ as well as with $\mathbf{x}_j$ and $\mathbf{z}^{comp}_j$. Hence, the regressors $(1, D_j, \mathbf{x}_j, \mathbf{z}^{comp}_j, \tilde{\mathbf{z}}^{part}_j)$ are mutually uncorrelated. Also, since $\tilde{\mathbf{z}}^{part}_j$ differs from $\mathbf{z}^{part}_j$ only by a linear function of $D_j$, replacing $\mathbf{z}^{part}_j$ with $\tilde{\mathbf{z}}^{part}_j$ leaves the coefficient vector $\bm{\lambda}_p$ unchanged. Consequently, the projection coefficient vectors $\bm{\lambda}_{o}$ and $\bm{\lambda}_{p}$ in \eqref{eqn: linear rep} coincide with the coefficients from the separate regressions of $p_{j}$ onto $(1, \mathbf{z}^{comp}_{j})$ and onto $(1, \tilde{\mathbf{z}}^{part}_{j})$, respectively. 
 
 Because $\mathbb{E}[\mathbf{z}^{comp}_{j} \mid D_{j}] = n_{o}\bm{\mu}_b$ is constant, the law of total covariance implies
 \begin{equation*}
    \mathrm{Cov}(\mathbf{z}^{comp}_{j}, p_{j}) = \mathbb{E}\big[\mathrm{Cov}(\mathbf{z}^{comp}_{j}, p_{j} \mid D_{j})\big] = \pi n_{o}\mathbf{g} + (1-\pi) n_{o}\mathbf{g}^N
 \end{equation*}
 and $\mathrm{Var}(\mathbf{z}^{comp}_{j}) = n_{o}\Sigma_b$, which yields $\bm{\lambda}_{o} = \Sigma_b^{-1}[\pi\mathbf{g} + (1-\pi)\mathbf{g}^N]$. Likewise, $\mathrm{Cov}(\tilde{\mathbf{z}}^{part}_{j}, p_{j}) = \pi \mathrm{Cov}(\mathbf{z}^{part}_{j}, p_{j} \mid D_{j} = 1) = \pi n_{p}\mathbf{g}$ and $\mathrm{Var}(\tilde{\mathbf{z}}^{part}_{j}) = \pi n_{p}\Sigma_b$, which yields $\bm{\lambda}_{p} = \Sigma_b^{-1}\mathbf{g}$. This establishes equation~\eqref{eqn: lambda pooled}; the equality $\bm{\lambda}_{p} = \bm{\lambda}_{o}$ follows immediately when $\pi = 1$ or $\mathbf{g}^N = \mathbf{g}$. \qed

\subsection{Proof of Proposition~\ref{prop: main}}\label{app: prop 1}

 All variables are transformed into deviations from their population linear projections onto $(1, D_{j})$ and marked with a tilde: $\tilde{p}_{j}$, $\tilde{\mathbf{x}}_{j}$, $\tilde{\mathbf{z}}^{comp}_{j}$, $\tilde{\mathbf{z}}^{part}_{j}$, and $\tilde{\mathbf{z}}^{coll}_{j} = \tilde{\mathbf{z}}^{comp}_{j} + \tilde{\mathbf{z}}^{part}_{j}$. Because $1$ and $D_{j}$ enter the projection \eqref{eqn: linear rep}, the coefficients on the remaining regressors are invariant to this transformation. The residual $e_{j}$ is already orthogonal to $(1, D_{j})$ by construction and is therefore unchanged. In this notation, \eqref{eqn: linear rep} reads
 \begin{equation*}
    \tilde{p}_{j} = \tilde{\mathbf{x}}_{j}\bm{\gamma} + \tilde{\mathbf{z}}^{comp}_{j}\bm{\lambda}_{o} + \tilde{\mathbf{z}}^{part}_{j}\bm{\lambda}_{p} + e_{j}.
 \end{equation*}
 
 Under Assumptions~\ref{ass: sampling} and~\ref{ass: symmetry}, $\mathbf{x}_{j}$ and $\mathbf{z}^{comp}_{j}$ are independent of $D_{j}$, so $\tilde{\mathbf{x}}_{j} = \mathbf{x}_{j} - \mathbb{E}[\mathbf{x}_{j}]$ and $\tilde{\mathbf{z}}^{comp}_{j} = \mathbf{z}^{comp}_{j} - n_{o}\bm{\mu}_b$ differ from the original variables only by constants. Therefore, $\mathrm{Var}(\tilde{\mathbf{z}}^{comp}_{j}) = n_o\Sigma_b$, $\mathrm{Var}(\tilde{\mathbf{z}}^{part}_{j}) = \pi n_p \Sigma_b$ and $\mathrm{Cov}(\tilde{\mathbf{x}}_{j},\tilde{\mathbf{z}}^{comp}_{j}) = \mathrm{Cov}(\tilde{\mathbf{x}}_{j},\tilde{\mathbf{z}}^{part}_{j}) = \mathrm{Cov}(\tilde{\mathbf{z}}^{comp}_{j},\tilde{\mathbf{z}}^{part}_{j}) = \mathbf{0}$, as shown in the proof of Lemma~\ref{lem: projection}. Also, by construction of the projection \eqref{eqn: linear rep}, $e_j$ is uncorrelated with $\tilde{\mathbf{x}}_{j}$, $\tilde{\mathbf{z}}^{comp}_{j}$ and $\tilde{\mathbf{z}}^{part}_{j}$, each a linear function of the original regressors and $D_{j}$, and hence uncorrelated with any linear combination of them.

 \textit{The competition regression.} Because $1$ and $D_{j}$ enter the regression \eqref{eqn: comp reg}, $Q_{comp}$ equals the mean squared residual from the projection of $\tilde{p}_j$ onto $(\tilde{\mathbf{x}}_j, \tilde{\mathbf{z}}^{comp}_{j})$. Let $v_j \equiv \tilde{\mathbf{z}}^{comp}_{j} \bm{\lambda}_o + \tilde{\mathbf{z}}^{part}_{j} \bm{\lambda}_p$ denote the conduct component. Since $\tilde{\mathbf{x}}_j$ is uncorrelated with $( \tilde{\mathbf{z}}^{comp}_{j}, \tilde{\mathbf{z}}^{part}_{j}, e_j )$, the projection of $\tilde{p}_j$ on $(\tilde{\mathbf{x}}_j, \tilde{\mathbf{z}}^{comp}_{j})$ has coefficient $\bm{\gamma}$ on $\tilde{\mathbf{x}}_j$, and the residual equals $e_j + v_j - L[v_j \vert \tilde{\mathbf{z}}^{comp}_{j}]$. Since $\tilde{\mathbf{z}}^{part}_{j}$ is uncorrelated with $\tilde{\mathbf{z}}^{comp}_{j}$, $L[v_j\,\vert\,\tilde{\mathbf{z}}^{comp}_{j}] = \tilde{\mathbf{z}}^{comp}_{j}\bm{\lambda}_o$, so that $\bm{\theta}^1 = \bm{\lambda}_o$ in \eqref{eqn: comp reg}, and the residual is $e_j + \tilde{\mathbf{z}}^{part}_{j}\bm{\lambda}_p$. Using the orthogonality of $e_j$,
\begin{equation}\label{eqn: Q comp}
	Q_{comp} = \mathrm{Var}(e_j) + \pi n_p\,\bm{\lambda}_p'\Sigma_b\bm{\lambda}_p.
\end{equation}

 \textit{The collusion regression.} $\mathrm{Var}(\tilde{\mathbf{z}}^{coll}_{j}) = (n_o + \pi n_p)\Sigma_b$ and $\mathrm{Cov}(\tilde{\mathbf{z}}^{coll}_{j},v_j) = \Sigma_b\big(n_o\bm{\lambda}_o + \pi n_p\bm{\lambda}_p\big)$. The projection coefficient of $v_j$ on $\tilde{\mathbf{z}}^{coll}_{j}$ is $(n_o\bm{\lambda}_o + \pi n_p\bm{\lambda}_p)/(n_o + \pi n_p) = (\bm{\lambda}_o + r\bm{\lambda}_p)/(1 + r)$, which is also the coefficient $\bm{\theta}^2$ on $\mathbf{z}^{coll}_j$ in \eqref{eqn: coll reg} because $\tilde{\mathbf{x}}_j$ is uncorrelated with $(\tilde{\mathbf{z}}^{coll}_{j}, v_j, e_j)$. The variance of the projection residual is
 \begin{equation}
 \begin{aligned}\label{eqn: residual variance}
    \mathrm{Var}\left(v_j - L[v_j \vert \tilde{\mathbf{z}}^{coll}_{j}]\right) &= 
	\mathrm{Var}(v_j) - \mathrm{Cov}(v_j,\tilde{\mathbf{z}}^{coll}_{j})\mathrm{Var}(\tilde{\mathbf{z}}^{coll}_{j})^{-1}\mathrm{Cov}(\tilde{\mathbf{z}}^{coll}_{j},v_j) \\
	&= n_o\,\bm{\lambda}_o'\Sigma_b\bm{\lambda}_o + \pi n_p\,\bm{\lambda}_p'\Sigma_b\bm{\lambda}_p - \frac{(n_o\bm{\lambda}_o + \pi n_p\bm{\lambda}_p)'\Sigma_b(n_o\bm{\lambda}_o + \pi n_p\bm{\lambda}_p)}{n_o + \pi n_p}.
 \end{aligned}
 \end{equation}

 Likewise, because $1$ and $D_{j}$ enter the regression \eqref{eqn: coll reg}, $Q_{coll}$ equals the mean squared residual from the projection of $\tilde{p}_j$ onto $(\tilde{\mathbf{x}}_j, \tilde{\mathbf{z}}^{coll}_{j})$. As in the competition regression, since $\tilde{\mathbf{x}}_j$ is uncorrelated with $(\tilde{\mathbf{z}}^{coll}_{j}, v_j, e_j)$, the projection of $\tilde{p}_j$ on $(\tilde{\mathbf{x}}_j, \tilde{\mathbf{z}}^{coll}_{j})$ has coefficient $\bm{\gamma}$ on $\tilde{\mathbf{x}}_j$, and its residual is $e_j + v_j - L[v_j \vert \tilde{\mathbf{z}}^{coll}_{j}]$. Since $e_j$ is orthogonal to $v_j - L[v_j \vert \tilde{\mathbf{z}}^{coll}_{j}]$, which is a linear combination of $\tilde{\mathbf{z}}^{comp}_{j}$ and $\tilde{\mathbf{z}}^{part}_{j}$, $Q_{coll}$ is the sum of $\mathrm{Var}(e_j)$ and $\mathrm{Var}\left(v_j - L[v_j \vert \tilde{\mathbf{z}}^{coll}_{j}]\right)$ in \eqref{eqn: residual variance}:
\begin{equation}\label{eqn: Q coll}
	Q_{coll} = \mathrm{Var}(e_j) + n_o\,\bm{\lambda}_o'\Sigma_b\bm{\lambda}_o + \pi n_p\,\bm{\lambda}_p'\Sigma_b\bm{\lambda}_p - \frac{(n_o\bm{\lambda}_o + \pi n_p\bm{\lambda}_p)'\Sigma_b(n_o\bm{\lambda}_o + \pi n_p\bm{\lambda}_p)}{n_o + \pi n_p}.
\end{equation}

 \textit{The differential.} Subtracting \eqref{eqn: Q coll} from \eqref{eqn: Q comp} and substituting $\pi n_p = r n_o$,
\begin{equation*}
	Q_{comp} - Q_{coll} = \frac{\Vert n_o\bm{\lambda}_o + \pi n_p\bm{\lambda}_p\Vert_{\Sigma_b}^2}{n_o + \pi n_p} - n_o\,\Vert\bm{\lambda}_o\Vert_{\Sigma_b}^2
	= n_o\left[\frac{\Vert \bm{\lambda}_o + r\bm{\lambda}_p\Vert_{\Sigma_b}^2}{1 + r} - \Vert\bm{\lambda}_o\Vert_{\Sigma_b}^2\right],
\end{equation*}
 which is \eqref{eqn: Q diff general}. In what follows we use the identity $1 + r\kappa_{+}(r) = \sqrt{1+r}$, which follows from \eqref{eqn: kappa star} because $r/(1+\sqrt{1+r}) = \sqrt{1+r} - 1$.

 \textit{Part (a).} If $\bm{\lambda}_p = \bm{\lambda}_o$, then $Q_{comp} - Q_{coll} = n_o\Vert\bm{\lambda}_o\Vert_{\Sigma_b}^2\big[(1+r) - 1\big] = n_o r\Vert\bm{\lambda}_o\Vert_{\Sigma_b}^2 > 0$ by Assumption~\ref{ass: regularity}(2) and (3).

 \textit{Part (b).} By the triangle inequality for the norm $\Vert\cdot\Vert_{\Sigma_b}$, $\Vert \bm{\lambda}_o + r\bm{\lambda}_p\Vert_{\Sigma_b} \leq \Vert\bm{\lambda}_o\Vert_{\Sigma_b} + r\Vert\bm{\lambda}_p\Vert_{\Sigma_b} < (1 + r\kappa_{+})\Vert\bm{\lambda}_o\Vert_{\Sigma_b} = \sqrt{1+r}\,\Vert\bm{\lambda}_o\Vert_{\Sigma_b}$. Squaring and dividing by $1 + r$ gives $\Vert \bm{\lambda}_o + r\bm{\lambda}_p\Vert_{\Sigma_b}^2/(1+r) < \Vert\bm{\lambda}_o\Vert_{\Sigma_b}^2$, so the differential is negative.

 \textit{Part (c).} Setting $\bm{\lambda}_p = \kappa\bm{\lambda}_o$ and $\bm{\lambda} = \bm{\lambda}_o$, the coefficient $\bm{\theta}^2$ becomes $(1 + r\kappa)\bm{\lambda}/(1 + r)$, and 
\begin{equation*}
	Q_{comp} - Q_{coll} = n_o\,\bm{\lambda}'\Sigma_b\bm{\lambda}\Big[\frac{(1 + r\kappa)^2}{1 + r} - 1\Big]
	= \bm{\lambda}'\Sigma_b\bm{\lambda}\cdot\frac{n_o r\big(r\kappa^2 + 2\kappa - 1\big)}{1 + r},
\end{equation*}
 which is \eqref{eqn: Q diff}. Since $\Sigma_b$ is positive definite, $\bm{\lambda}\neq \mathbf{0}$, and $n_o r/(1+r) > 0$, the sign of the differential is the sign of $q(\kappa) \equiv r\kappa^2 + 2\kappa - 1$. The quadratic $q$ is convex, and its roots are $\kappa = \big(-1 \pm \sqrt{1+r}\big)/r$. Rationalizing, $(-1+\sqrt{1+r})/r = 1/(1+\sqrt{1+r}) = \kappa_{+}(r)$ and $(-1-\sqrt{1+r})/r = -1/(\sqrt{1+r}-1) = \kappa_{-}(r)$, and $\kappa_{+} < 1/2$ because $\sqrt{1+r} > 1$. Hence $q(\kappa) < 0$ on $(\kappa_{-},\kappa_{+})$, $q(\kappa) = 0$ at the roots, and $q(\kappa) > 0$ otherwise.
 
 \textit{Omitting $D_j$.} If $D_j$ is not included in the regressions, all variables are centered at their unconditional means. The centered partner sum has variance $V \equiv \pi n_p\Sigma_b + \pi(1-\pi)n_p^2\bm{\mu}_b\bm{\mu}_b'$, which is positive definite, and it remains uncorrelated with the centered $\mathbf{x}_j$ and $\mathbf{z}^{comp}_j$. Repeating the computation with $V$ in place of $\pi n_p\Sigma_b$ yields $Q_{comp} - Q_{coll} =(n_o\Sigma_b\bm{\lambda}_o + V\bm{\lambda}_p)'(n_o\Sigma_b + V)^{-1}(n_o\Sigma_b\bm{\lambda}_o + V\bm{\lambda}_p) - n_o\bm{\lambda}_o'\Sigma_b\bm{\lambda}_o$, where $\bm{\lambda}_o$ and $\bm{\lambda}_p$ now denote the coefficients from the linear projection of $p_j$ onto $(1, \mathbf{x}_j, \mathbf{z}^{comp}_j, \mathbf{z}^{part}_j)$. At $\bm{\lambda}_p = \bm{\lambda}_o$, the differential equals $\bm{\lambda}_o'V\bm{\lambda}_o > 0$, and at $\bm{\lambda}_p = \mathbf{0}$, it equals $n_o^2\bm{\lambda}_o'\Sigma_b\big[(n_o\Sigma_b + V)^{-1} - (n_o\Sigma_b)^{-1}\big]\Sigma_b\bm{\lambda}_o < 0$ because $V$ is positive definite. Hence, part~(a) holds as stated, and by continuity, part~(b) holds for a threshold that no longer admits the closed form in \eqref{eqn: kappa star}. Moreover, without $D_j$, the covariance between the partner sum and price becomes
\begin{equation*}
    \mathrm{Cov}(\mathbf{z}^{part}_j, p_j) = \pi n_p \mathbf{g} + \pi(1-\pi) n_p \Delta\, \bm{\mu}_b, \qquad \Delta \equiv \mathbb{E}[p_j \mid D_j = 1] - \mathbb{E}[p_j \mid D_j = 0],
\end{equation*}
so that $\bm{\lambda}_p$ also reflects the average price premium $\Delta$ of suspected products. \qed

\subsection{Proof of Corollary~\ref{cor: logit}}\label{app: cor 2}

The proof proceeds in two steps. We first approximate equilibrium markups for small market shares. We then use this approximation to characterize the covariances that determine $\bm{\lambda}_o$ and $\bm{\lambda}_p$. Throughout, the number of products and cells is fixed. The small-share expansions are
understood along a common scaling $a_k\mapsto\tau a_k$ with $\tau\to0$, with sufficiently high finite moments for the stated remainder bounds. In Step~1, $s\equiv\max_{\mathscr{C}'}s_{\mathscr{C}'}$ is the largest realized cell share, which is of order $\tau$, and $O(s^k)$ denotes a random term bounded by a constant times $s^k$. When applied to moments (expectations and covariances), $O(s^k)$ means $O(\tau^k)$. For the relative covariance expansions, we also require a nonzero leading instrument signal, $\mathrm{Cov}(b(\mathbf{x}_k),a_k)/\mathbb{E}[a_k]\neq\mathbf{0}$, which is unchanged by this scaling.

\textit{Step 1: Markups for small market shares.}
Define
\begin{equation*}
    H_0 \equiv 1+e^{-1}\sum_{l\in\mathscr{J}}a_l,
    \qquad
    \tilde{s}_{\mathscr{C}}\equiv\frac{e^{-1}A_{\mathscr{C}}}{H_0}.
\end{equation*}
By the definition of $a_l$ in \eqref{eqn:a_index},
$\exp(\delta_l)=a_l e^{-\alpha\mu_l}$. As market shares vanish,
\eqref{eqn: cell markup} implies that each markup converges to $1/\alpha$.
Thus, $H_0$ and $\tilde{s}_{\mathscr{C}}$ are the inclusive value and cell
market share evaluated at markups equal to $1/\alpha$. Let $u_{\mathscr{C}}\equiv s_{\mathscr{C}}/(1-s_{\mathscr{C}})$.
Since $\alpha\mu_{\mathscr{C}}=1+u_{\mathscr{C}}$,
\eqref{eqn: s_C and H} becomes
\begin{equation*}
    s_{\mathscr{C}}
    =\frac{e^{-1}A_{\mathscr{C}}e^{-u_{\mathscr{C}}}}{H},
    \qquad
    H=1+e^{-1}\sum_{\mathscr{C}'}A_{\mathscr{C}'}e^{-u_{\mathscr{C}'}}.
\end{equation*}
Define $ h(x)\equiv\frac{e^{-x/(1-x)}}{1-x}$. Then, we have
\[u_{\mathscr{C}}=\tilde{s}_{\mathscr{C}}h(s_{\mathscr{C}})\frac{H_0}{H},\quad \frac{H_0}{H}=1+\sum_{\mathscr{C}'} s_{\mathscr{C}'}\big(e^{u_{\mathscr{C}'}} - 1\big)\]
where the second identity follows by substituting
$e^{-1}A_{\mathscr{C}'}=Hs_{\mathscr{C}'}e^{u_{\mathscr{C}'}}$
into $H_0-H$.

Because $h(0)=1$ and $h'(0)=0$, by a second-order Taylor expansion, we have
$h(s_{\mathscr{C}})=1+O(s^2)$. Moreover,
$u_{\mathscr{C}'}=O(s)$, so that
$s_{\mathscr{C}'}(e^{u_{\mathscr{C}'}}-1)=O(s^2)$.
With a fixed number of cells, $H_0/H=1+O(s^2)$.
It follows that
\begin{equation*}
    u_{\mathscr{C}}
    =\tilde{s}_{\mathscr{C}}\big(1+O(s^2)\big)
    =\tilde{s}_{\mathscr{C}}+O(s^3),
    \qquad
    \tilde{s}_{\mathscr{C}}=s_{\mathscr{C}}+O(s^2).
\end{equation*}
Therefore,
\begin{equation}\label{eqn: small share markup}
    \mu_{\mathscr{C}}
    =\frac{1}{\alpha}\Big[1+\tilde{s}_{\mathscr{C}}+O(s^3)\Big].
\end{equation}

\textit{Step 2: Covariances.}
Fix distinct products $j$ and $k$, and write
$\mathscr{C}=\mathscr{C}(j)$.
By Lemma~\ref{lem: logit},
$p_j=\mathbf{x}_j\gamma+\omega_j+\mu_{\mathscr{C}}$.
Assumption~\ref{ass: sampling} implies that
$b(\mathbf{x}_k)$ is independent of $(\mathbf{x}_j,\omega_j)$.
Hence \eqref{eqn: small share markup} gives
\begin{equation*}
    \mathrm{Cov}\big(b(\mathbf{x}_k),p_j\big)
    =\frac{1}{\alpha}\,
     \mathrm{Cov}\big(b(\mathbf{x}_k),\tilde{s}_{\mathscr{C}}\big)
     +O(s^3).
\end{equation*}
We evaluate the leading covariance by expanding the explicit expression for
$\tilde{s}_{\mathscr{C}}$. Let
\begin{equation*}
    q_l\equiv e^{-1}a_l,\quad
    Q_{\mathscr{C}}\equiv\sum_{l\in\mathscr{C}}q_l,\quad
    Q\equiv\sum_{l\in\mathscr{J}}q_l,
    \qquad
    \bar q\equiv\mathbb{E}[q_k].
\end{equation*}
Writing $J\equiv|\mathscr{J}|$ and
$n_{\mathscr{C}}\equiv|\mathscr{C}|$, we have
\begin{equation*}
    \tilde{s}_{\mathscr{C}}
    =\frac{Q_{\mathscr{C}}}{1+Q}
    =Q_{\mathscr{C}}-Q_{\mathscr{C}}Q+O(s^3).
\end{equation*}
Define the common covariance vectors,  
\(\mathbf{G}_1\equiv\mathrm{Cov}\big(b(\mathbf{x}_k),q_k\big) \) and \(\mathbf{G}_2\equiv\mathrm{Cov}\big(b(\mathbf{x}_k),q_k^2\big)\). By independence across products, if $k\in\mathscr{C}$,
\[\mathrm{Cov}\big(b(\mathbf{x}_k),Q_{\mathscr{C}}\big)=\mathbf{G}_1,\quad
    \mathrm{Cov}\big(b(\mathbf{x}_k),Q_{\mathscr{C}}Q\big)=\mathbf{G}_2+(J+n_{\mathscr{C}}-2)\bar q\,\mathbf{G}_1.
\]
If $k\notin\mathscr{C}$, the corresponding covariances are
$\mathbf{0}$ and $n_{\mathscr{C}}\bar q\,\mathbf{G}_1$.
Thus,
\begin{equation*}
    \mathrm{Cov}\big(b(\mathbf{x}_k),p_j\big)
    =\frac{1}{\alpha}
    \begin{cases}
      \mathbf{G}_1-\mathbf{G}_2
      -(J+n_{\mathscr{C}}-2)\bar q\,\mathbf{G}_1+O(s^3),
      &k\in\mathscr{C},\\[3pt]
      -n_{\mathscr{C}}\bar q\,\mathbf{G}_1+O(s^3),
      &k\notin\mathscr{C}.
    \end{cases}
\end{equation*}
By Step~1, the expected cell market share satisfies
$\mathbb{E}[s_{\mathscr{C}}]=n_{\mathscr{C}}\bar q+O(s^2)$.
From here on, and in Corollary~\ref{cor: logit}, $s_{\mathscr{C}}$, $s_f$, $s_{f_c}$, and $s_k$ denote expected market shares. Define
\begin{equation*}
    \bar{\mathbf{c}}
    \equiv\frac{1}{\alpha}
    \big[\mathbf{G}_1-\mathbf{G}_2-(J-2)\bar q\,\mathbf{G}_1\big].
\end{equation*}
Because $\mathbf{G}_1=O(s)$ and $\mathbf{G}_2=O(s^2)$,
the preceding expressions imply
\begin{equation}\label{eqn: cov small share}
    \mathrm{Cov}\big(b(\mathbf{x}_k),p_j\big)
    =\begin{cases}
       (1-s_{\mathscr{C}(j)})\bar{\mathbf{c}}+O(s^3),
       &k\in\mathscr{C}(j),\\[3pt]
       -s_{\mathscr{C}(j)}\bar{\mathbf{c}}+O(s^3),
       &k\notin\mathscr{C}(j).
     \end{cases}
\end{equation}
The leading vector is common to inside- and outside-cell products and does
not depend on $j$ or $k$. Moreover, since
$s_k=\bar q+O(s^2)$,
\(\bar{\mathbf{c}} =\frac{s_k}{\alpha} \frac{\mathrm{Cov}\big(b(\mathbf{x}_k),a_k\big)}{\mathbb{E}[a_k]} +O(s^2)\). Under the stated nondegeneracy condition, $\|\bar{\mathbf{c}}\|$ is of order $s$.
In particular, the $O(s^3)$ terms in
\eqref{eqn: cov small share} are of relative order $O(s^2)$
compared with $\bar{\mathbf{c}}$.

\textit{Under collusive conduct ($\phi=1$).}
If $\pi=1$, Lemma~\ref{lem: projection} gives
$\bm{\lambda}_p=\bm{\lambda}_o$,
and Proposition~\ref{prop: main}(a) implies
$Q_{comp}>Q_{coll}$. Suppose $\pi<1$.
For products of colluding firms, the cell is the coalition, with expected
market share $s_{f_c}$; for non-colluding firms, it is the firm, with expected
market share $s_f$. By \eqref{eqn: cov small share},
\begin{equation*}
    \mathbf{g}=(1-s_{f_c})\bar{\mathbf{c}}+O(s^3),
    \qquad
    \mathbf{g}^N=(1-s_f)\bar{\mathbf{c}}+O(s^3).
\end{equation*}
Substituting into \eqref{eqn: lambda pooled} yields
\[\bm{\lambda}_o=\big[1-\pi s_{f_c}-(1-\pi)s_f\big] \Sigma_b^{-1}\bar{\mathbf{c}}+O(s^3),\quad
\bm{\lambda}_p=(1-s_{f_c})\Sigma_b^{-1}\bar{\mathbf{c}}+O(s^3).\]
Thus, to first order in market shares,
\[    \bm{\lambda}_p=\kappa\bm{\lambda}_o+O(s^3),\quad
    \kappa \equiv
    \frac{1-s_{f_c}}{1-\pi s_{f_c}-(1-\pi)s_f}
    =1-(1-\pi)(s_{f_c}-s_f)+O(s^2).
\]
Since $\kappa\to1$ as $s\to0$ and $\kappa_+<1/2$,
Proposition~\ref{prop: main}(c) gives a positive fit differential
for the leading coefficient vectors.
The differential in \eqref{eqn: Q diff general} is continuous and
homogeneous of degree two; after normalizing the coefficients by
$\|\bar{\mathbf{c}}\|$, the $O(s^3)$ errors vanish.
Therefore $Q_{comp}>Q_{coll}$ for sufficiently small market shares.

\textit{Under competitive conduct ($\phi=0$).}
Each firm constitutes its own cell. By
Assumptions~\ref{ass: sampling} and~\ref{ass: symmetry},
all firms have the same expected market share $s_f$.
The argument in Appendix~\ref{app: lemma 2} gives
$\bm{\lambda}_o=\Sigma_b^{-1}\mathbf{g}_{own}$ and
$\bm{\lambda}_p=\Sigma_b^{-1}\mathbf{g}_{part}$,
where $\mathbf{g}_{own}$ and $\mathbf{g}_{part}$ are the
covariances with an own-firm and a suspected partner product,
respectively.\footnote{Under competition, the expected share of every firm
is the same, so the own-product covariance is common to suspected and
non-suspected firms.}
Since partner products lie outside the firm's cell,
\eqref{eqn: cov small share} implies
\begin{equation*}
    \mathbf{g}_{own}=(1-s_f)\bar{\mathbf{c}}+O(s^3),
    \qquad
    \mathbf{g}_{part}=-s_f\bar{\mathbf{c}}+O(s^3).
\end{equation*}
Hence,
\begin{equation*}
    \bm{\lambda}_p=\kappa\bm{\lambda}_o+O(s^3),
    \qquad
    \kappa\equiv-\frac{s_f}{1-s_f}<0,
    \qquad
    \frac{\|\bm{\lambda}_p\|_{\Sigma_b}}
         {\|\bm{\lambda}_o\|_{\Sigma_b}}
    =\frac{s_f}{1-s_f}+O(s^2).
\end{equation*}
By Proposition~\ref{prop: main}(c), the fit differential for the leading
coefficient vectors has the sign of $q(\kappa)=r\kappa^2+2\kappa-1$, which is
negative if and only if $\kappa\in(\kappa_-,\kappa_+)$. Because
$\kappa<0<\kappa_+$, this reduces to $\kappa>\kappa_-$, that is,
$s_f/(1-s_f)<1/(\sqrt{1+r}-1)$, or equivalently $s_f\sqrt{1+r}<1$,
which is \eqref{eqn: logit competition condition}.
As in the collusive case, the differential in \eqref{eqn: Q diff general}
is continuous and homogeneous of degree two, and the $O(s^3)$ errors vanish
after normalizing the coefficients by $\|\bar{\mathbf{c}}\|$. Since
$\kappa\to0$ as $s\to0$ and $q(0)=-1<0$, $Q_{comp}<Q_{coll}$ for
sufficiently small market shares, and
\eqref{eqn: logit competition condition} is the first-order condition
under which this ranking holds. Proposition~\ref{prop: main}(b) alone would
yield the stricter sufficient condition $s_f/(1-s_f)<\kappa_+$. \qed

\subsection{Unequal numbers of products across firms} \label{subsec: asymmetry}

Assumption \ref{ass: symmetry} simplifies the analysis by imposing a common portfolio size across firms. We discuss how Lemma \ref{lem: projection}, Proposition \ref{prop: main}, and Corollary \ref{cor: logit} extend when firms offer different numbers of products or suspected coalitions differ in size. For each product \(j\), let
\[
n_{oj}=|\mathscr{J}_{f(j)}|-1,\qquad
n_{pj}=|\mathscr{J}_{f_c(j)}\setminus\mathscr{J}_{f(j)}|,
\]
where \(n_{pj}=0\) for firms not suspected of collusion. Further define
\[
\bar n_o=\mathbb E[n_{oj}],\qquad
\bar n_p=\mathbb E[n_{pj}],\qquad
\bar r=\frac{\bar n_p}{\bar n_o},
\]
and suppose \(\bar n_o>0\) and \(\bar n_p>0\).

\subsubsection{Extension of Proposition \ref{prop: main}} When portfolio sizes vary, the instrument sums may be correlated through their varying numbers of components, even under Assumption \ref{ass: sampling}. To account for this variation, consider including \(n_{oj}\) and \(n_{pj}\) as additional controls in both price regressions (\ref{eqn: comp reg}) and (\ref{eqn: coll reg}).\footnote{Under Assumption \ref{ass: symmetry}, \(n_{oj}\) is constant and \(n_{pj}=n_pD_j\), so these controls are collinear with \((1,D_j)\) and add nothing to the projection; in estimation, they are simply dropped. More generally, only the components of \((n_{oj},n_{pj})\) that vary independently of \((1,D_j)\) need to be included, and fixed effects that already absorb portfolio sizes make them redundant.} Under Assumption \ref{ass: sampling}, partialling out the portfolio-size controls yields
\[
\widetilde {\mathbf{z}}_j^{comp}
=\mathbf{z}_j^{comp}-n_{oj}\bm{\mu}_b,\quad
\widetilde{\mathbf{z}}_j^{part}
=\mathbf{z}_j^{part}-n_{pj}\bm{\mu}_b.
\]
Because product-level primitives are i.i.d. and independent of product assignment, these re-centered instruments satisfy
\[
\operatorname{Var}(\widetilde{ \mathbf{z}}_j^{comp})
=\bar n_o\Sigma_b,\quad
\operatorname{Var}(\widetilde{ \mathbf{z}}_j^{part})
=\bar n_p\Sigma_b,\quad
\operatorname{Cov}(\widetilde{\mathbf{z}}_j^{comp},
\widetilde{\mathbf{z}}_j^{part})=0.
\]

Let \(\bm{\lambda}_o\) and \(\bm{\lambda}_p\) denote the coefficients on the two instrument vectors in the population linear projection of \(p_j\) that also includes the common controls. Repeating the projection argument in Appendix \ref{app: prop 1} yields
\[
Q_{comp}-Q_{coll}
=
\bar n_o
\left(
\frac{\|\bm{\lambda}_o+\bar r \bm{\lambda}_p\|_{\Sigma_b}^2}
{1+\bar r}
-\| \bm{\lambda}_o\|_{\Sigma_b}^2
\right).
\]
Thus, Proposition \ref{prop: main} continues to hold with \(n_o\) and \(r\) replaced by \(\bar n_o\) and \(\bar r\), respectively. In particular, the corresponding thresholds are
\[
\kappa_+
=\frac{1}{1+\sqrt{1+\bar r}},
\qquad
\kappa_-
=-\frac{1}{\sqrt{1+\bar r}-1}.
\]

\subsubsection{Extension of Lemma \ref{lem: projection}} Unequal portfolio sizes affect the interpretation of the projection coefficients. Let \(\textbf g_{oj}\) and \( \textbf g_{pj}\) denote the conditional covariance vectors between \(p_j\) and the characteristics of an individual own-firm or partner product, respectively, conditional on the portfolio-size and coalition configuration. Exchangeability implies that \(\textbf g_{oj}\) is common across the firm's other products, conditional on that configuration. The same holds for \(\textbf g_{pj}\) under full profit internalization, when all partner products lie in product \(j\)'s cell. Under competition, partner products of different partner firms need not be exchangeable when those firms differ in portfolio size; we then define \(\textbf g_{pj}\) as the average of these covariances over product \(j\)'s partner products. In either case, \(\operatorname{Cov}(\widetilde{\mathbf z}_j^{part},p_j\mid\text{configuration})=n_{pj}\textbf g_{pj}\), and likewise for \(\textbf g_{oj}\). The pooled projection coefficients satisfy
\[
\bm \lambda_o
=
\Sigma_b^{-1}
\frac{\mathbb E[n_{oj} \textbf g_{oj}]}{\bar n_o},
\qquad
 \bm \lambda_p
=
\Sigma_b^{-1}
\frac{\mathbb E[n_{pj} \textbf g_{pj}]}{\bar n_p}.
\]
Hence, these coefficients are portfolio-size weighted averages of conditional price effects. Under full profit internalization, own and partner products remain exchangeable within a coalition, so that \(\textbf g_{oj}= \textbf g_{pj}\) for products of colluding firms. Nevertheless, this conditional equality does not necessarily imply \( \bm \lambda_o= \bm \lambda_p\), for two reasons. First, as in Lemma \ref{lem: projection} with \(\pi<1\), products of non-suspected firms enter \(\bm\lambda_o\) through their own-firm covariances but do not enter \(\bm\lambda_p\), because \(n_{pj}=0\) for those products. Second, the two pooled coefficients weight products by \(n_{oj}\) and \(n_{pj}\), respectively, and therefore place different weights on suspected coalitions that differ in size and hence in their price responses. Neither reason applies if all firms belong to a single coalition in a common market environment with a fixed total number of products: the common covariance is then identical across own and partner products. Consequently, \(\bm \lambda_o= \bm \lambda_p\) continues to hold regardless of differences in individual firms' portfolio sizes, and Proposition \ref{prop: main}(a) applies.

\subsubsection{Extension of Corollary \ref{cor: logit}} Under logit demand, the covariance approximation in (\ref{eqn: cov small share}) continues to apply conditional on the portfolio-size and coalition configuration. Because the outside-cell covariance in (\ref{eqn: cov small share}) depends only on the focal product's cell, partner covariances under competition are common to first order, so the averaging in the definition of \(\textbf g_{pj}\) is immaterial here. For a fixed market environment with a common leading covariance vector \(\bar{\mathbf{c}}\), define the weighted expected cell shares
\[
\bar s_o^C
=
\frac{\mathbb E[n_{oj}s_{\mathscr{C}(j)}]}{\bar n_o},
\qquad
\bar s_p^C
=
\frac{\mathbb E[n_{pj}s_{\mathscr{C}(j)}]}{\bar n_p},
\]
where $s_{\mathscr{C}(j)}$ here and $s_{f(j)}$ below denote expected shares conditional on the portfolio-size and coalition configuration. Under collusive conduct, (\ref{eqn: cov small share}) implies, to first order in market shares,
\[
\bm \lambda_o\simeq
(1-\bar s_o^C)\Sigma_b^{-1}\bar{\mathbf{c}},
\qquad
\bm \lambda_p\simeq
(1-\bar s_p^C)\Sigma_b^{-1}\bar{\mathbf{c}},
\]
where $\simeq$ denotes equality up to terms of second order in market shares. Therefore,
\[
\bm \lambda_p\simeq\kappa \bm \lambda_o,
\qquad
\kappa\simeq
1-(\bar s_p^C-\bar s_o^C).
\]
As the maximum cell market share approaches zero, \(\kappa\) approaches one. Thus, the collusion IVs continue to provide a better fit for sufficiently small cell market shares.

Under competitive conduct, each firm constitutes its own cell. Defining
\[
\bar s_o^F
=
\frac{\mathbb E[n_{oj}s_{f(j)}]}{\bar n_o},
\qquad
\bar s_p^F
=
\frac{\mathbb E[n_{pj}s_{f(j)}]}{\bar n_p},
\]
equation (\ref{eqn: cov small share}) yields
\[
\bm \lambda_o\simeq
(1-\bar s_o^F)\Sigma_b^{-1}\bar{\mathbf{c}},
\qquad
\bm \lambda_p\simeq
-\bar s_p^F\Sigma_b^{-1}\bar{\mathbf{c}}.
\]
Accordingly, the analogue of condition (\ref{eqn: logit competition condition}) is \(\kappa > \kappa_-\) with \(\kappa \simeq -\bar s_p^F/(1-\bar s_o^F)\) and \(\bar r\) in place of \(r\), that is,
\(
\frac{\bar s_p^F}{1-\bar s_o^F}<\frac{1}{\sqrt{1+\bar r}-1}
\). Under Assumption~\ref{ass: symmetry}, we have \(\bar r=r\) and \(\bar s_o^F=\bar s_p^F=s_f\), recovering Proposition \ref{prop: main} and Corollary \ref{cor: logit}.

These extensions show that equal portfolio sizes are not essential to the comparison of instrument fit. Rather, relaxing Assumption \ref{ass: symmetry} requires accounting for variation in portfolio sizes and replacing common market share effects with appropriately weighted averages. The main exchangeability mechanism remains unchanged, although heterogeneity across coalitions can affect the coefficients of pooled price projections.

\newpage
\section{The role of the demand system}\label{app: demand system}

 This appendix discusses the role of the demand system in our testing framework. Specifically, we examine the two channels shown in \eqref{eqn: foc general} of Section~\ref{subsec: demand system} under nested logit and random coefficients logit demand. We argue that the magnitudes of both channels are governed by how closely the focal product $j$ and a partner product $\ell$ substitute for each other, which in turn hinges on the underlying demand system.
 
 To see this point, consider a random utility model with a constant price coefficient $\alpha$, holding prices (and hence markups) fixed. Through the demand channel, a change in $\delta_\ell$ alters $s_j$ by $-\alpha^{-1}\,\partial s_\ell/\partial p_j$.\footnote{Since the matrix of share derivatives with respect to mean utilities is symmetric, $\partial s_{j}/\partial\delta_{\ell} = \partial s_{\ell}/\partial\delta_{j} = -\alpha^{-1}\,\partial s_{\ell}/\partial p_{j}$.} The change also shifts $\partial s_j/\partial p_j$ through the demand channel and $\partial s_\ell/\partial p_j$ through the internalization channel by an amount proportional to $\partial s_\ell/\partial p_j$. The effect on other terms (the cross-price derivatives $\partial s_k/\partial p_j$, $k \in \mathscr{J}_f\setminus\{j\}$ through the demand channel and $\partial s_{\ell'}/\partial p_{j}$ where $\ell' \neq \ell$ through the internalization channel) is of higher order in market shares.\footnote{\label{fn: rc derivative}For random coefficients logit demand, consider two products $j$ and $k \neq j$ of the same firm and two partner products $\ell$ and $\ell' \neq \ell$. With consumer types $\nu$ and individual choice probabilities $s_{ij}(\nu)$, $\partial s_{m}/\partial p_{j} = \alpha\int s_{ij}s_{im}\,dF(\nu)$ for $m \in \{k, \ell\}$ and $\partial s_{j}/\partial p_{j} = -\alpha\int s_{ij}(1-s_{ij})\,dF(\nu)$. Differentiating with respect to $\delta_{\ell}$ gives $\alpha\int s_{ij}s_{i\ell}(1-2s_{ij})\,dF(\nu)$ for $\partial s_{j}/\partial p_{j}$, $\alpha\int s_{ij}s_{i\ell}(1-2s_{i\ell})\,dF(\nu)$ for $\partial s_{\ell}/\partial p_{j}$, and $-2\alpha\int s_{ij}s_{ik}s_{i\ell}\,dF(\nu)$ for $\partial s_{k}/\partial p_{j}$, and likewise for any partner product $\ell' \neq \ell$. The first two equal $\partial s_{\ell}/\partial p_{j}$ times a factor $1+O(s)$ and the third equals $\partial s_{\ell}/\partial p_{j}$ times $O(s)$, where $s$ stands for individual choice probabilities.} The demand system therefore determines how partner characteristics shift $p_{j}$ and whether a single sum over partner products, $\mathbf{z}^{part}_{j}$, captures that shift.
 
 We now examine how these channels operate under nested logit and random coefficients logit.

\subsection{Nested logit}

  Under a nested logit demand structure,
\begin{equation}\label{eqn: nested logit dsdp}
\frac{\partial s_k}{\partial p_j} =
	\begin{cases}
	\frac{\alpha}{1-\sigma}s_j\big( (1-\sigma)s_k + \sigma s_{k\vert g} - 1\big) & \quad \text{if } j = k \\
	\frac{\alpha}{1-\sigma}s_j\big( (1-\sigma)s_k + \sigma s_{k\vert g}\big) & \quad \text{if } j \in \mathscr{J}_{g(k)},j\neq k \\
	\alpha s_js_k & \quad \text{if } j \notin \mathscr{J}_{g(k)}
	\end{cases}
\end{equation}
 where the nesting parameter $\sigma \in [0,1)$ measures the degree of taste correlation within a nest, $s_{k \vert g}$ denotes the market share of product $k$ within nest $g$, and \( g(k) \) indexes the nest to which product \( k \) belongs. By \eqref{eqn: nested logit dsdp}, the cross-price derivative $\partial s_\ell/\partial p_j$ that governs both channels equals $\alpha s_j s_\ell$ if partner product $\ell$ lies in a different nest from product $j$ but $\alpha s_j\big(s_\ell + \sigma s_{\ell\vert g}/(1-\sigma)\big)$ if it lies in product $j$'s nest $g$. Therefore, the influence of a different-nest partner product relative to that of a same-nest partner product vanishes as $\sigma \to 1$ regardless of conduct. This two-level structure has distinct consequences under each conduct, which we examine in turn.

 \paragraph{Under collusive conduct.} 
Lemma~\ref{lem: projection} continues to apply because, with nest membership treated as a product characteristic, own and partner products are equally likely to be placed in product $j$'s nest under Assumptions~\ref{ass: specification}(1) and \ref{ass: sampling}. As a result, when $\pi = 1$, $\bm{\lambda}_p = \bm{\lambda}_o$ and the collusion IVs fit better by Proposition~\ref{prop: main}(a). If, however, partner products lie systematically in different nests from the focal firm's products, in violation of Assumption~\ref{ass: sampling}, they are no longer exchangeable with the firm's own products. Partner products then enter only through the different-nest derivative $\alpha s_j s_\ell$, which lacks the within-nest term $\sigma s_{\ell\vert g}/(1-\sigma)$ that own products in product $j$'s nest carry, so that $\kappa$ falls as $\sigma$ increases and may drop below the threshold $\kappa_{+}$, in which case the competition IVs fit better although firms collude.

 \paragraph{Under competitive conduct.} We proceed as in Corollary~\ref{cor: inside outside}. Let $\mathscr{J}_{g}$ denote the set of products in nest $g \in \mathscr{G} \equiv \{1, \ldots, G\}$, with the outside option forming its own nest. Write $e_{k} \equiv \exp(\delta_{k}/(1-\sigma))$, $D_{g} \equiv \sum_{k\in\mathscr{J}_{g}} e_{k}$ and $H \equiv 1 + \sum_{g'} D_{g'}^{1-\sigma}$, so that, for a product $k$ in nest $g$, $s_{k\vert g} = e_{k}/D_{g}$, $s_{g} = D_{g}^{1-\sigma}/H$ and $s_{k} = s_{k\vert g}\,s_{g}$. Let $s_{f \vert g}$ denote the market share of firm $f$ within nest $g$. We impose the following assumption for tractability.
\begin{assumption}[Symmetric nest structure]\label{ass: symmetric nests}
For each firm $f$, the within-nest share $s_{f \vert g}$ is identical across all nests $g \in \mathscr{G}$.
\end{assumption}
 This assumption implies that the markup of firm $f$ is common across all of its products in every nest. Substituting \eqref{eqn: nested logit dsdp} into \eqref{eqn: foc general} with $\phi = 0$ and solving for this common markup gives $\mu_{fg}(S_{f}) = (1-\sigma)/[\alpha(1-S_{f})]$, where the index $S_{f} \equiv (1-\sigma)s_{f} + \sigma s_{f\vert g}$ does not depend on $g$ under Assumption~\ref{ass: symmetric nests}. The competitive markup thus depends on the prices and characteristics of all products only through $S_{f}$, with $\mu_{fg}'(S) = (1-\sigma)/[\alpha(1-S)^{2}] > 0$, so that, as in Corollary~\ref{cor: inside outside}, the effect of product $k \neq j$ on $p_{j}$ is proportional to its effect on $S_{f}$, scaled by the common term $\mu_{fg}'(S_{f})$.\footnote{Note that
 \begin{align*}
 \frac{\partial p_{j}}{\partial \mathbf{x}_{k}} = \frac{\mathrm{d}\mu_{fg}}{\mathrm{d}S_{f}} \frac{\partial S_{f}}{\partial \log e_{k}} \frac{\partial \log e_{k}}{\partial \mathbf{x}_{k}} = \mu_{fg}'(S_{f})\, \frac{\partial S_{f}}{\partial \log e_{k}}\, \frac{c}{1-\sigma},
 \end{align*}
 where $\mu_{fg}'(S_{f})$ is a common multiplier identical across all products $k$, $\partial \log e_{k}/\partial \mathbf{x}_{k} = c/(1-\sigma)$ holding markups fixed, and $c \equiv \beta - \alpha\gamma$.}
 
 \begin{corollary}[Relative price effects of own and rival products under nested logit]\label{cor: nl inside outside}
 Suppose that firms compete ($\phi = 0$) and that Assumption~\ref{ass: symmetric nests} holds. Holding markups fixed, for any product $j \in \mathscr{J}_{f}\cap\mathscr{J}_{g}$ and any product $k \neq j$,
 \begin{equation}\label{eqn: nl inside outside}
	\frac{\partial S_{f}}{\partial \log e_{k}} =
	\begin{cases}
		(1-\sigma)\,s_{k}(1 - S_{f}) + \sigma\, s_{k\vert g}(1 - s_{f\vert g}) & k \in \mathscr{J}_{f}\cap\mathscr{J}_{g}, \\[4pt]
		(1-\sigma)\,s_{k}(1 - S_{f}) & k \in \mathscr{J}_{f}\setminus\mathscr{J}_{g}, \\[4pt]
		-(1-\sigma)\,s_{k}S_{f} - \sigma\, s_{k\vert g}s_{f\vert g} & k \in \mathscr{J}_{g}\setminus\mathscr{J}_{f}, \\[4pt]
		-(1-\sigma)\,s_{k}S_{f} & k \notin \mathscr{J}_{f}\cup\mathscr{J}_{g}.
	\end{cases}
 \end{equation}
 Hence, for products of equal market share, the effect on $p_{j}$ of a rival firm's product in $j$'s nest relative to that of an own product in $j$'s nest is
 \begin{equation}\label{eqn: nl ratio}
	-\frac{(1-\sigma)s_{g}S_{f} + \sigma s_{f\vert g}}{(1-\sigma)s_{g}(1-S_{f}) + \sigma(1 - s_{f\vert g})},
 \end{equation}
 which equals $-s_{f}/(1-s_{f})$ at $\sigma = 0$ and approaches $-s_{f\vert g}/(1-s_{f\vert g})$ as $\sigma \to 1$. The corresponding ratio for a rival firm's product in a different nest is $-(1-\sigma)s_{g}S_{f}/[(1-\sigma)s_{g}(1-S_{f}) + \sigma(1-s_{f\vert g})]$, which vanishes as $\sigma \to 1$.
 \end{corollary}

 \begin{proof}
 Let $E_{fg} \equiv \sum_{k\in\mathscr{J}_{f}\cap\mathscr{J}_{g}} e_{k}$, so that $s_{f\vert g} = E_{fg}/D_{g}$ and $s_{f} = \sum_{g'} E_{fg'}D_{g'}^{-\sigma}/H$. First, with $g$ and $g(k)$ denoting the nests of products $j$ and $k$, respectively, 
\begin{equation*}
	\frac{\partial s_{f\vert g}}{\partial \log e_{k}} = \frac{e_{k}\mathbf{1}\{k \in \mathscr{J}_{f}\}}{D_{g}} - \frac{E_{fg}\,e_{k}}{D_{g}^{2}} = s_{k\vert g}\big[\mathbf{1}\{k \in \mathscr{J}_{f}\} - s_{f\vert g}\big]
\end{equation*}
 for $g = g(k)$, and $\partial s_{f\vert g}/\partial \log e_k = 0$ otherwise. Second, $\partial H/\partial \log e_{k} = (1-\sigma)D_{g(k)}^{-\sigma}e_{k}$, and writing $N \equiv \sum_{g'} E_{fg'}D_{g'}^{-\sigma}$ for the numerator of $s_f$, so that $N = s_{f}H$, $\partial N/\partial \log e_{k} = e_{k}D_{g(k)}^{-\sigma}\big[\mathbf{1}\{k \in \mathscr{J}_{f}\} - \sigma s_{f\vert g(k)}\big]$. Since $s_{f} = N/H$ and $e_{k}D_{g(k)}^{-\sigma}/H = s_{k}$,
 \begin{equation*}
	\frac{\partial s_{f}}{\partial \log e_{k}} = \frac{1}{H}\frac{\partial N}{\partial \log e_{k}} - \frac{s_{f}}{H}\frac{\partial H}{\partial \log e_{k}} = s_{k}\big[\mathbf{1}\{k \in \mathscr{J}_{f}\} - \sigma s_{f\vert g(k)} - (1-\sigma)s_{f}\big] = s_{k}\big[\mathbf{1}\{k \in \mathscr{J}_{f}\} - S_{f}\big],
 \end{equation*}
 where the last equality holds under Assumption~\ref{ass: symmetric nests}. Third, since $S_{f} = (1-\sigma)s_{f} + \sigma s_{f\vert g}$,
 \begin{equation*}
	\frac{\partial S_{f}}{\partial \log e_{k}} = (1-\sigma)\,s_{k}\big[\mathbf{1}\{k \in \mathscr{J}_{f}\} - S_{f}\big] + \sigma\, s_{k\vert g}\big[\mathbf{1}\{k \in \mathscr{J}_{f}\} - s_{f\vert g}\big]\mathbf{1}\{k \in \mathscr{J}_{g}\},
 \end{equation*}
 which gives the four cases of \eqref{eqn: nl inside outside}. Finally, for products with equal market shares $s_{k}$, substituting $s_{k\vert g} = s_{k}/s_{g}$ for those in nest $g$ and dividing the third case by the first gives \eqref{eqn: nl ratio}. Dividing the fourth case by the first gives the ratio for a rival's product in a different nest. At $\sigma = 0$, $S_{f} = s_{f}$ and \eqref{eqn: nl ratio} reduces to $-s_{f}/(1-s_{f})$, while as $\sigma \to 1$, \eqref{eqn: nl ratio} approaches $-s_{f\vert g}/(1-s_{f\vert g})$. \qed
 \end{proof}


 Corollary~\ref{cor: nl inside outside} shows that the price effect of a same-nest rival product relative to that of a same-nest own product of equal market share equals $-s_f/(1-s_f)$ at $\sigma = 0$ and approaches $-s_{f\vert g}/(1-s_{f\vert g})$ as $\sigma \to 1$, where $s_{f\vert g}$ denotes the firm's within-nest market share. Under Assumption~\ref{ass: sampling}, as $\sigma \to 1$, the effects of different-nest products vanish, so that $\Vert\bm{\lambda}_p\Vert_{\Sigma_b}/\Vert\bm{\lambda}_o\Vert_{\Sigma_b}$ approaches $s_{f\vert g}/(1-s_{f\vert g})$ (by an argument analogous to Appendix~\ref{app: cor 2}) and the analogue of condition~\eqref{eqn: logit competition condition} involves the firm's within-nest share rather than its market share, a more demanding requirement since the former exceeds the latter: under Assumption~\ref{ass: symmetric nests}, $s_{f\vert g} = s_f/(1-s_0) > s_f$ in every market, and hence also in expectation. If, in violation of Assumption~\ref{ass: sampling}, the firm's products are spread across nests so that each shares its nest with partner products rather than with the firm's other products, $\bm{\lambda}_o$ is diluted by own products whose effect on $p_j$ vanishes as $\sigma \to 1$ whereas $\bm{\lambda}_p$ is not, so that $\kappa$, which is negative under competition, can fall below $\kappa_{-}$ and the collusion IVs fit better by Proposition~\ref{prop: main}(c) even though firms compete.

 \subsection{Random coefficients logit}

 With random coefficients on observed characteristics, the cross-price derivative $\partial s_{\ell}/\partial p_{j}$ depends on the proximity of product $j$ and partner product $\ell$ in characteristic space as well as on their market shares.\footnote{\label{fn: cross price derivative} The cross-price derivative in footnote~\ref{fn: rc derivative} decomposes as $\partial s_{\ell}/\partial p_{j} = \alpha\big[s_{j}s_{\ell} + \mathrm{Cov}_{\nu}(s_{ij}, s_{i\ell})\big]$, where the first term coincides with the logit cross-price derivative and depends only on market shares, whereas the covariance term, zero under logit, is positive when the two products are close in characteristic space and thus appeal to the same consumer types.} Since both the demand and the internalization channels for partner product $\ell$ are governed by this derivative, the price response of $p_{j}$ to $\mathbf{x}_{\ell}$ varies with the distance $\vert \mathbf{x}_{\ell} - \mathbf{x}_{j}\vert$, within and across cells. Other products then enter $p_{j}$ through pairwise terms in $(\mathbf{x}_{\ell}, \mathbf{x}_{j})$ rather than through a sum of functions of $\mathbf{x}_{\ell}$ alone; consequently, the price response is no longer a function of sums of powers of the characteristics alone, as it is under logit demand by \eqref{eqn: sum expansion}. The breakdown of this symmetric sum structure affects the fit comparison differently under each conduct.\footnote{What remains is a weaker symmetry: products with the same mean utility at the same distance from product $j$, with distance weighted by the dispersion of tastes, affect $p_{j}$ identically, which is the structure that differentiation IVs exploit \citep{gandhi2019measuring}.}

 \paragraph{Under collusive conduct.} Lemma~\ref{lem: projection} continues to hold because own and partner products are equally likely to be located near product $j$ in characteristic space under Assumptions~\ref{ass: specification}(1) and~\ref{ass: sampling}. Consequently, $\bm{\lambda}_p = \bm{\lambda}_o$ when $\pi = 1$, and the collusion IVs fit better by Proposition~\ref{prop: main}(a). The fit comparison can fail when the suspected partners' products are located systematically far from the firm's own products, a configuration that violates Assumption~\ref{ass: sampling}, so that own and partner products are not exchangeable and $\Vert\bm{\lambda}_p\Vert_{\Sigma_b}$ falls short of $\Vert\bm{\lambda}_o\Vert_{\Sigma_b}$; once it falls below $\kappa_{+}\Vert\bm{\lambda}_o\Vert_{\Sigma_b}$, the competition IVs fit better by Proposition~\ref{prop: main}(b) although firms collude. Because such a configuration also alters the second moments of the instruments, the thresholds of Proposition~\ref{prop: main} apply approximately in this case. If the partner products are so distant that the cross-price derivatives vanish, the third term in \eqref{eqn: foc general} is zero. Collusion then leaves no trace in prices, and no price-based test can detect it.\footnote{The markup-based RV test of Section~\ref{sec: markup testing} fails in this case as well, because collusion leaves markups unchanged and the two conduct models become observationally equivalent.}

 \paragraph{Under competitive conduct.} A rival product located close to product $j$ affects $p_{j}$ through local competition, not merely through a market aggregate. Hence $\bm{\lambda}_p$ is governed by how close the suspected partner products are to product $j$. Under Assumption~\ref{ass: sampling}, own and partner products are equally likely to be located near product $j$, and the ratio $\Vert\bm{\lambda}_p\Vert_{\Sigma_b}/\Vert\bm{\lambda}_o\Vert_{\Sigma_b}$ depends on the firm's share among the close substitutes of product $j$, the analogue of the within-nest share under nested logit; the competition IVs then fit better as long as this local share is modest. If suspected partner products constitute the closest substitutes of the firm's products, a configuration that violates Assumption~\ref{ass: sampling}, $\Vert\bm{\lambda}_p\Vert_{\Sigma_b}$ can exceed the bound in Proposition~\ref{prop: main}(b), so that the competition IVs are no longer guaranteed to fit better; if $\bm{\lambda}_p = \kappa\bm{\lambda}_o$ for a scalar $\kappa$, the collusion IVs fit better by Proposition~\ref{prop: main}(c) once $\kappa$, which is negative under competition, falls below $\kappa_{-}$, even though firms compete.

\newpage
\section{Own-firm versus other-firm instrument}\label{sec: own vs other}

 Our test utilizes $\mathbf{z}^{comp}$ and $\mathbf{z}^{coll}$, which are functions of the attributes of products produced by the firm and its suspected colluding partners (e.g., summation of attributes or squared attributes). Here, we set aside the issue of collusion and focus on the role of the standard own- and other-firm instruments, $\mathbf{z}^{own}$ and $\mathbf{z}^{other}$, in demand estimation. The simulation study presented in this appendix provides additional intuition and aids in interpreting the previous simulation results related to our proposed testing procedure.

 The Monte Carlo setup is as follows: We fix the number of products at $J = 36$ and vary the number of firms $F \in \{1, 2, \ldots, 36\}$. Since we do not consider any form of collusive behavior among firms, $\phi$ is fixed at 0. We generate instruments $\mathbf{z}^{own}$ and $\mathbf{z}^{other}$ as defined in equation \eqref{eqn: own and other iv}. To investigate their roles in demand estimation, we report the median F-statistics from the price regression and the median absolute errors of the estimated utility parameters across 500 simulations under the following instrument configurations: (i) $\mathbf{z}^{own}$ only, (ii) $\mathbf{z}^{other}$ only, and (iii) $\mathbf{z}^{both} = (\mathbf{z}^{own}, \mathbf{z}^{other})$.

\begin{figure}[htbp]
\centering
\caption{Own-firm vs other-firm instruments}
    \begin{subfigure}[b]{\textwidth}
    \centering
        \vspace{.15in}
    \caption*{\underline{\textit{Panel A: without a random coefficient}}}
        \begin{subfigure}[b]{0.47\textwidth}
        \vspace{.1in}
        \caption*{(a) Median first stage F-stat}
        \includegraphics[width=\textwidth]{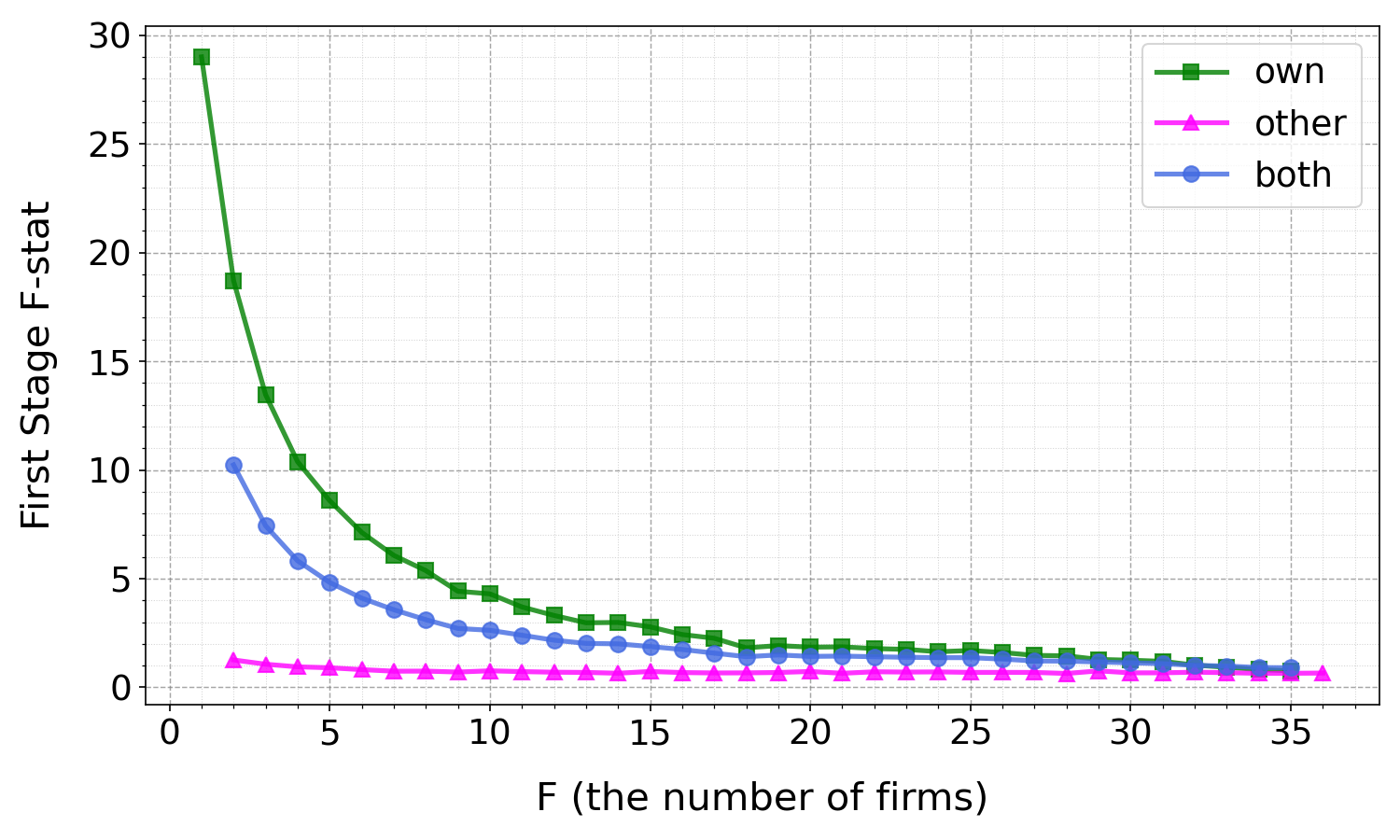}
        \end{subfigure}
        \qquad
        \begin{subfigure}[b]{0.47\textwidth}
        \caption*{(b) Median $\vert\hat{\alpha}-\alpha\vert$}
        \includegraphics[width=\textwidth]{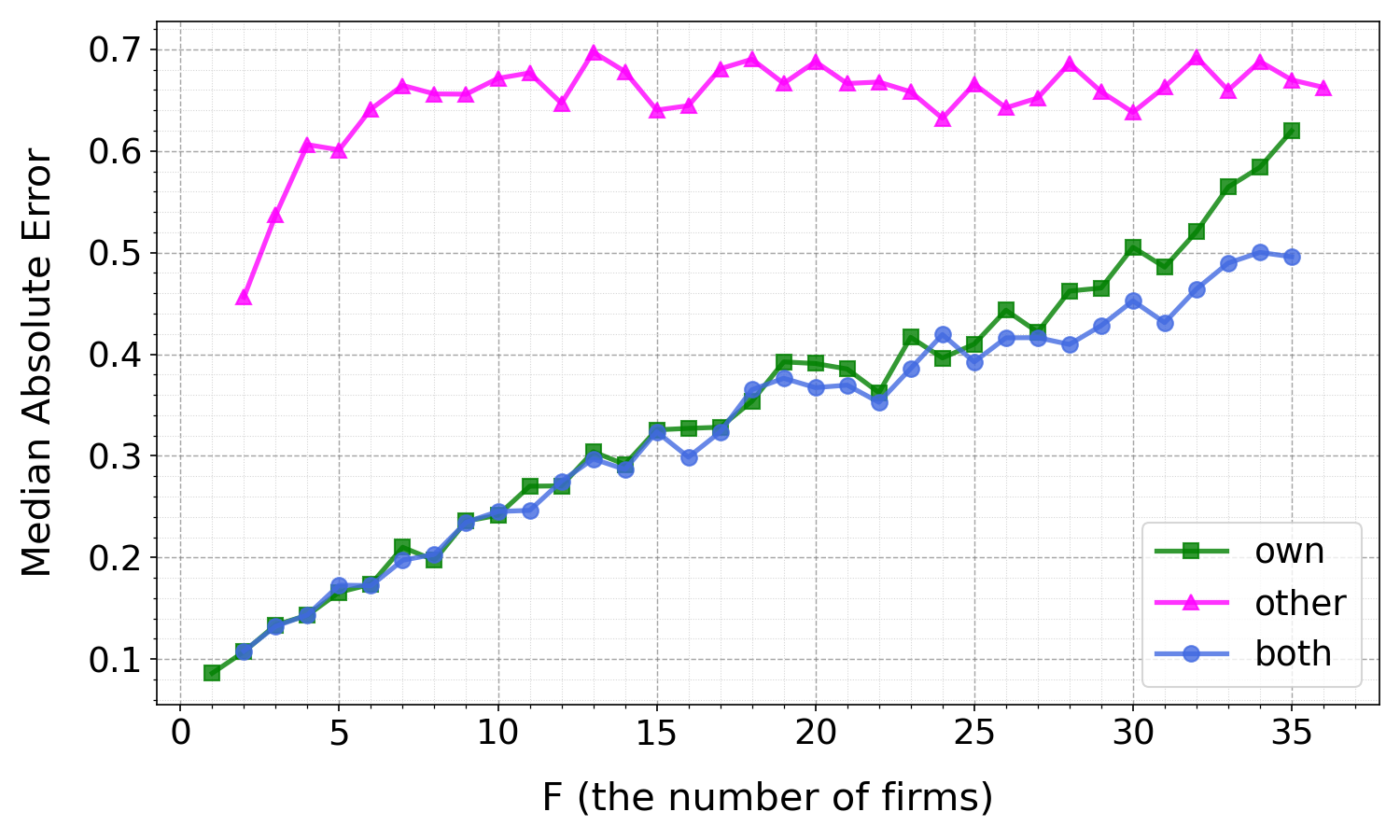}
        \end{subfigure}
    \end{subfigure}
    \begin{subfigure}[b]{\textwidth}
    \centering
    \vspace{.15in}
    \caption*{\underline{\textit{Panel B: with a random coefficient}}}
        \begin{subfigure}[b]{0.47\textwidth}
        \vspace{.1in}
        \caption*{(c) Median first stage F-stat}
        \includegraphics[width=\textwidth]{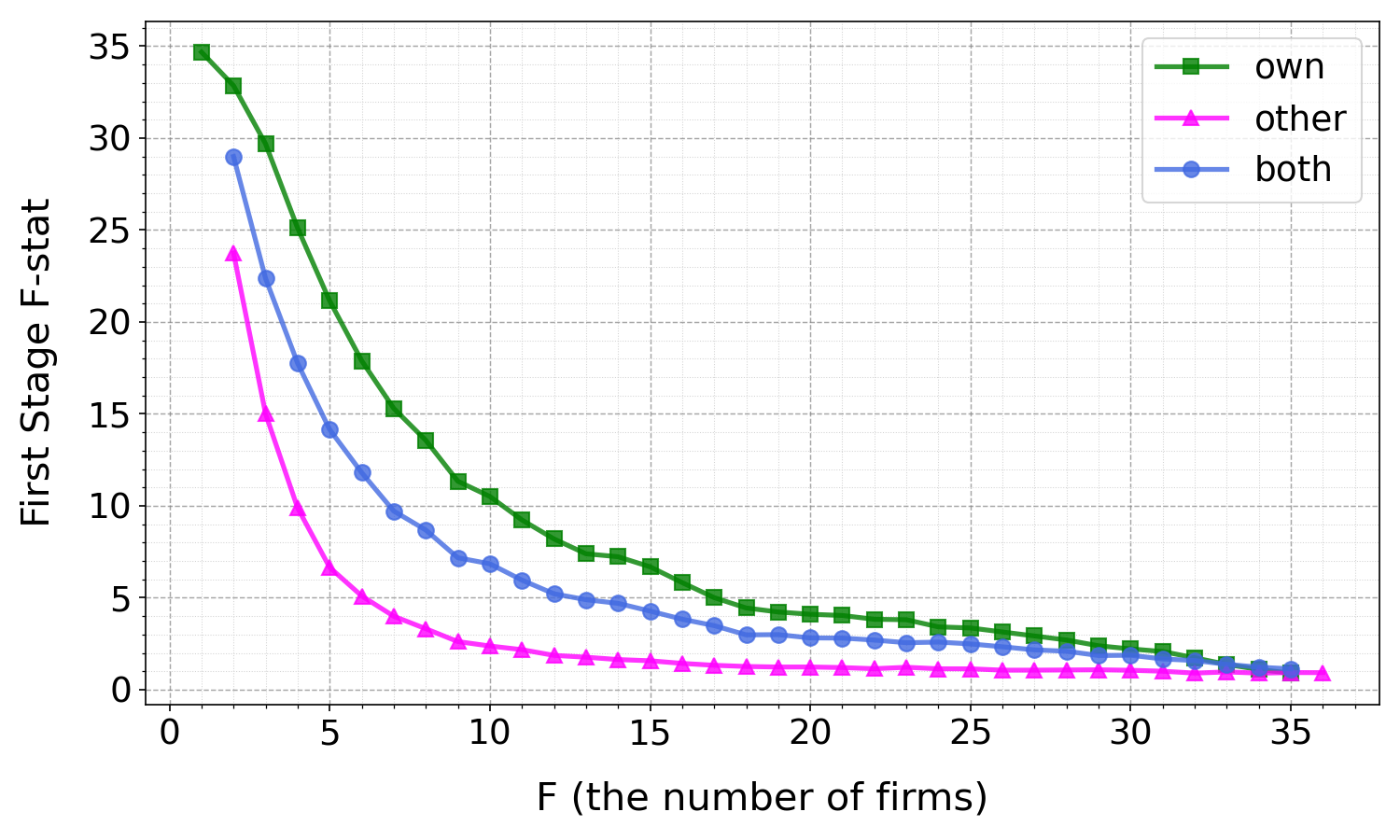}
        \end{subfigure}
        \qquad
        \begin{subfigure}[b]{0.47\textwidth}
        \caption*{(d) Median $\vert\hat{\alpha}-\alpha\vert$}
        \includegraphics[width=\textwidth]{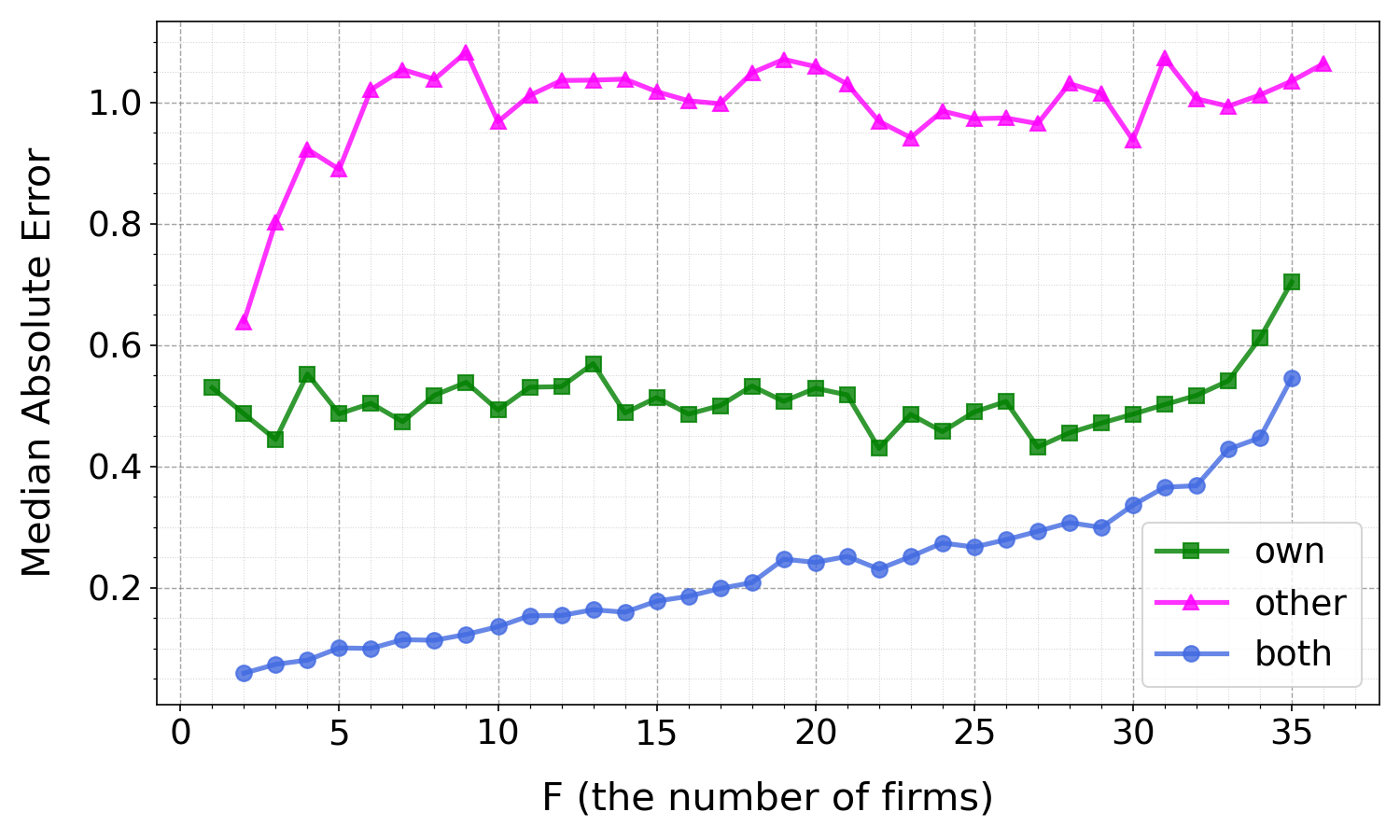}
        \end{subfigure}
        \begin{subfigure}[b]{0.47\textwidth}
        \vspace{.05in}
        \caption*{(e) Median $\vert\hat{\sigma}_x-\sigma_x\vert$}
        \includegraphics[width=\textwidth]{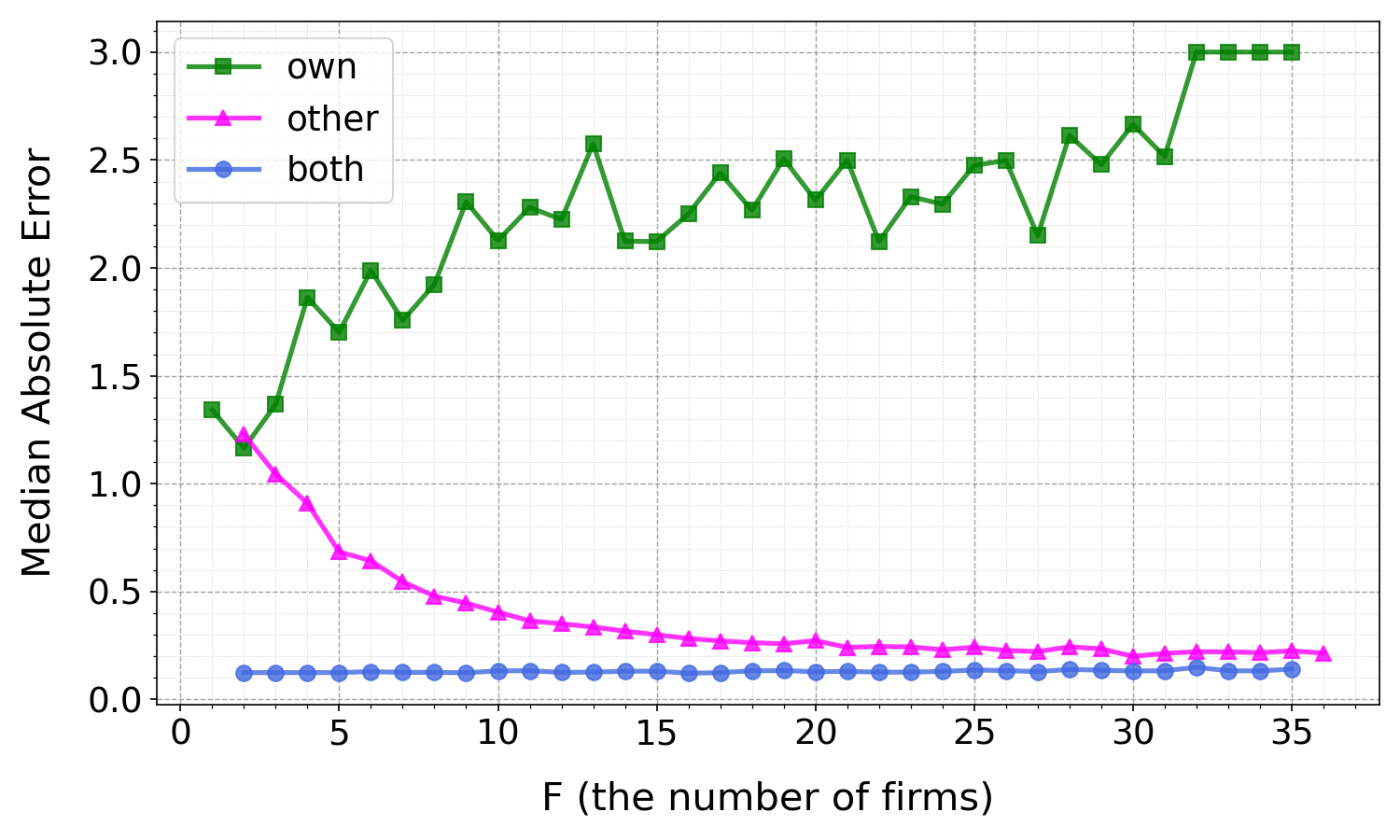}
    	\end{subfigure}
    \end{subfigure}
\label{fig: own vs other}
\tablenotes The figure shows the median F-statistic from the price regression and the median absolute errors of the estimated utility parameters, $\hat{\alpha}$ and $\hat{\sigma}_x$, across 500 simulated datasets for each Monte Carlo configuration ($J=36, F, T=100$). The upper panel presents results for the utility specification without a random coefficient, while the bottom panel shows results with a random coefficient.
\end{figure}

 The results for the utility specification without (with) a random coefficient, illustrated in the upper panel (bottom panel, respectively) of Figure \ref{fig: own vs other}, reveal several interesting points.\footnote{Tables \ref{tab: own vs other full} and \ref{tab: own vs other full rc} provide the full numeric results, including the median root-mean-squared errors (RMSE) of the estimated coefficients.} First, own-firm instruments exhibit stronger relevance than other-firm instruments in the price regression, as shown in Figures \ref{fig: own vs other}(a) and (c). Consequently, the median absolute error of the estimated price coefficient is lower when $\mathbf{z}^{own}$ is used as the instrument, as shown in Figures \ref{fig: own vs other}(b) and (d). These observations are consistent with the argument in Section \ref{sec: conceptual framework} and previous literature (e.g., \citealp{bresnahan1987competition}; \citealp{Berry95}) that firms internalize the cross-price elasticities of their own products when setting profit-maximizing prices, resulting in greater explanatory power of own-firm product attributes in determining equilibrium prices. They also align with our simulation results presented in the main text and echo the importance of correctly indexing firms based on the true conduct model.

 Our test uses only own-firm instruments, $\mathbf{z}^{comp}$ and $\mathbf{z}^{coll}$, because they are more relevant than other-firm instruments in the price regression. In fact, our testing procedure can be extended to incorporate other-firm instruments constructed in the same manner as the own-firm instruments, based on the original firm and the suspected colluding firm indices. Excluding the product attributes of colluding partners from the instruments for other firms enhances their relevance.\footnote{Simulation results available from the authors upon request are both quantitatively and qualitatively similar to those obtained using only the own-firm instruments. Additionally, it is worth noting that when testing whether all firms in the market collude ($F_c = 1$), constructing other-firm instruments under the suspected colluding firm index is impossible, making it more likely for our testing procedure to diagnose that the model of competition provides a better fit.}

 Second, the bottom panel of the figure shows that while own-firm instruments are more relevant in the price regression and more informative about the price coefficient, other-firm instruments are more useful for identifying the non-linear coefficient $\sigma_x$.\footnote{For every $F \geq 3$, using $\mathbf{z}^{other}$ instead of $\mathbf{z}^{own}$ results in a lower median absolute error and a lower median RMSE for the estimated non-linear coefficient, as shown in Table \ref{tab: own vs other full rc}.} These observations help explain the results in the bottom panel of Table \ref{tab: Fstat evidence}, where using $\mathbf{z}^{coll}$ leads to a smaller median absolute error for the estimated non-linear coefficient, even when the degree of profit internalization is low. Specifically, variations in product attributes of other firms unintentionally included in $\mathbf{z}^{coll}$ may assist in identifying the non-linear coefficient.

 Third, the less concentrated the market, the lower the explanatory power of the two BLP-style instruments. This occurs because, as competition among firms intensifies, equilibrium markups converge, which in turn weakens the correlation between markups and instruments. For instance, assuming a simple logit demand, when the market is competitive, the cell markups in equation \eqref{eqn: cell markup} differ little across products in the market. In fact, this finding aligns with the observation made by \cite{armstrong2016large} in the extreme case of $J \rightarrow \infty$.\footnote{Specifically, \citet{armstrong2016large} examined cases where markets consist of many single-product firms ($J_t = F_t$), and demonstrated that as $J_t \to \infty$, the explanatory power of BLP-style instruments for equilibrium markups diminishes. Cross-firm or market-level variations in equilibrium markups can help maintain relevance in such cases. See \citet{armstrong2016large} for further details.} On the other hand, in a more concentrated market, greater dispersion in markup sizes makes their correlation with the instruments less sensitive to unobserved shocks.

\newpage
\clearpage

\newpage
\begin{table}[htbp]
\scriptsize
  \centering
  \caption{Own-firm vs other-firm instruments without a random coefficient: full results}
    \begin{tabular}{lrrrrrrrrrrrrr}
    \toprule
          & \multicolumn{3}{c}{Median Fstat} &       & \multicolumn{3}{c}{Median $\vert\alpha-\hat{\alpha}\vert$} &       & \multicolumn{3}{c}{Median RMSE ($\hat{\alpha}$)} &       &  \\
\cmidrule{2-4}\cmidrule{6-8}\cmidrule{10-12}    $F$     & \multicolumn{1}{l}{own} & \multicolumn{1}{l}{other} & \multicolumn{1}{l}{both} &       & \multicolumn{1}{l}{own} & \multicolumn{1}{l}{other} & \multicolumn{1}{l}{both} &       & \multicolumn{1}{l}{own} & \multicolumn{1}{l}{other} & \multicolumn{1}{l}{both} &       & \multicolumn{1}{l}{$s_0$} \\
    \midrule
    1     & 29.00 &       &       &       & 0.09  &       &       &       & 0.16  &       &       &       & 0.72 \\
    2     & 18.71 & 1.26  & 10.23 &       & 0.11  & 0.46  & 0.11  &       & 0.20  & 0.84  & 0.19  &       & 0.68 \\
    3     & 13.46 & 1.06  & 7.43  &       & 0.13  & 0.54  & 0.13  &       & 0.24  & 0.95  & 0.24  &       & 0.67 \\
    4     & 10.38 & 0.94  & 5.82  &       & 0.14  & 0.61  & 0.14  &       & 0.28  & 1.02  & 0.26  &       & 0.66 \\
    5     & 8.59  & 0.90  & 4.82  &       & 0.17  & 0.60  & 0.17  &       & 0.30  & 1.06  & 0.30  &       & 0.66 \\
    6     & 7.13  & 0.81  & 4.10  &       & 0.17  & 0.64  & 0.17  &       & 0.33  & 1.13  & 0.31  &       & 0.65 \\
    7     & 6.07  & 0.74  & 3.57  &       & 0.21  & 0.66  & 0.20  &       & 0.37  & 1.12  & 0.36  &       & 0.65 \\
    8     & 5.37  & 0.75  & 3.11  &       & 0.20  & 0.66  & 0.20  &       & 0.38  & 1.15  & 0.36  &       & 0.65 \\
    9     & 4.42  & 0.70  & 2.71  &       & 0.24  & 0.66  & 0.23  &       & 0.43  & 1.17  & 0.40  &       & 0.65 \\
    10    & 4.30  & 0.76  & 2.62  &       & 0.24  & 0.67  & 0.25  &       & 0.44  & 1.15  & 0.40  &       & 0.65 \\
    11    & 3.70  & 0.72  & 2.40  &       & 0.27  & 0.68  & 0.25  &       & 0.47  & 1.13  & 0.43  &       & 0.65 \\
    12    & 3.31  & 0.70  & 2.17  &       & 0.27  & 0.65  & 0.28  &       & 0.50  & 1.16  & 0.45  &       & 0.65 \\
    13    & 2.97  & 0.68  & 2.02  &       & 0.30  & 0.70  & 0.30  &       & 0.53  & 1.21  & 0.48  &       & 0.65 \\
    14    & 2.99  & 0.65  & 2.00  &       & 0.29  & 0.68  & 0.29  &       & 0.52  & 1.21  & 0.48  &       & 0.65 \\
    15    & 2.78  & 0.73  & 1.88  &       & 0.33  & 0.64  & 0.32  &       & 0.55  & 1.14  & 0.51  &       & 0.64 \\
    16    & 2.43  & 0.68  & 1.74  &       & 0.33  & 0.64  & 0.30  &       & 0.57  & 1.13  & 0.50  &       & 0.64 \\
    17    & 2.26  & 0.66  & 1.57  &       & 0.33  & 0.68  & 0.32  &       & 0.60  & 1.18  & 0.52  &       & 0.64 \\
    18    & 1.82  & 0.66  & 1.41  &       & 0.35  & 0.69  & 0.37  &       & 0.67  & 1.18  & 0.57  &       & 0.64 \\
    19    & 1.92  & 0.68  & 1.49  &       & 0.39  & 0.67  & 0.38  &       & 0.68  & 1.17  & 0.60  &       & 0.64 \\
    20    & 1.85  & 0.73  & 1.42  &       & 0.39  & 0.69  & 0.37  &       & 0.67  & 1.18  & 0.58  &       & 0.64 \\
    21    & 1.86  & 0.65  & 1.44  &       & 0.39  & 0.67  & 0.37  &       & 0.68  & 1.20  & 0.58  &       & 0.64 \\
    22    & 1.78  & 0.72  & 1.41  &       & 0.36  & 0.67  & 0.35  &       & 0.69  & 1.14  & 0.58  &       & 0.64 \\
    23    & 1.75  & 0.70  & 1.38  &       & 0.42  & 0.66  & 0.39  &       & 0.71  & 1.15  & 0.60  &       & 0.64 \\
    24    & 1.64  & 0.71  & 1.37  &       & 0.40  & 0.63  & 0.42  &       & 0.76  & 1.17  & 0.62  &       & 0.64 \\
    25    & 1.69  & 0.69  & 1.36  &       & 0.41  & 0.67  & 0.39  &       & 0.73  & 1.20  & 0.62  &       & 0.64 \\
    26    & 1.60  & 0.69  & 1.30  &       & 0.44  & 0.64  & 0.42  &       & 0.76  & 1.21  & 0.64  &       & 0.64 \\
    27    & 1.47  & 0.68  & 1.21  &       & 0.42  & 0.65  & 0.42  &       & 0.76  & 1.15  & 0.63  &       & 0.64 \\
    28    & 1.45  & 0.64  & 1.20  &       & 0.46  & 0.69  & 0.41  &       & 0.79  & 1.19  & 0.64  &       & 0.64 \\
    29    & 1.28  & 0.76  & 1.17  &       & 0.47  & 0.66  & 0.43  &       & 0.84  & 1.12  & 0.64  &       & 0.64 \\
    30    & 1.25  & 0.67  & 1.13  &       & 0.51  & 0.64  & 0.45  &       & 0.86  & 1.13  & 0.69  &       & 0.64 \\
    31    & 1.20  & 0.67  & 1.09  &       & 0.49  & 0.66  & 0.43  &       & 0.86  & 1.18  & 0.68  &       & 0.64 \\
    32    & 1.02  & 0.70  & 1.01  &       & 0.52  & 0.69  & 0.46  &       & 0.95  & 1.18  & 0.70  &       & 0.64 \\
    33    & 0.92  & 0.67  & 0.98  &       & 0.56  & 0.66  & 0.49  &       & 0.98  & 1.15  & 0.73  &       & 0.64 \\
    34    & 0.82  & 0.64  & 0.91  &       & 0.58  & 0.69  & 0.50  &       & 1.08  & 1.14  & 0.78  &       & 0.64 \\
    35    & 0.76  & 0.65  & 0.90  &       & 0.62  & 0.67  & 0.50  &       & 1.12  & 1.16  & 0.78  &       & 0.64 \\
    36    &       & 0.66  &       &       &       & 0.66  &       &       &       & 1.14  &       &       & 0.64 \\
    \bottomrule
    \end{tabular}%
  \label{tab: own vs other full}%
  \tablenotes The table reports the median values of the absolute error and RMSE of the estimated price coefficient, as well as the median F-statistics, when $\mathbf{z}^{own}$, $\mathbf{z}^{other}$, and ($\mathbf{z}^{own}, \mathbf{z}^{other}$) are used as instruments individually. A random coefficient is excluded from the indirect utility function \eqref{eqn: dgp utility} in the DGP. The median outside option share across 500 simulated datasets for each Monte Carlo configuration is denoted by $s_0$.
\end{table}%
\clearpage

\newpage
\begin{landscape}
\begin{table}[htbp]
\scriptsize
  \centering
  \caption{Own-firm vs other-firm instruments with a random coefficient: full results}
    \begin{tabular}{lrrrrrrrrrrrrrrrrrrrrr}
    \toprule
          & \multicolumn{3}{c}{Median Fstat} &       & \multicolumn{3}{c}{Median $\vert\alpha-\hat{\alpha}\vert$} &       & \multicolumn{3}{c}{Median RMSE($\hat{\alpha}$)} &       & \multicolumn{3}{c}{Median $\vert\sigma_x-\hat{\sigma}_x\vert$} &       & \multicolumn{3}{c}{Median RMSE($\hat{\sigma}_x$)} &       &  \\
\cmidrule{2-4}\cmidrule{6-8}\cmidrule{10-12}\cmidrule{14-16}\cmidrule{18-20}    $F$   & \multicolumn{1}{l}{own} & \multicolumn{1}{l}{other} & \multicolumn{1}{l}{both} &       & \multicolumn{1}{l}{own} & \multicolumn{1}{l}{other} & \multicolumn{1}{l}{both} &       & \multicolumn{1}{l}{own} & \multicolumn{1}{l}{other} & \multicolumn{1}{l}{both} &       & \multicolumn{1}{l}{own} & \multicolumn{1}{l}{other} & \multicolumn{1}{l}{both} &       & \multicolumn{1}{l}{own} & \multicolumn{1}{l}{other} & \multicolumn{1}{l}{both} &       & \multicolumn{1}{l}{$s_0$} \\
    \midrule
    1     & 34.68 &       &       &       & 0.53  &       &       &       & 0.88  &       &       &       & 1.34  &       &       &       & 2.22  &       &       &       & 0.74 \\
    2     & 32.85 & 23.75 & 29.00 &       & 0.49  & 0.64  & 0.06  &       & 0.86  & 1.06  & 0.11  &       & 1.16  & 1.23  & 0.12  &       & 1.98  & 2.54  & 0.23  &       & 0.64 \\
    3     & 29.68 & 15.00 & 22.36 &       & 0.44  & 0.80  & 0.07  &       & 0.79  & 1.47  & 0.13  &       & 1.37  & 1.04  & 0.13  &       & 2.56  & 1.95  & 0.23  &       & 0.61 \\
    4     & 25.10 & 9.88  & 17.75 &       & 0.55  & 0.92  & 0.08  &       & 0.83  & 1.70  & 0.15  &       & 1.86  & 0.91  & 0.12  &       & 3.00  & 1.63  & 0.23  &       & 0.60 \\
    5     & 21.15 & 6.65  & 14.15 &       & 0.49  & 0.89  & 0.10  &       & 0.82  & 1.83  & 0.17  &       & 1.70  & 0.68  & 0.13  &       & 2.97  & 1.34  & 0.23  &       & 0.60 \\
    6     & 17.89 & 5.10  & 11.80 &       & 0.50  & 1.02  & 0.10  &       & 0.84  & 2.04  & 0.18  &       & 1.99  & 0.64  & 0.13  &       & 3.00  & 1.25  & 0.24  &       & 0.59 \\
    7     & 15.30 & 4.00  & 9.72  &       & 0.47  & 1.05  & 0.11  &       & 0.81  & 2.11  & 0.20  &       & 1.76  & 0.55  & 0.13  &       & 3.00  & 1.16  & 0.23  &       & 0.59 \\
    8     & 13.56 & 3.32  & 8.68  &       & 0.52  & 1.04  & 0.11  &       & 0.86  & 2.20  & 0.22  &       & 1.92  & 0.48  & 0.13  &       & 3.00  & 1.06  & 0.24  &       & 0.59 \\
    9     & 11.34 & 2.62  & 7.18  &       & 0.54  & 1.08  & 0.12  &       & 0.91  & 2.24  & 0.24  &       & 2.31  & 0.45  & 0.12  &       & 3.00  & 0.97  & 0.24  &       & 0.59 \\
    10    & 10.51 & 2.38  & 6.85  &       & 0.49  & 0.97  & 0.14  &       & 0.81  & 2.18  & 0.25  &       & 2.13  & 0.41  & 0.13  &       & 3.01  & 0.89  & 0.24  &       & 0.59 \\
    11    & 9.23  & 2.18  & 5.96  &       & 0.53  & 1.01  & 0.15  &       & 0.90  & 2.22  & 0.27  &       & 2.28  & 0.36  & 0.13  &       & 3.12  & 0.88  & 0.24  &       & 0.59 \\
    12    & 8.20  & 1.87  & 5.22  &       & 0.53  & 1.04  & 0.15  &       & 0.94  & 2.35  & 0.29  &       & 2.22  & 0.35  & 0.13  &       & 3.36  & 0.77  & 0.24  &       & 0.59 \\
    13    & 7.39  & 1.77  & 4.89  &       & 0.57  & 1.04  & 0.16  &       & 0.98  & 2.22  & 0.31  &       & 2.58  & 0.34  & 0.13  &       & 3.22  & 0.74  & 0.25  &       & 0.59 \\
    14    & 7.23  & 1.64  & 4.70  &       & 0.49  & 1.04  & 0.16  &       & 0.84  & 2.34  & 0.31  &       & 2.12  & 0.32  & 0.13  &       & 3.29  & 0.76  & 0.25  &       & 0.59 \\
    15    & 6.68  & 1.57  & 4.28  &       & 0.51  & 1.02  & 0.18  &       & 0.87  & 2.26  & 0.33  &       & 2.12  & 0.30  & 0.13  &       & 3.26  & 0.67  & 0.25  &       & 0.59 \\
    16    & 5.83  & 1.43  & 3.84  &       & 0.49  & 1.00  & 0.19  &       & 0.88  & 2.29  & 0.34  &       & 2.25  & 0.28  & 0.12  &       & 3.35  & 0.68  & 0.25  &       & 0.58 \\
    17    & 5.01  & 1.33  & 3.48  &       & 0.50  & 1.00  & 0.20  &       & 0.96  & 2.45  & 0.38  &       & 2.44  & 0.27  & 0.12  &       & 3.45  & 0.65  & 0.25  &       & 0.58 \\
    18    & 4.44  & 1.26  & 2.98  &       & 0.53  & 1.05  & 0.21  &       & 1.02  & 2.39  & 0.40  &       & 2.27  & 0.26  & 0.13  &       & 3.73  & 0.66  & 0.25  &       & 0.58 \\
    19    & 4.23  & 1.23  & 3.00  &       & 0.51  & 1.07  & 0.25  &       & 0.99  & 2.12  & 0.43  &       & 2.51  & 0.26  & 0.13  &       & 3.69  & 0.66  & 0.25  &       & 0.58 \\
    20    & 4.10  & 1.24  & 2.82  &       & 0.53  & 1.06  & 0.24  &       & 0.98  & 2.32  & 0.42  &       & 2.31  & 0.27  & 0.13  &       & 3.72  & 0.61  & 0.25  &       & 0.58 \\
    21    & 4.04  & 1.20  & 2.81  &       & 0.52  & 1.03  & 0.25  &       & 0.97  & 2.35  & 0.43  &       & 2.50  & 0.24  & 0.13  &       & 3.75  & 0.63  & 0.25  &       & 0.58 \\
    22    & 3.83  & 1.14  & 2.70  &       & 0.43  & 0.97  & 0.23  &       & 0.85  & 2.26  & 0.43  &       & 2.12  & 0.25  & 0.13  &       & 3.25  & 0.63  & 0.25  &       & 0.58 \\
    23    & 3.81  & 1.22  & 2.56  &       & 0.49  & 0.94  & 0.25  &       & 0.89  & 2.18  & 0.43  &       & 2.33  & 0.24  & 0.13  &       & 3.84  & 0.62  & 0.25  &       & 0.58 \\
    24    & 3.42  & 1.14  & 2.59  &       & 0.46  & 0.99  & 0.27  &       & 0.93  & 2.25  & 0.46  &       & 2.30  & 0.23  & 0.13  &       & 3.37  & 0.60  & 0.25  &       & 0.58 \\
    25    & 3.36  & 1.14  & 2.49  &       & 0.49  & 0.97  & 0.27  &       & 0.95  & 2.06  & 0.48  &       & 2.47  & 0.24  & 0.14  &       & 3.54  & 0.57  & 0.25  &       & 0.58 \\
    26    & 3.14  & 1.06  & 2.33  &       & 0.51  & 0.97  & 0.28  &       & 0.91  & 2.25  & 0.49  &       & 2.50  & 0.23  & 0.13  &       & 3.63  & 0.53  & 0.25  &       & 0.58 \\
    27    & 2.92  & 1.06  & 2.17  &       & 0.43  & 0.97  & 0.29  &       & 0.86  & 2.23  & 0.50  &       & 2.15  & 0.22  & 0.13  &       & 3.23  & 0.55  & 0.26  &       & 0.58 \\
    28    & 2.70  & 1.07  & 2.09  &       & 0.46  & 1.03  & 0.31  &       & 0.87  & 2.22  & 0.51  &       & 2.61  & 0.24  & 0.14  &       & 3.83  & 0.58  & 0.26  &       & 0.58 \\
    29    & 2.39  & 1.08  & 1.86  &       & 0.47  & 1.02  & 0.30  &       & 0.99  & 2.10  & 0.54  &       & 2.48  & 0.23  & 0.14  &       & 3.97  & 0.57  & 0.26  &       & 0.58 \\
    30    & 2.22  & 1.06  & 1.87  &       & 0.49  & 0.94  & 0.34  &       & 0.94  & 2.15  & 0.59  &       & 2.66  & 0.20  & 0.13  &       & 3.74  & 0.55  & 0.26  &       & 0.58 \\
    31    & 2.08  & 1.01  & 1.67  &       & 0.50  & 1.07  & 0.37  &       & 0.95  & 2.41  & 0.60  &       & 2.52  & 0.21  & 0.13  &       & 4.18  & 0.56  & 0.26  &       & 0.58 \\
    32    & 1.73  & 0.90  & 1.59  &       & 0.52  & 1.01  & 0.37  &       & 1.05  & 2.37  & 0.64  &       & 3.00  & 0.22  & 0.15  &       & 4.43  & 0.56  & 0.27  &       & 0.58 \\
    33    & 1.38  & 0.97  & 1.38  &       & 0.54  & 0.99  & 0.43  &       & 1.14  & 2.25  & 0.70  &       & 3.00  & 0.22  & 0.13  &       & 4.17  & 0.54  & 0.27  &       & 0.58 \\
    34    & 1.10  & 0.90  & 1.22  &       & 0.61  & 1.01  & 0.45  &       & 1.22  & 2.26  & 0.78  &       & 3.00  & 0.22  & 0.13  &       & 4.75  & 0.52  & 0.27  &       & 0.58 \\
    35    & 0.90  & 0.93  & 1.13  &       & 0.70  & 1.04  & 0.55  &       & 1.37  & 2.31  & 0.84  &       & 3.00  & 0.23  & 0.14  &       & 5.29  & 0.53  & 0.28  &       & 0.58 \\
    36    &       & 0.93  &       &       &       & 1.06  &       &       &       & 2.32  &       &       &       & 0.21  &       &       &       & 0.52  &       &       & 0.58 \\
    \bottomrule
    \end{tabular}%
  \label{tab: own vs other full rc}%
  \tablenotes The table reports the median values of the absolute errors and RMSEs of the estimated utility parameters, $\hat{\alpha}$ and $\hat{\sigma}_x$, as well as the median F-statistics when $\mathbf{z}^{own}$, $\mathbf{z}^{other}$, and ($\mathbf{z}^{own}, \mathbf{z}^{other}$) are used as instruments individually. A random coefficient is included in the indirect utility function \eqref{eqn: dgp utility} in the DGP. The median outside option share across 500 simulated datasets for each Monte Carlo configuration is denoted by $s_0$.
\end{table}%
\end{landscape}
\clearpage

\newpage
\section{Appendix for empirical applications}\label{app: data appendix}
In this section, we describe data used in Section \ref{sec: application} in detail and present some descriptive statistics of the South Korean automobile and instant noodles industries.

\subsection{Car data}\label{app: car data}

\subsubsection*{Data description}

 Our raw data contain year/province-level information (12 years from 2012 to 2023 and 17 provinces) on the total number of new registrations, sales revenue, and various attributes -- nameplate, model year, engine displacement, fuel type, fuel efficiency, size, and other miscellaneous physical characteristics that vary across trim levels or consumer-selected options -- of passenger vehicles produced by 13 brands under nine parent companies. These brands, including six domestic ones, accounted for approximately 93\% of all private passenger cars registered in South Korea during the sample period. We define a market as a unique combination of year and province, and a product as a unique combination of nameplate and fuel type; for example, Toyota Camry Hybrid. Our sample includes 540 nameplates and five fuel types -- gasoline, diesel, LPG, hybrid, and electric -- resulting in 776 unique products.

 Product attributes used in the empirical application include fuel economy (km per 1,000 Won), acceleration (horsepower/curb weight), and size (width $\times$ length $\times$ height). For each product sold in a given market, we calculate fuel economy by dividing fuel efficiency (measured in kilometers per liter or kilometers per kilowatt-hour (kWh) for electric vehicles (EVs)) by the per-liter fuel price (in 1,000 Won) of the corresponding fuel type (gasoline, diesel, or LPG) in that market, obtained from the Oil Price Information Network (Opinet).\footnote{Opinet's web address is \url{https://www.opinet.co.kr/user/main/mainView.do}.} We assume that hybrid electric vehicles (HEVs) are fueled by gasoline, as diesel HEVs are rare.\footnote{Unlike HEVs, plug-in hybrid electric vehicles (PHEVs) can be recharged from an external power source (e.g., an EV charging station). Given that PHEVs have a very low sales share (0.24\%) in our sample, we exclude them from our empirical application to simplify fuel economy calculations for EVs. Similarly, we also exclude fuel-cell electric vehicles (FCEVs), which are powered by hydrogen and have a sales share of approximately 0.24\%.} Additionally, we collect yearly per-kWh EV charging prices at quick-charging stations from the Ministry of Environment in South Korea to calculate fuel economy for electric vehicles.\footnote{Prices since 2016 are available at \url{https://mcee.go.kr/home/web/board/read.do?boardId=1539980&boardMasterId=939&menuId=10598}. We assume that prices prior to 2016 are identical to those in 2016.}

 We aggregate sales figures and attribute values at the market/product level, using the number of units registered in the market as weights. Consequently, the same product may exhibit different fuel economy, acceleration, and size across markets. Additionally, we calculate a product's price in a market by dividing total sales revenue by the number of registrations in that market. This measure differs from the list price or the Manufacturer's Suggested Retail Price (MSRP) commonly used in the literature. Specifically, sales in our data reflect acquisition prices, which account for manufacturer promotions at the point of sale as well as the costs of miscellaneous vehicle options selected by consumers. As a result, our price measure more accurately reflects market conditions than the MSRP. Our final data consist of 41,716 product-market-level observations in total.

\subsubsection*{Descriptive statistics}

 Table \ref{tab: auto mkt structure} presents the sales share and average product price for each of the 13 brands and nine parent companies. Hyundai and Kia are the two leading brands in our sample, each accounting for approximately 32.5\% of new passenger car sales during this period. Genesis, initially part of Hyundai, was established as an independent luxury brand in late 2015, specializing in high-end vehicles. The average price of a Genesis vehicle is 62 million Won, more than twice the average price of Hyundai (30 million Won) and Kia (28 million Won). Collectively, these three brands, owned by the Hyundai Motor Group, hold a 68.4\% sales share, solidifying the group's position as the dominant market leader.

 There are seven foreign brands in the sample, four of which are German, accounting for the majority (82.5\%) of foreign brand sales. While Mercedes-Benz vehicles are the most expensive, with an average price of 74 million Won, Volkswagen offers the cheapest models, with an average price of 37.5 million Won. Additionally, there are two Japanese brands, Toyota and Lexus, both owned by the Toyota Group. Lexus targets the luxury segment alongside Mercedes-Benz, BMW, and Audi, whereas Toyota focuses on affordable vehicles. Tesla, the only U.S. brand in our sample, produces only EVs.

\begin{figure}[htbp]
	\centering
	\caption{Sales share composition of new private passenger cars}
	\begin{subfigure}[b]{0.70\textwidth}	
    	\caption{By brand}
    	\includegraphics[width=\textwidth]{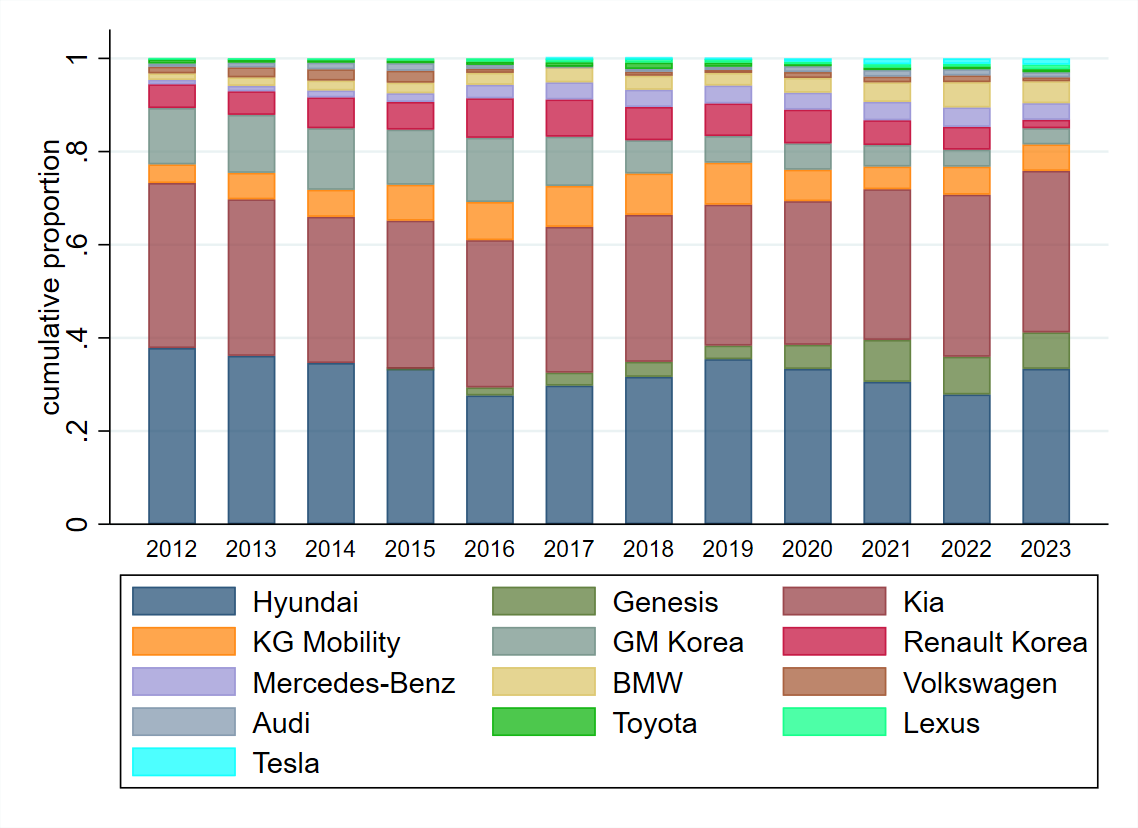}
    	\end{subfigure}
    \qquad
	\begin{subfigure}[b]{0.70\textwidth}	
    	\caption{By parent company}
    	\includegraphics[width=\textwidth]{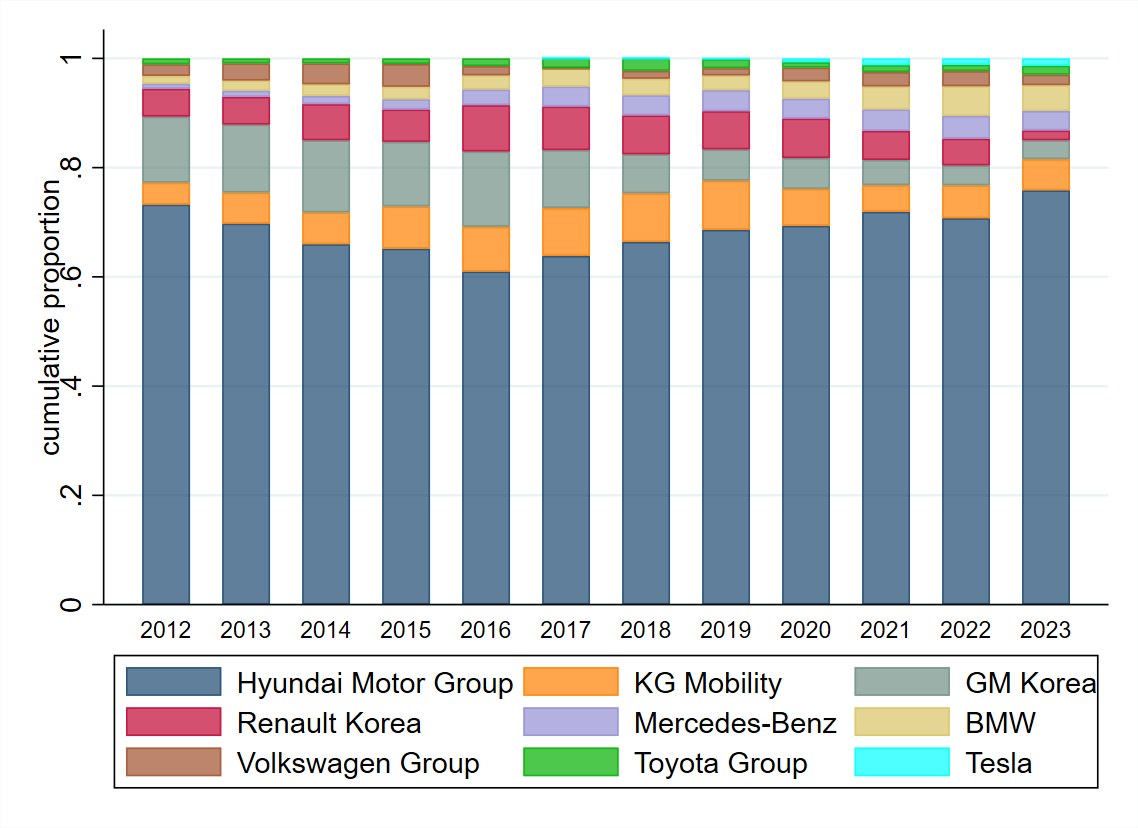}
    	\end{subfigure}
\label{fig: auto mkt structure}
\tablenotes The upper and bottom panels of the figure depict the trend in sales share composition by brand and parent company, respectively, from 2012 to 2023.
\end{figure}

 Figure \ref{fig: auto mkt structure} illustrates the yearly changes in sales share composition at the brand level (upper panel) and parent company level (bottom panel). The sales shares of German brands gradually increased in the early 2010s following the implementation of the Free Trade Agreement between South Korea and the European Union in 2011, which progressively reduced import tariffs on German automobiles. However, this upward trend slowed in the mid-2010s, particularly after the Dieselgate scandal in 2015. As shown in the bottom panel, the sales share of Hyundai Motor Group declined gradually in the early 2010s but rebounded after Dieselgate and the launch of its luxury brand, Genesis. Meanwhile, the sales shares of the other three domestic companies steadily contracted over the same period.

\subsection{Instant noodles data}

\subsubsection*{Data description}

\begin{figure}[htbp]
	\centering
	\caption{Six regions of South Korea}
    	\includegraphics[width=0.40\textwidth]{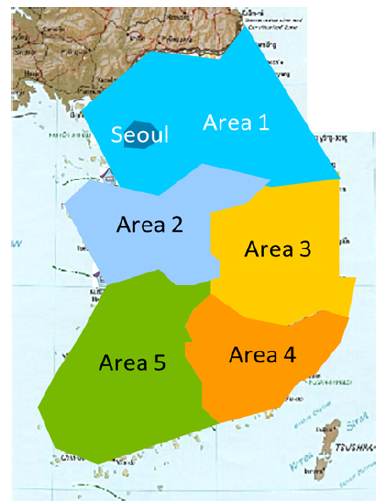}
\label{fig: nielsen market}
\tablenotes The figure shows the six geographical regions of South Korea classified by NielsenIQ.
\end{figure}

 The data, acquired from NielsenIQ, contain monthly sales volume and prices for instant noodle products offered by four major firms -- Nongshim, Ottogi, Samyang, and Paldo -- across six regions of South Korea from January 2010 to December 2019. Each region comprises multiple adjacent provinces, with one exception: the capital, Seoul, which constitutes a region consisting of a single province. The remaining five regions geographically partition South Korea into (i) North, (ii) Mid-West, (iii) Mid-East, (iv) South-East, and (v) South-West, which correspond to Areas 1--5 in Figure \ref{fig: nielsen market}.\footnote{The data cover 16 out of 17 provinces of South Korea; Jeju, an island located in the south, is not included.} Instant noodles are available in two package types: (i) pouch and (ii) cup. For example, Nongshim's Shin Ramyun, the most popular brand in South Korea, is available in both package types. We define a product as a unique combination of brand and package type, and a market as a unique combination of region and year-month pair.

 We analyze the 70 best-selling products, which account for approximately 90\% of total sales in the instant noodles market during the sample period. Cold noodle products are excluded from the analysis because their demand exhibits seasonality distinct from that of typical instant noodle products, which are consumed warm or hot. We obtain attributes of the 70 products from two databases: (i) the Ministry of Food and Drug Safety and (ii) FatSecret.\footnote{We primarily gather data from the Ministry of Food and Drug Safety database: \url{https://various.foodsafetykorea.go.kr/nutrient/}. When attributes are unavailable for some products in this database, we utilize FatSecret as an alternative: \url{https://platform.fatsecret.com/platform-api}. We also verify that the attributes of each product available on both websites are consistent with each other.} Specifically, we collect data on serving size, calorie content, and the quantities of key nutrients, namely sugar, fat, protein, sodium, and cholesterol, for each product. In our main analysis, nutrient quantities are divided by serving size. Our final dataset consists of a total of 44,670 product/market-level observations.

\subsubsection*{Descriptive statistics}

 Table \ref{tab: noodle stat} presents the number of products, sales shares, and average product prices by firm (top panel), package type (middle panel), and soup type (bottom panel). Of the 70 products, 29 are owned by Nongshim, which holds a sales share of nearly two-thirds (63.7\%) during the sample period. The other three firms are distant followers, with sales shares of 19\% for Ottogi, 12.3\% for Samyang, and 5.1\% for Paldo. Additionally, Ottogi's products are the cheapest (710 KRW), while Paldo's are the most expensive (943 KRW), on average.

 There are 44 products packaged in pouches and 26 in cups in our sample. The pouch type is the more common choice among consumers, accounting for 71.2\% of total sales volume. On average, cup-type noodles are approximately 25\% (or 200 KRW in absolute terms) more expensive than pouch-type noodles. Red-colored soup is the most popular type of instant noodle in South Korea, with 47 products in the sample categorized under this type. These products make up around four-fifths of the total sales volume, followed by soupless noodles (14\%) and noodles served with white-colored soup (6\%). On average, products in the latter two categories are approximately 200 KRW more expensive than those with red-colored soup.

\begin{figure}[htbp]
	\centering
	\caption{Sales share composition of instant noodles}
	\begin{subfigure}[b]{0.50\textwidth}
	\caption{By firm}
    	\includegraphics[width=\textwidth]{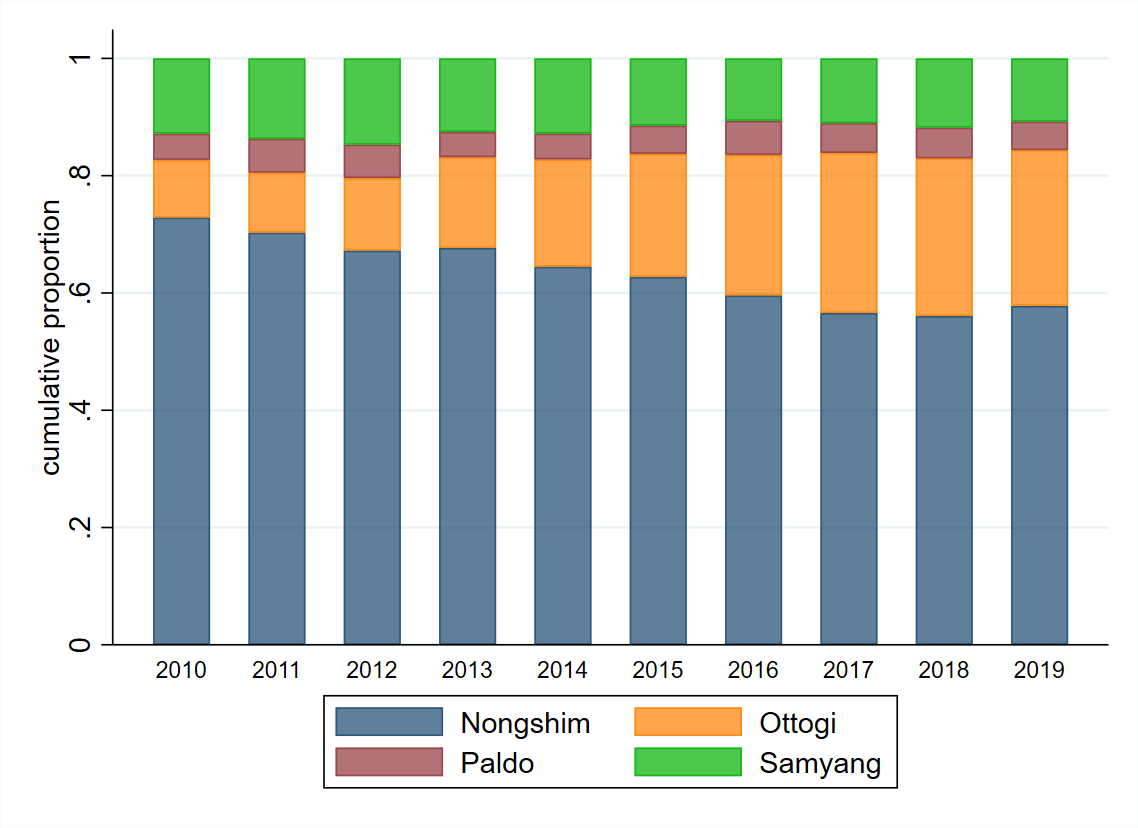}
    	\end{subfigure}
     	\begin{subfigure}[b]{0.50\textwidth}
	\caption{By package type}
    	\includegraphics[width=\textwidth]{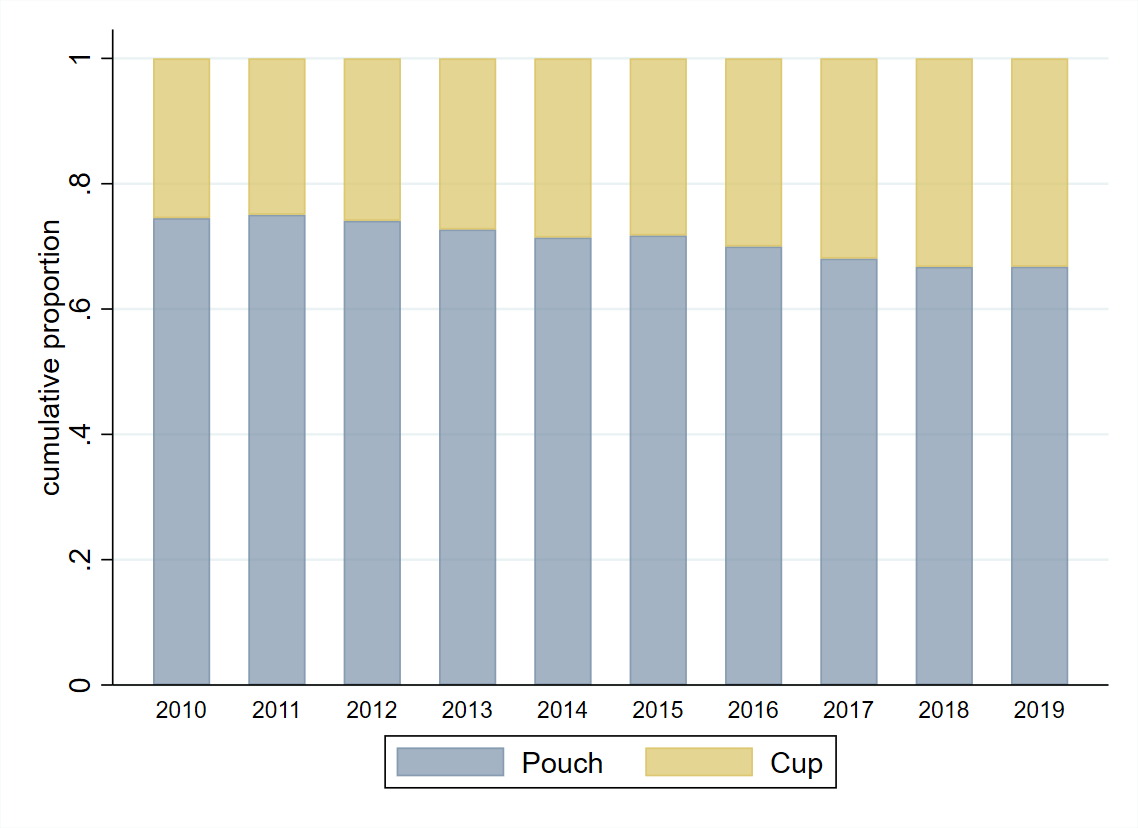}
    	\end{subfigure}
    	\begin{subfigure}[b]{0.50\textwidth}
	\caption{By soup type}
    	\includegraphics[width=\textwidth]{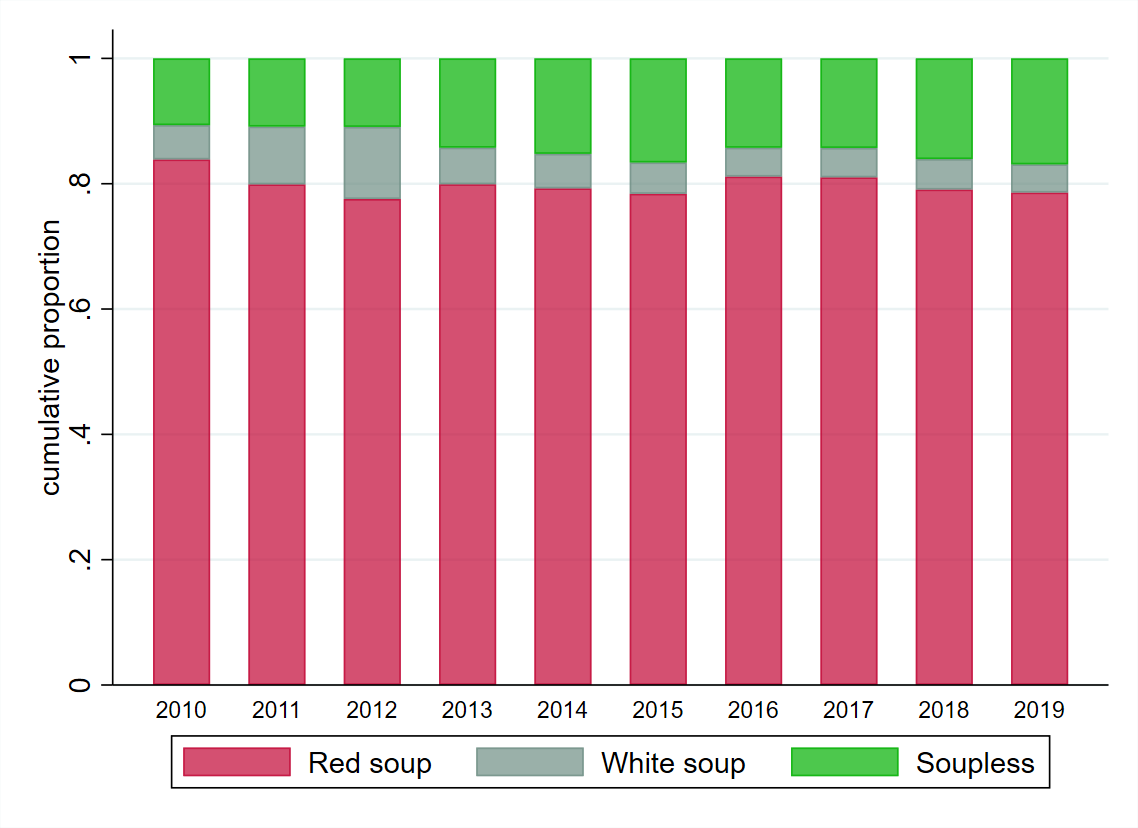}
    	\end{subfigure}
\label{fig: noodle mkt structure}
\tablenotes The figure depicts the trend in sales share composition by firm (top panel), package type (middle panel), and soup type (bottom panel) from 2010 to 2019.
\end{figure}

 Figure \ref{fig: noodle mkt structure} depicts the trend in sales share composition by firm (top panel), package type (middle panel), and soup type (bottom panel) during the sample period. Nongshim, while maintaining its position as the dominant firm, gradually lost sales throughout the 2010s, mostly to Ottogi, whose sales share steadily expanded during the decade. Over time, cup-type noodles gained popularity, increasing their combined sales share by eight percentage points during the sample period. While red-colored soup has consistently been the most preferred choice among consumers, white-colored soup experienced a temporary surge in popularity in the early 2010s. Additionally, the sales share of soupless products rose by roughly one half, from about 10 percent in the early 2010s to about 15 percent from 2013 onward.

\newpage
\section{Additional tables and figures}\label{app: additional table and figure}

\begin{table}[htbp]
\footnotesize
  \centering
  \caption{Sales shares and prices: the new passenger car market (2012--2023)}
    \begin{tabular}{llrrrrr}
    \toprule
          &       & \multicolumn{2}{c}{Sales share} &       & \multicolumn{2}{c}{Average price} \\
          &       &       &       &       & \multicolumn{2}{c}{(100 million Won)} \\
\cmidrule{3-4}\cmidrule{6-7}    Parent Company & Brand & \multicolumn{1}{l}{by Brand} & \multicolumn{1}{l}{by Company} &       & \multicolumn{1}{l}{by Brand} & \multicolumn{1}{l}{by Company} \\
    \midrule
    \textit{Domestic} &       &       &       &       &       &  \\
    \hspace{0.1in}Hyundai Motor Group & Hyundai & 0.325 & 0.684 &       & 0.299 & 0.305 \\
          & Genesis & 0.033 &       &       & 0.622 &  \\
          & Kia   & 0.325 &       &       & 0.279 &  \\
    \hspace{0.1in}KG Mobility & KG Mobility & 0.069 & 0.069 &       & 0.280 & 0.280 \\
    \hspace{0.1in}GM Korea & GM Korea & 0.088 & 0.088 &       & 0.208 & 0.208 \\
    \hspace{0.1in}Renault Korea & Renault Korea & 0.062 & 0.062 &       & 0.265 & 0.265 \\
          &       &       &       &       &       &  \\
    \textit{Foreign} &       &       &       &       &       &  \\
    \hspace{0.1in}Mercedes-Benz & Mercedes-Benz & 0.028 & 0.028 &       & 0.740 & 0.740 \\
    \hspace{0.1in}BMW & BMW   & 0.029 & 0.029 &       & 0.632 & 0.632 \\
    \hspace{0.1in}Volkswagen Group & Volkswagen & 0.013 & 0.023 &       & 0.375 & 0.470 \\
          & Audi  & 0.010 &       &       & 0.594 &  \\
    \hspace{0.1in}Toyota Group & Toyota & 0.007 & 0.013 &       & 0.387 & 0.486 \\
          & Lexus & 0.006 &       &       & 0.606 &  \\
    \hspace{0.1in}Tesla & Tesla & 0.004 & 0.004 &       & 0.681 & 0.681 \\
    \bottomrule
    \end{tabular}%
  \label{tab: auto mkt structure}%
 \tablenotes The table presents the sales share and average product price for each of the 13 brands and nine parent companies. The average price per product is obtained by dividing the total sales revenue by registrations by each firm and parent company during the sample period (2012--2023). Prices are adjusted for inflation using the 2020 Consumer Price Index (CPI) as the base year.
\end{table}%
\clearpage

\newpage
\begin{table}[htbp]
  \centering
  \caption{Product counts, sales shares, and prices: the instant noodles market (2010--2019)}
    \begin{tabular}{lrrr}
    \toprule
          & \multicolumn{1}{c}{\# products} & \multicolumn{1}{c}{Sales share} & \multicolumn{1}{c}{Average price} \\
          &       &       & \multicolumn{1}{c}{(1,000 KRW)} \\
    \midrule
    \multicolumn{4}{c}{\underline{\textit{Panel A: by firm}}} \\
          &       &       &  \\
    Nongshim & 29    & 0.637 & 0.806 \\
    Ottogi & 19    & 0.190 & 0.710 \\
    Paldo & 10    & 0.051 & 0.943 \\
    Samyang & 12    & 0.123 & 0.787 \\
    \midrule
    \multicolumn{4}{c}{\underline{\textit{Panel B: by package type}}} \\
          &       &       &  \\
    Pouch & 44    & 0.712 & 0.736 \\
    Cup   & 26    & 0.288 & 0.930 \\
    \midrule
    \multicolumn{4}{c}{\underline{\textit{Panel C: by soup type}}} \\
          &       &       &  \\
    Red soup & 47    & 0.799 & 0.754 \\
    White soup & 11    & 0.062 & 0.954 \\
    Soupless & 12    & 0.140 & 0.937 \\
    \bottomrule
    \end{tabular}%
  \label{tab: noodle stat}%
  \tablenotes The table presents the number of products, sales shares, and average product prices by firm (top panel), package type (middle panel), and soup type (bottom panel).
\end{table}%
\clearpage


\newpage
\begin{figure}[htbp]
	\centering
	\caption{F-stat evidence}
	\begin{subfigure}[b]{0.47\textwidth}	
    	\caption{$F=3$ without a random coefficient}
    	\includegraphics[width=\textwidth]{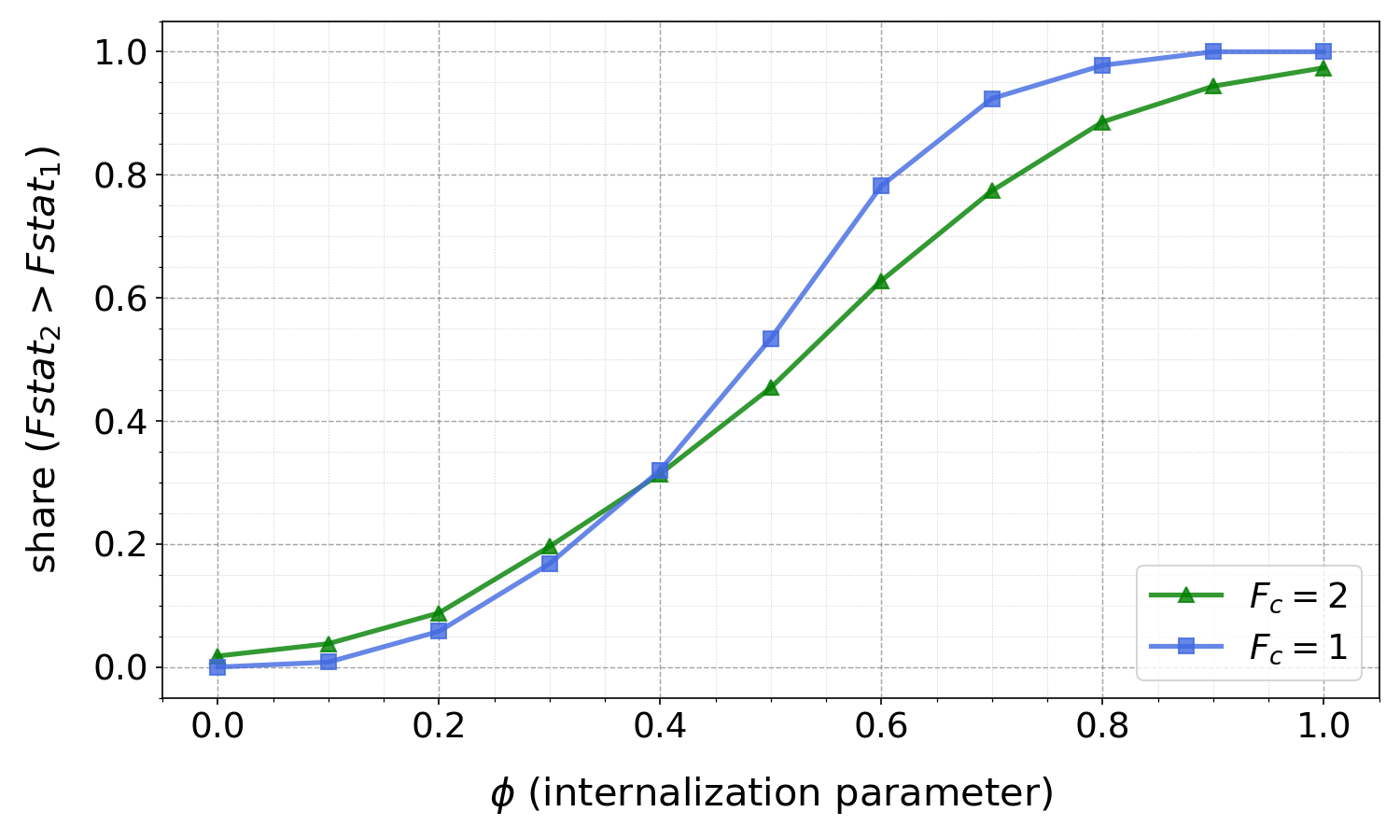}
    	\end{subfigure}
    \qquad
	\begin{subfigure}[b]{0.47\textwidth}	
    	\caption{$F=3$ with a random coefficient}
    	\includegraphics[width=\textwidth]{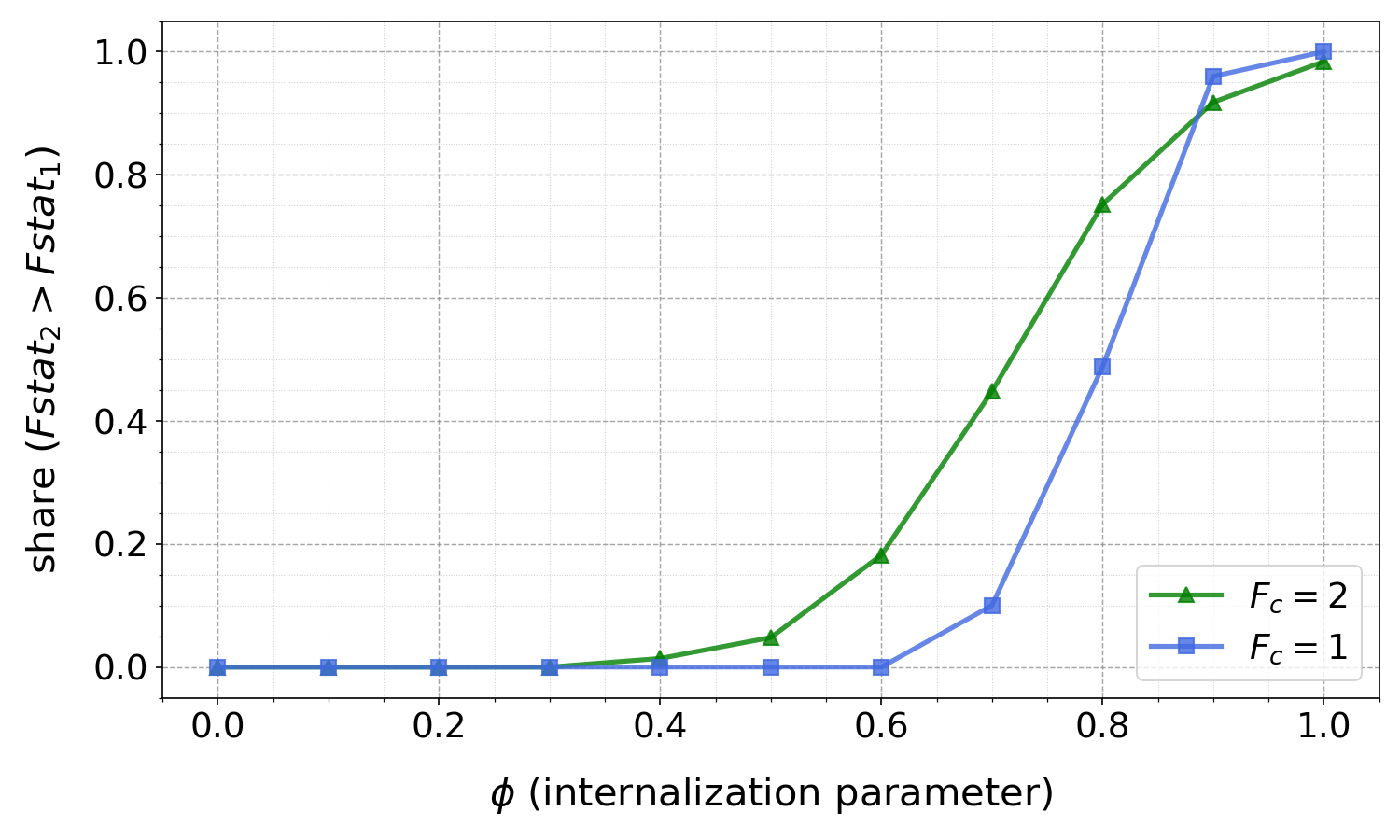}
    	\end{subfigure}
    	\begin{subfigure}[b]{0.47\textwidth}
    \vspace{0.2in}
    	\caption{$F=4$ without a random coefficient}
    	\includegraphics[width=\textwidth]{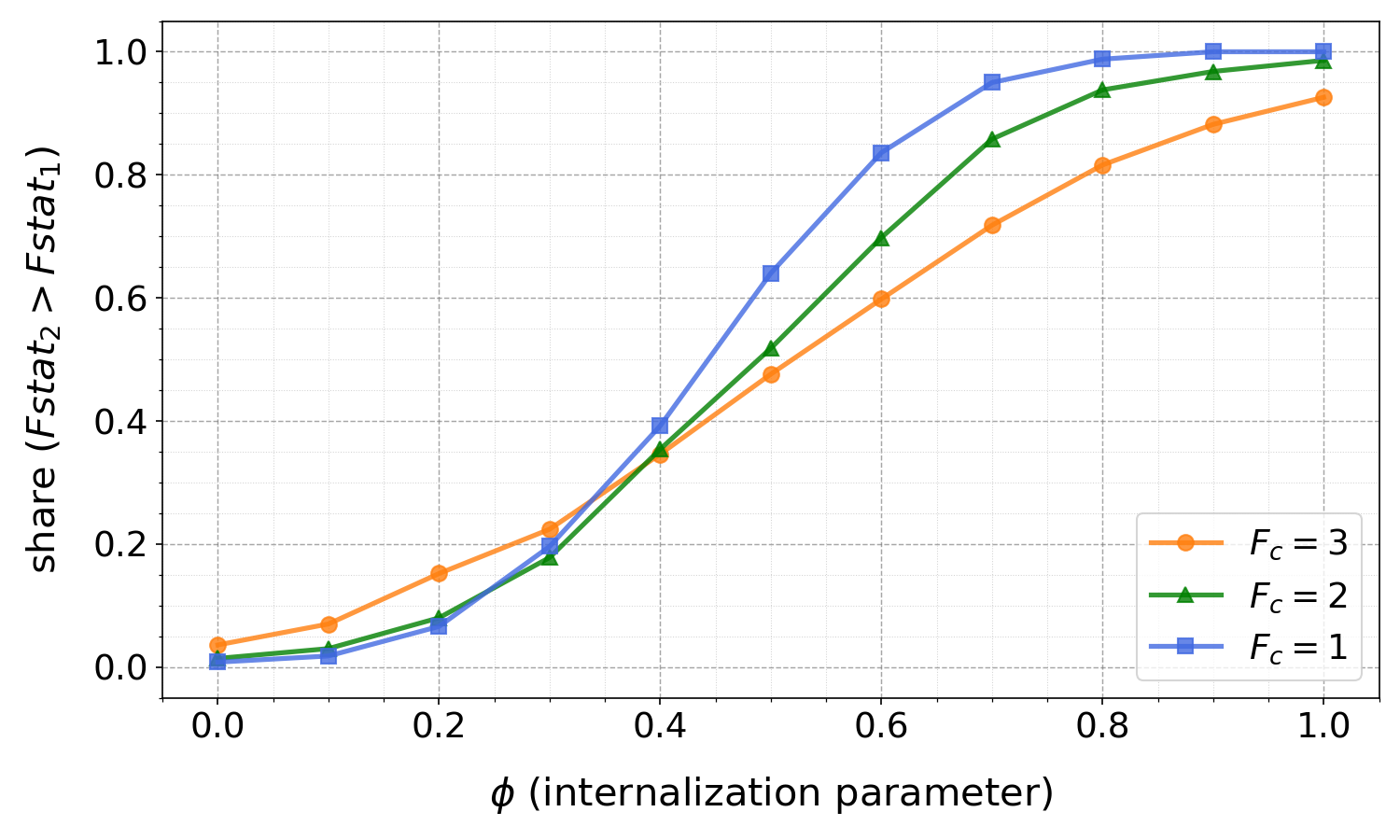}
    	\end{subfigure}
        \qquad
    	\begin{subfigure}[b]{0.47\textwidth}
    	\caption{$F=4$ with a random coefficient}
    	\includegraphics[width=\textwidth]{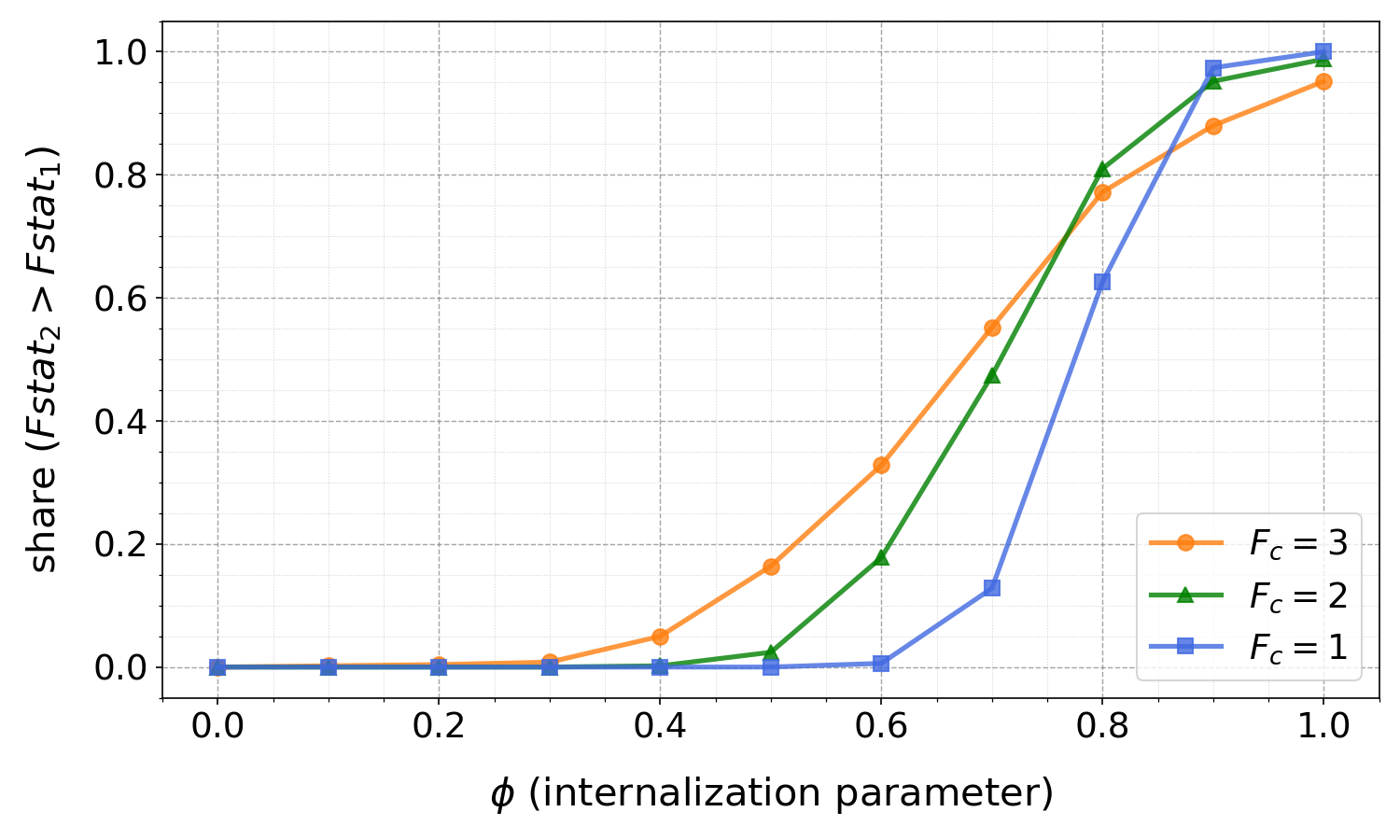}
    	\end{subfigure}
    	\begin{subfigure}[b]{0.47\textwidth}
    \vspace{0.2in}
    	\caption{$F=6$ without a random coefficient}
    	\includegraphics[width=\textwidth]{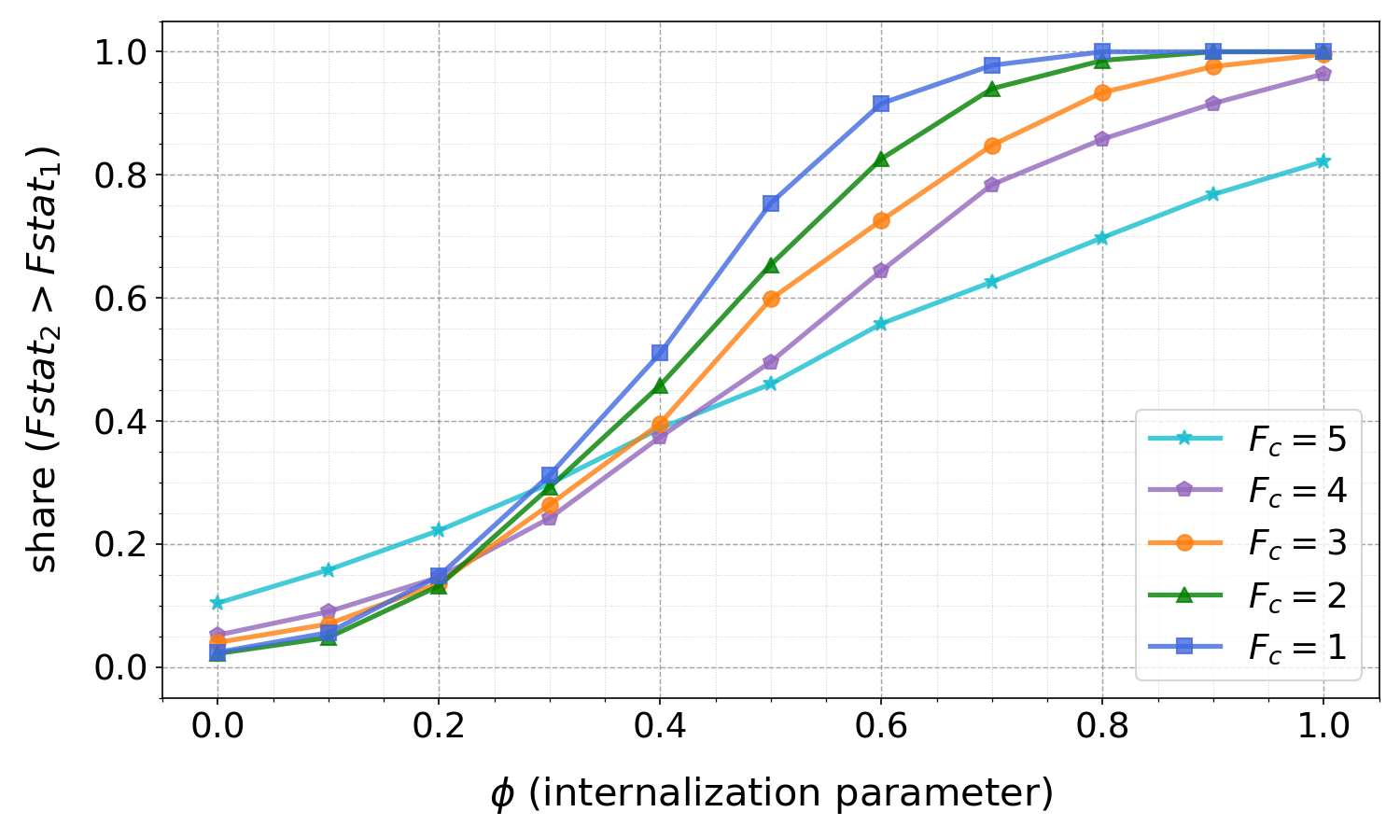}
    	\end{subfigure}
        \qquad
    	\begin{subfigure}[b]{0.47\textwidth}
    	\caption{$F=6$ with a random coefficient}
    	\includegraphics[width=\textwidth]{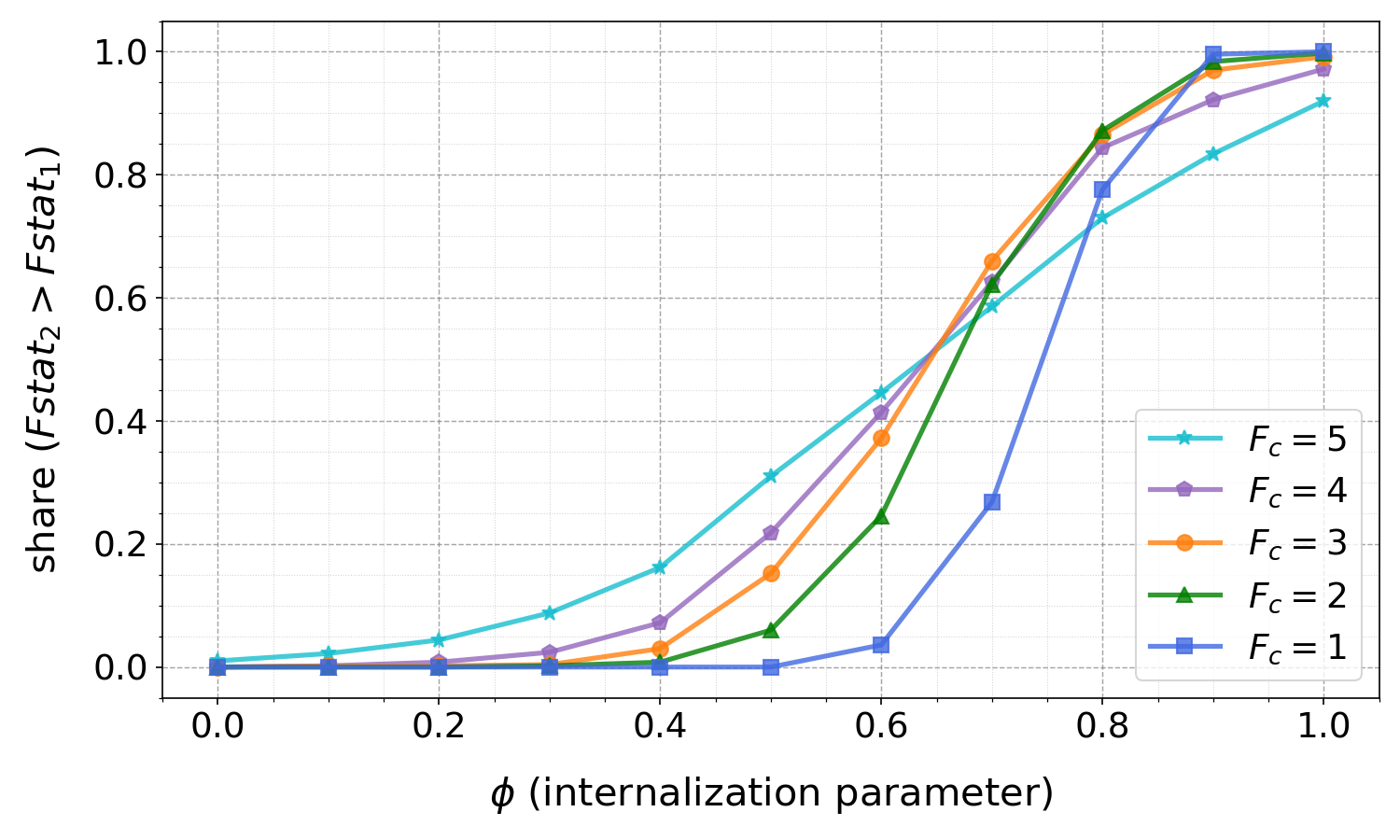}
    	\end{subfigure}
\label{fig: Fstat evidence}
\tablenotes The figure shows the share of cases where Fstat$_2$ $>$ Fstat$_1$ in 500 simulation results for each Monte Carlo configuration ($J=36, F, T=100, \phi, F_c$). The left panel illustrates the results for the utility specification without a random coefficient, while the right panel presents the results with a random coefficient.
\end{figure}
\clearpage

\newpage
\begin{figure}[htbp]
	\centering
	\caption{P-values of $T_{IV}^{RV}$ under the alternative hypothesis $H_2: Q_{comp} > Q_{coll}$}
	\begin{subfigure}[b]{0.47\textwidth}	
    	\caption{$F=3$ without a random coefficient}
    	\includegraphics[width=\textwidth]{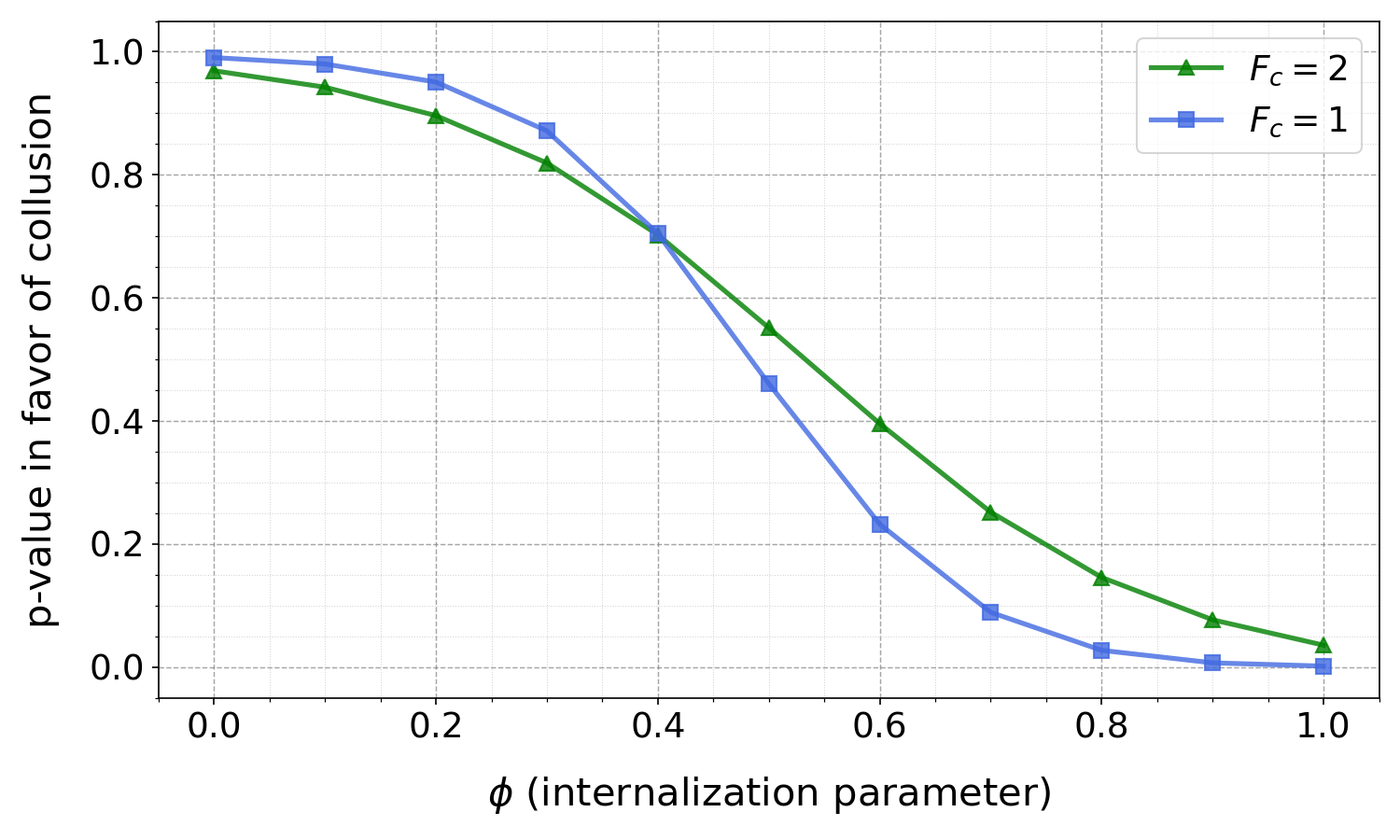}
    	\end{subfigure}
    \qquad
	\begin{subfigure}[b]{0.47\textwidth}	
    	\caption{$F=3$ with a random coefficient}
    	\includegraphics[width=\textwidth]{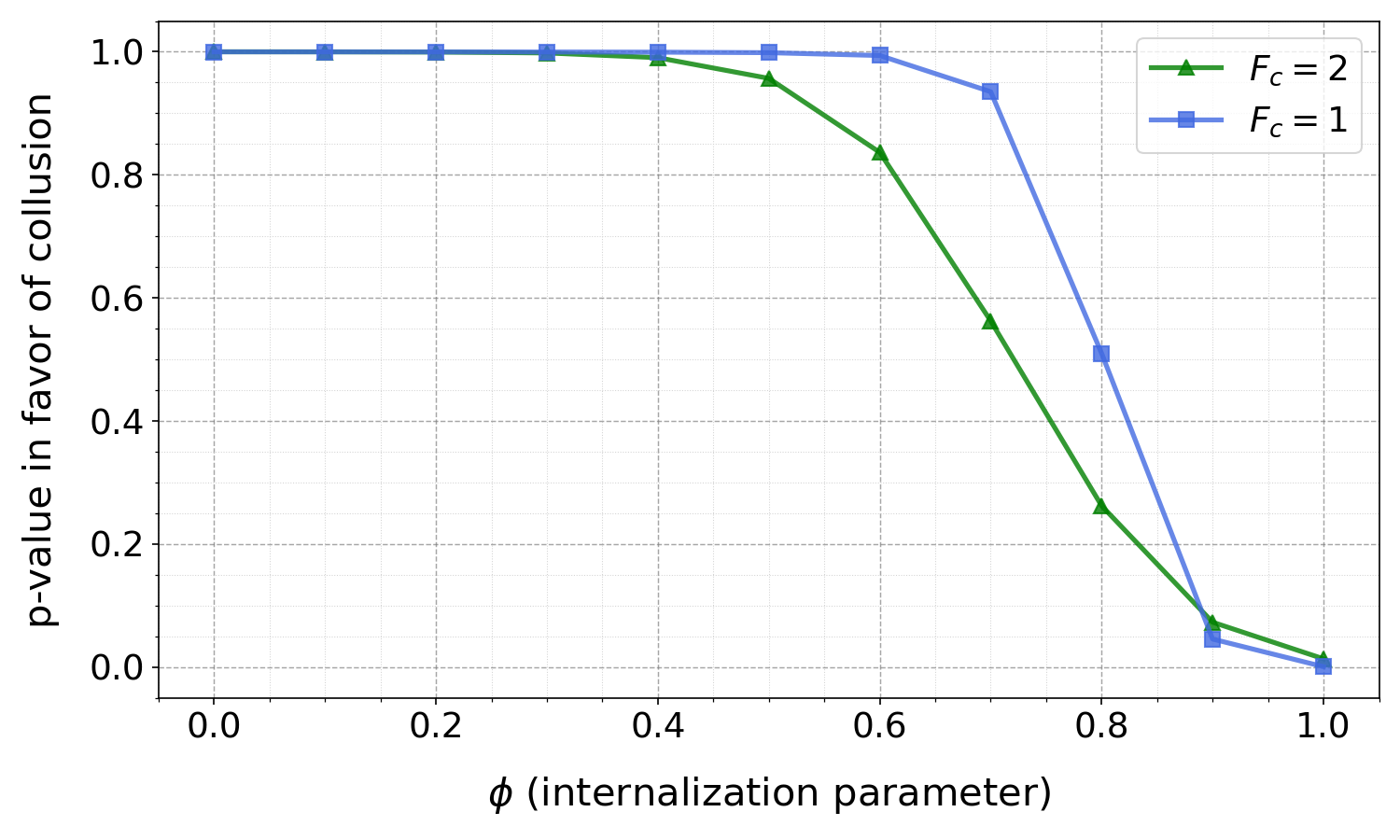}
    	\end{subfigure}
    	\begin{subfigure}[b]{0.47\textwidth}
        \vspace{.2in}
    	\caption{$F=4$ without a random coefficient}
    	\includegraphics[width=\textwidth]{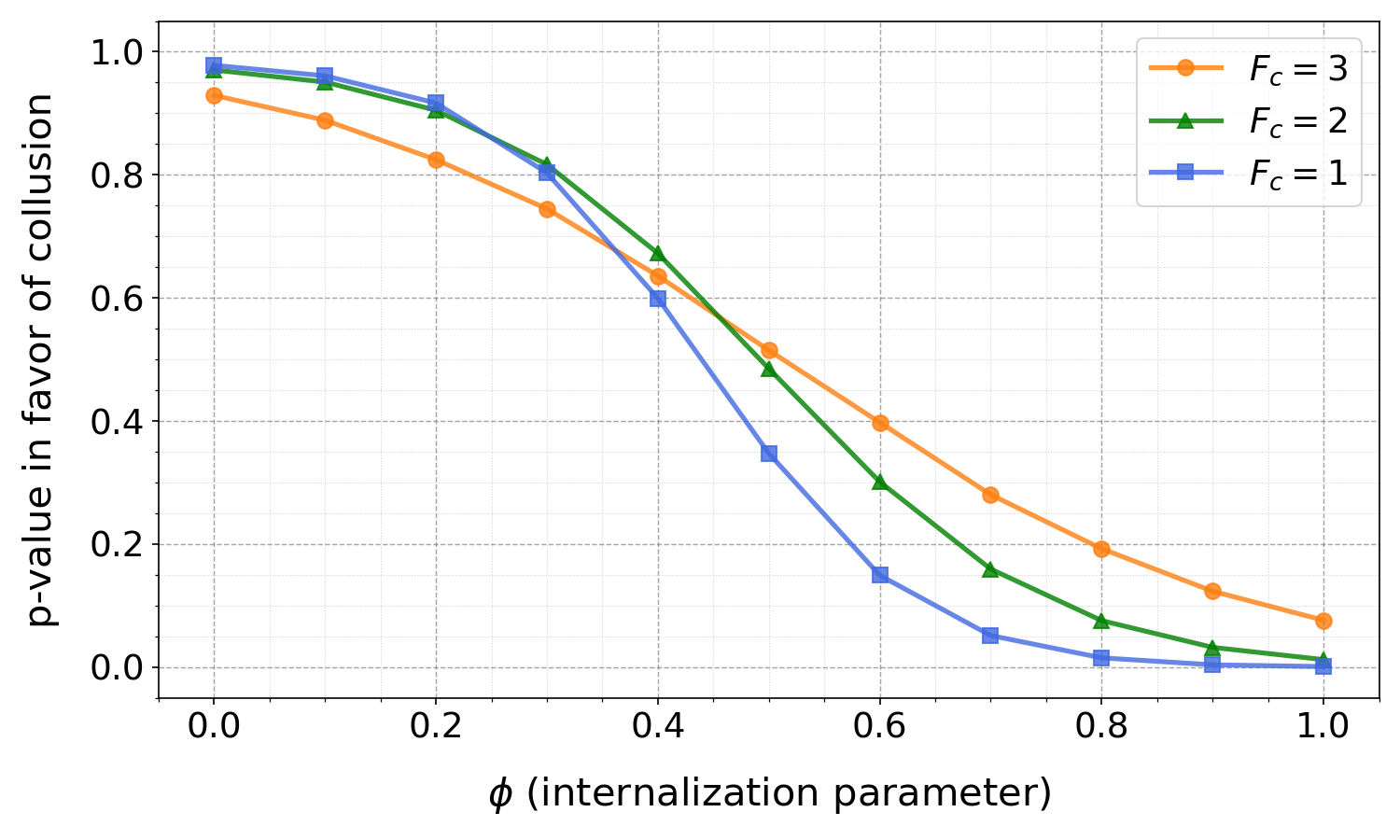}
    	\end{subfigure}
        \qquad
    	\begin{subfigure}[b]{0.47\textwidth}
    	\caption{$F=4$ with a random coefficient}
    	\includegraphics[width=\textwidth]{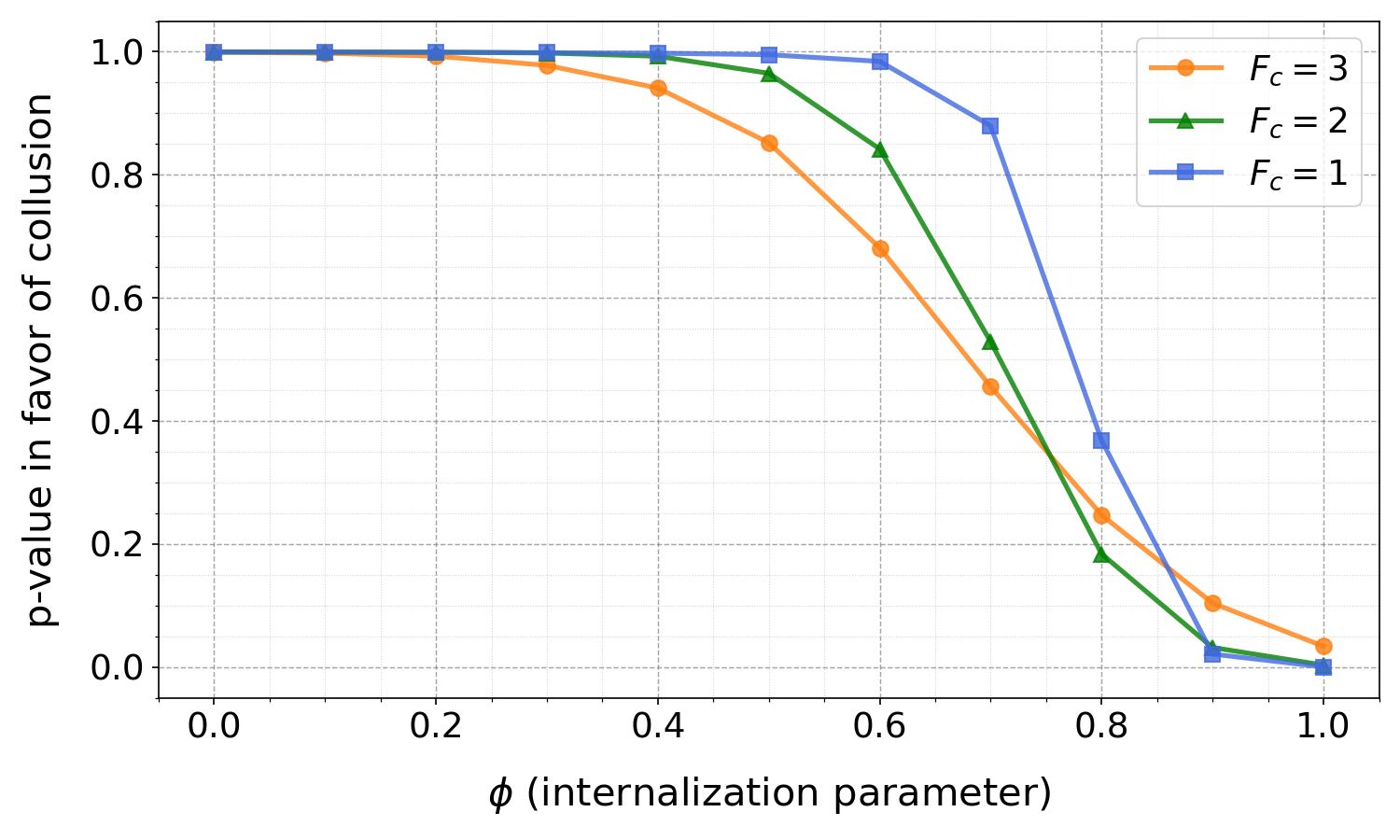}
    	\end{subfigure}
    	\begin{subfigure}[b]{0.47\textwidth}
            \vspace{.2in}
    	\caption{$F=6$ without a random coefficient}
    	\includegraphics[width=\textwidth]{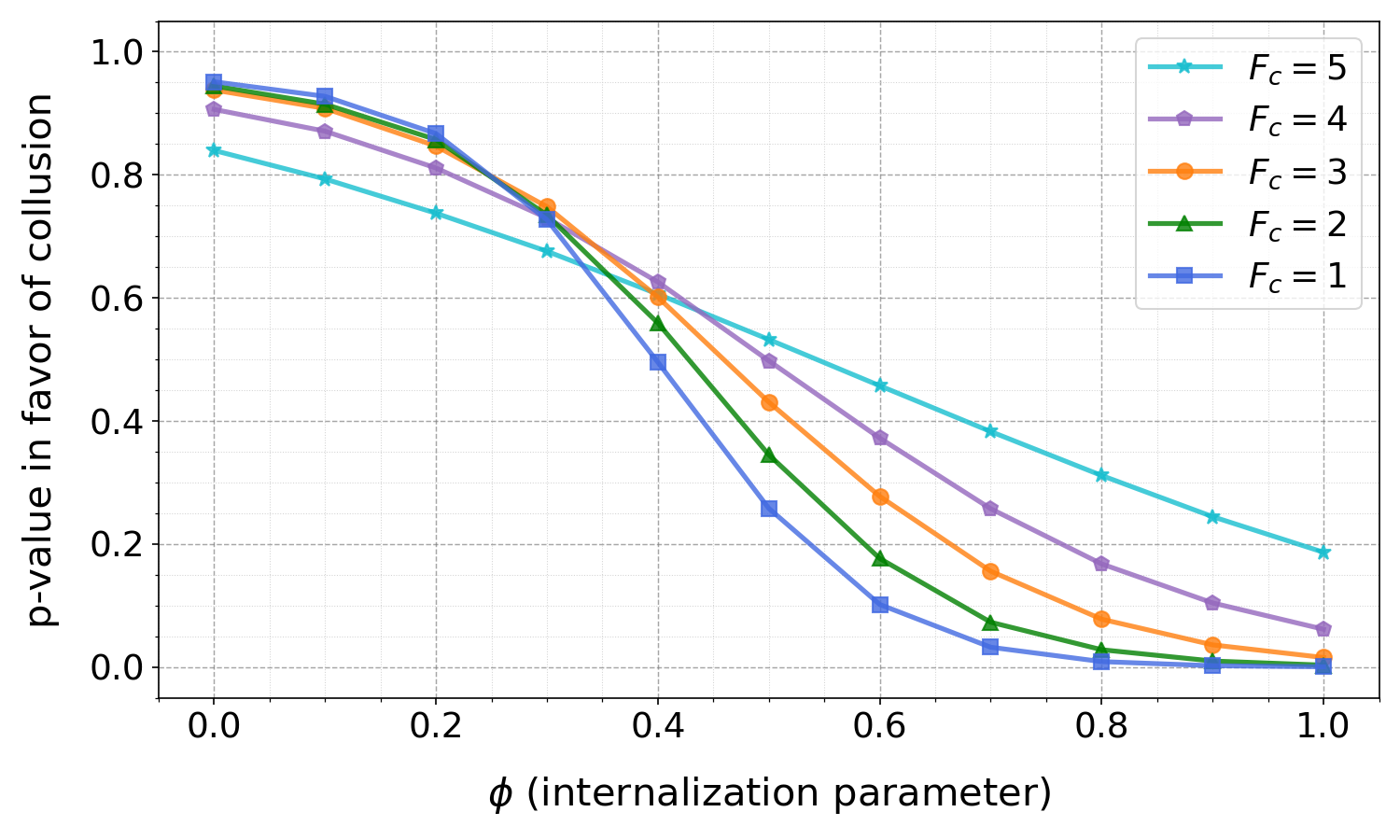}
    	\end{subfigure}
        \qquad
    	\begin{subfigure}[b]{0.47\textwidth}
    	\caption{$F=6$ with a random coefficient}
    	\includegraphics[width=\textwidth]{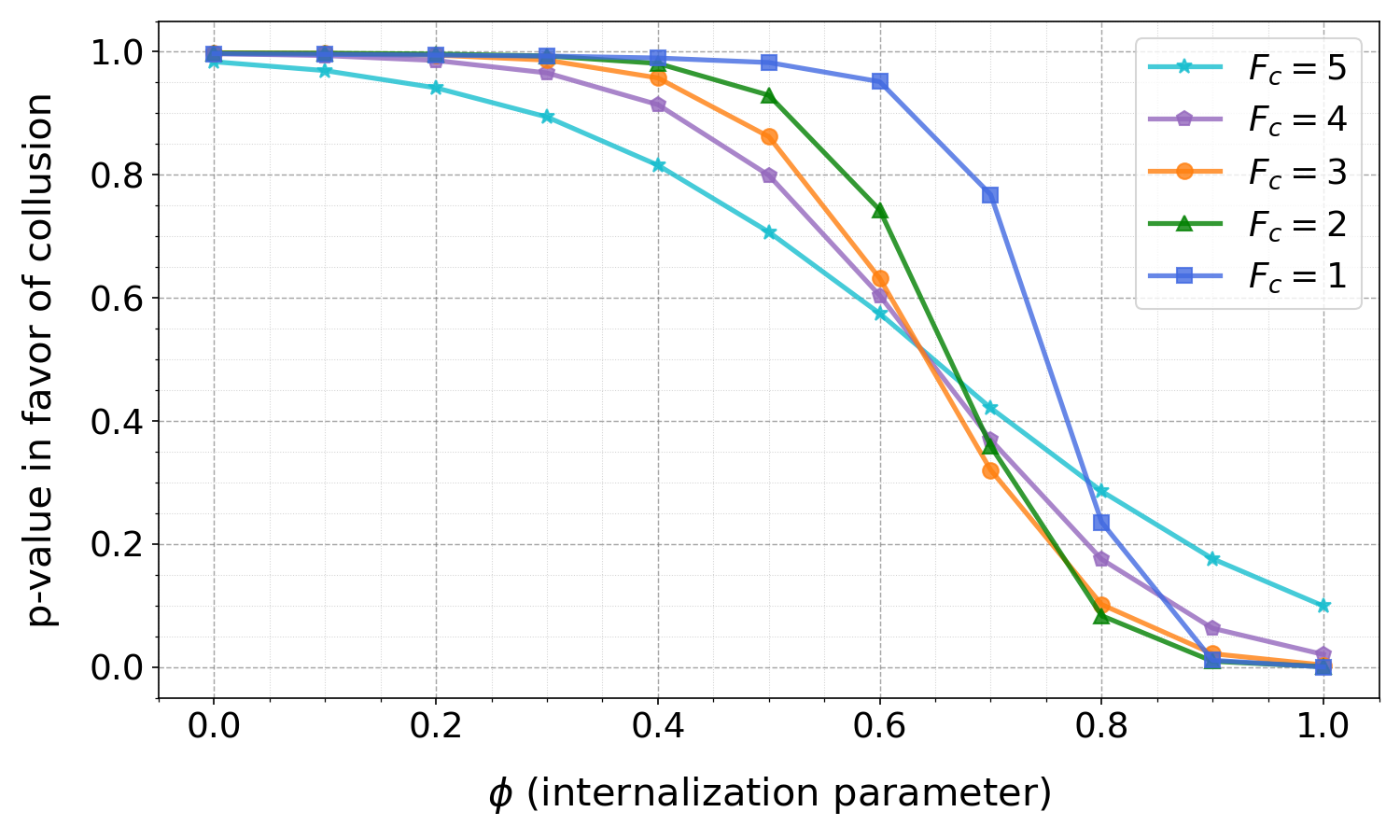}
    	\end{subfigure}
\label{fig: p-value iv vuong}
\tablenotes The figure shows the median p-value of $T_{IV}^{RV}$ under the alternative hypothesis, $H_2: Q_{comp} > Q_{coll}$, across 500 simulated datasets for each Monte Carlo configuration ($J=36, F, T=100, \phi, F_c$). The left panel illustrates the results for the utility specification without a random coefficient, while the right panel presents the results with a random coefficient.
\end{figure}
\clearpage

\newpage
\clearpage

\newpage
\clearpage

\newpage
\clearpage
\end{document}